%% file: TwoCjet2D_HD_Acc2_arxiv.tex
\documentclass{aa} 
\pdfoutput=1 
\usepackage{txfonts}
\usepackage{graphicx}
\usepackage{natbib}
\usepackage[dvipsnames]{xcolor}

\bibpunct{(}{)}{;}{a}{}{,}

\newcommand{\FIG}[1]{}
\usepackage{float}
\usepackage{dblfloatfix}
\usepackage[normalem]{ulem} 
\usepackage{color} 
  \definecolor{gray}{rgb}{0.6,0.6,0.6}
  \definecolor{green}{rgb}{0,0.6,0}

\begin{document}
\title{Shocks in relativistic transverse stratified jets} 
\subtitle{A new paradigm for radio-loud AGN}

  \titlerunning{Relativistic two-component jets}
  \authorrunning{O. Hervet, Z. Meliani et al.}
  \author{O. Hervet \inst{1}$^a$, Z. Meliani \inst{2}$^b$, A. Zech\inst{2}, C. Boisson\inst{2}, V. Cayatte\inst{2}, C. Sauty\inst{2}, H. Sol\inst{2}}


\institute{
Santa Cruz Institute for Particle Physics and Department of Physics, University of California at Santa Cruz, Santa Cruz,
CA 95064, USA\\
\and
 LUTH, Observatoire de Paris, PSL Research University, CNRS, Universit\'e Paris Diderot, Sorbonne Paris Cit\'e, 5 place Jules Janssen, 92195 Meudon, France\\
  $^a$\email{ohervet@ucsc.edu}\\  
  $^b$\email{Zakaria.Meliani@obspm.fr}
 }

\date{Received 08 March 2017 / Accepted 18 May 2017}

\abstract{
The transverse stratification of active galactic nuclei (AGN) jets is suggested by observations and theoretical arguments, as a consequence of intrinsic properties of the central engine (accretion disc + black hole) and external medium. 
On the other hand, the one-component jet approaches are heavily challenged by the various observed properties of plasmoids in radio jets (knots), often associated with internal shocks.
Given that such a transverse stratification plays an important role on the jets acceleration, stability, and  interaction with the external medium, it should also induce internal shocks with various strengths and configurations, able to describe the observed knots behaviours.}
{By establishing a relation between the  transverse stratification of the jets, the internal shock properties, and the multiple observed AGN jet morphologies and behaviours, our aim is to provide a consistent global scheme of the various AGN jet structures.
}
{
Working on a large sample of AGN radio jets monitored in very long baseline interferometry (VLBI) by the MOJAVE collaboration, we determined the consistency of a systematic association of the multiple knots with successive re-collimation shocks. We then investigated the re-collimation shock formation and the influence of different transverse stratified structures by parametrically exploring the two relativistic outflow components with the specific relativistic hydrodynamic (SRHD) code AMRVAC.}
{
We were able to link the different spectral classes of AGN with specific stratified jet characteristics, in good accordance with their VLBI radio properties and their accretion regimes. High-frequency synchrotron peaked BL Lacs, associated with advection-dominated accretion flow (ADAF) discs, are consistent with the simulations of a very weak outer jet ($<\,1\%$ of the total energy) and reproduce stationary equal-distance internal shocks damped within a short distance from the central object. Flat spectrum radio quasars, associated with standard discs (optically thick, geometrically thin), have properties that are nicely reproduced by simulations of two jet components with similar energy fluxes by presenting a very strong first internal shock. Finally, low- and intermediate-frequency synchrotron peaked BL Lacs, associated with hybrid discs (ADAF in the centre and standard outer), are characterized by simulations with a relatively powerful outer jet enhancing inner stationary shocks near the core with moving shocks at large distance, accompanied by an increase in the jet aperture angle.
}
{%
}

\maketitle

\section{Introduction}

There is  growing evidence of transverse stratification of relativistic astrophysical jets with clear indication of a fast inner jet (spine) embedded in a slower outer flow (layer). 
In many active galactic nuclei (AGN) a limb-brightened jet morphology is observed on pc scales 
\citep{Edwardsetal00,Giovanninietal01,Girolettietal04a},
 which is interpreted as an outcome of the differential Doppler boosting between the jet spine and layer \citep{Giovanni03}. 
This scenario is supported by multiple radio-loud AGN observations, such as the well-known radio galaxy M87 which presents hints of a two-component jet from its polarized emission (\citet{Attridgeetal99} ; \citet{Perlmanetal99}) and from the detection of different motions from inner and outer jets \citep{Mertens&Lobanov16}.
In addition, the high-energy observations of AGN indicate the presence of emission processes implying a transverse structure of jet with a fast spine and a slow layer \citep{Ghisellinietal05,Hardcastle06,Jesteretal06,Jesteretal07,Siemiginowskaetal07,Kataokaetal08,Hervet_2015}.

Other astrophysical jets reveal similar structures, such as X-Ray binaries, where the multi-wavelength observations attest of different jet velocities component with an ultra-relativistic core along an extended mildly relativistic flow \citep{Fenderetal04,Migliarietal05}. 
This scenario is supported by theoretical arguments and numerical simulations addressing the physics of jet launching and collimation mechanisms \citep{Soletal89,Laing96,BogovalovTsinganos01, Meier03,Melianietal06a,Melianietal06b,Ferreiraetal06, Hardee07,Mizunoetal07,
Aloy&Mimica08,Fendt09,Matsakosetal09, Porth&Komissarov15}.

Numerous standing and moving radio knots in AGN jets have been observed over the last decades in very long baseline interferometry (VLBI) thanks to dedicated long-term monitoring programs such as MOJAVE 
or TANAMI. 
From these observations, stationary knots are often interpreted as re-collimation shocks resulting from the propagation of overpressured super-Alfvenic jets through the external medium. 
This pressure difference between the jet and the external medium is caused by the large distances covered by relativistic AGN jets in the galactic medium. Indeed, with distance the external medium pressure decreases faster than the jet pressure, which gives rise to rarefaction waves and shock waves within the jet.  
This phenomena was studied using the characteristic methods (\cite{Daly&Marscher88}; \cite{Nalewajkoetal11}, and also numerically in the hydrodynamical case \citep{Falle91,Gomezetal97, Komissaor&Falle97,Mimicaetal09}, and in the magnetohydrodynamical case \citep{Mizunoetal15}.
Other recent works on this field can also be highlighted, such as the study of the interaction between the travelling shock waves with these re-collimation shocks by \cite{Frommetal16}, the effect of the external medium on jets surrounded by shear flow by \cite{Porth&Komissarov15}, or the effect of transition on the external medium disc wind-cocoon by \cite{Tchekhovskoy&Bromberg16}.

Most of the internal shock models adopt   a uniform one-component jet. 
Such models are able to reproduce strings of stationary equal-distance shocks, as proven by \cite{Gomezetal97} and \cite{Mizunoetal15}. However, the diversity of the kinematic behaviour of knots is such that this one-component approach is insufficient. In particular, it cannot explain  the change in knot strength with distance or the moving knots observed in numerous jets \citep{Listeretal13, Jorstadetal05, Cohenetal14}.
As suggested in \cite{Hervet_2016}, taking into account the radial structure of jets gives a key to understanding the observation of knot properties.

In order to study the effects of the jet transverse stratification, we elaborate a  two-component jet model  according to the jet formation scenarios \citep{BlandfordZnajek77,BP82,Sauty&Tsinganos94,Melianietal06a} and numerical simulation \citep{Koideetal00,Nishikawaetal05,McKinney&Blandford09}. We adopt a two-component jet model with  various kinetic energy flux distributions between the  inner-outer jets. We aim to investigate how this energy distribution  influences the overall jet stability, the re-collimation shock development, and the local jet acceleration. Also, we assume various configurations. The first case is an overpressured uniform jet propagating in the external medium. In all the other investigated cases, we set  a transverse structured jet with an overpressured inner jet. This assumption results from the intrinsic properties of the launching region, as described above. 
Moreover, the transverse stratification of the jet increases the pressure difference between the jet and the external medium. 
Indeed, the contrast between the two components can  induce  a pressure difference  between them. The inner jet is fast and is also isolated from the external medium by the outer jet. Moreover, the inner component can be subject to an energy deposit that sustains its pressure  \citep{Soletal89}. 
All these configurations are used to study the effects of the energy distribution between the two components  on the shock states.


We also investigate how the jet transverse stratification affects the efficiency of the rarefaction-acceleration mechanism to re-boost the flow.
According to the jet transverse stratification this mechanism increases the jet Lorentz factor considerably  and could explain the wide variety of Lorentz factors deduced from AGN jets observations. 
The rarefaction-acceleration mechanism is well established in uniform jets and wind \citep{Aloy&Rezzolla06,Sapountzis&Vlahakis13,Sapountzis&Vlahakis14}. This mechanism is triggered when the external medium pressure drops faster than the jet pressure and develops a locally powerful flow acceleration, as has been observed in many studies \citep[e.g.][]{Mizunoetal08,Tchekhovskoyetal10, Komissarovetal10}.

The paper is  organized as follows. In Sect.~\ref{Sec:Link_Obs} we use radio VLBI observations to study the association of stationary radio knots with multiple re-collimation shocks, following theoretical assumptions. Working on a wide blazar sample we also study the link between blazar spectral classes with a change of the inner-jets apertures.
In Sect.~\ref{s-model} we  discuss the settings of the two-component model from the observed properties of AGN jets and we introduce the numerical configurations used. 
In Sect.~\ref{Sec:NS_Results}, we present the parameter study, concentrating our effort on analysing the kinetic/pressure aspect of jets, which we discuss in Sect.~\ref{Sec:Discussion}. In Sect.~\ref{Sec:ImproveClassAGN}, a global scheme linking AGN spectral classes, stratified jet structure, radio knot kinematics, and accretion regime is elaborated.
Throughout this paper we consider a non-structured small-scale magnetic field which contributes to the total pressure.

\section{Jet structures from radio VLBI observations}\label{Sec:Link_Obs}

 Radio VLBI imaging, whose resolution reaches the fraction of milliarcseconds, is currently the most efficient tool for studying the dynamics and the geometrical structures of AGN jets. These VLBI measurements have been proven highly valuable in  recent years; they  strengthen the association of the jet  overdensities, called knots, with a phenomenon of re-collimation shocks such as for BL~Lacertae (\cite{Cohenetal14} and \cite{Gomez_2016}) or M87 \citep{Hadaetal13}. 
These associations of peculiar knots in peculiar sources are still lacking  a global view, which could answer some long-standing questions, for example whether all the knots are re-collimation shocks or whether they are tracers of various different processes, such as Kelvin-Helmholtz instabilities, shocks induced by various flow injection speeds, shocks with the external medium, {etc}.
A global association of re-collimation shocks with all the VLBI radio knots observed in AGN jets implies several specific constraints on the jet structure  inconsistent with other explanations, which can be observationally checked. 

This section is hence dedicated to testing this association in a wide sample of VLBI jets studied by the MOJAVE collaboration, where the knot size and dynamics are described for many years of monitoring \citep{Listeretal13}.

\subsection{Consistency of multiple stationary shocks structures}
Assuming that stationary knotty radio structures correspond to multiple successive re-collimation shocks, the  positions of the knots in jets should not be randomly distributed. In the re-collimation shock scenario, the gap between two successive knots is constrained by the radius of the jet and by the reflection angle of shock waves on the jet edge. 
Since this angle is depending on the Mach number, some theoretical assumptions can be considered here. 
We consider an isothermal conical expending jet with a constant sound speed $v_s$,

\begin{equation}
v_{\rm s}(Z) \, \propto  \, \sqrt{\frac{p(Z)}{\rho(Z)\,h(Z)}}\,=\,{\rm cst}\,,
\end{equation}where  $Z$ is the distance from the core with $p(Z)$, $\rho(Z)$ and $h(Z)$ the profiles of pressure, density and enthalpy respectively. We consider the density profile such as $\rho(Z) \propto Z^{-2}$. In the following we assume a non-structured magnetic field $B$ contributing to the jet dynamics as the main effective pressure. From radio polarimetric observations, the magnetic field strength profile can be written as $B(Z) \propto Z^{-1} $\citep{Gabuzda_2014}. Hence, the pressure shows a profile $p(Z) \propto B(Z)^2 \propto Z^{-2}$ consistent with the isothermal approximation.
We also consider  the enthalpy $h=1+e+p/\rho$, where $e$ is the specific internal energy, as constant.

From these considerations, the Alfv\`en speed $v_A$ follows a steady state along the jet propagation
\begin{equation}\label{Eq:Alfven_speed}
v_{\rm A}(Z) \, \propto  \, \frac{B(Z)}{\sqrt{\rho(Z)\,h(Z)+B^2(Z)}}\,=\,{\rm cst}\,.
\end{equation}

\begin{figure}[t!]
\begin{center}
	\includegraphics[width= 9.0cm]{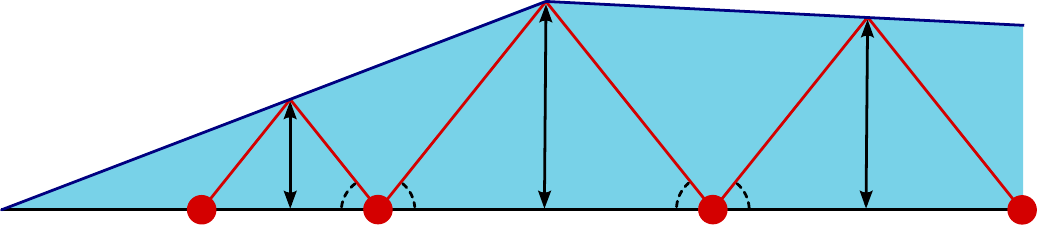}
 		\put(-209,-9){\makebox(0,0)[lb]{$k_1$}}
 		\put(-165,-9){\makebox(0,0)[lb]{$k_2$}}
 		\put(-82,-9){\makebox(0,0)[lb]{$k_3$}}
 		\put(-7,-9){\makebox(0,0)[lb]{$k_4$}}
 		\put(-187,-4){\makebox(0,0)[lb]{$x_1$}}
 		\put(-123,-4){\makebox(0,0)[lb]{$x_2$}}
 		\put(-44,-4){\makebox(0,0)[lb]{$x_3$}}
 		\put(-192,12){\makebox(0,0)[lb]{$r_1$}}
 		\put(-130,25){\makebox(0,0)[lb]{$r_2$}}
 		\put(-50,25){\makebox(0,0)[lb]{$r_3$}}
 		\put(-176,5){\makebox(0,0)[lb]{$\beta$}}
 		\put(-152,5){\makebox(0,0)[lb]{$\beta$}}
 		\put(-93,5){\makebox(0,0)[lb]{$\beta$}}
 		\put(-69,5){\makebox(0,0)[lb]{$\beta$}}
 		\caption{Geometric scheme of a jet assuming that radio knots at positions $k_{\rm n}$ are stationary re-collimation shocks with an incident angle of shock waves $\beta$ that is constant.}
 		\label{Fig::reflexion_chocs}
 \end{center}
\end{figure}
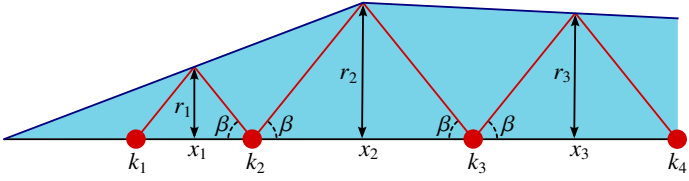

Admitting an average constant flow speed, the Mach number is also constant, leading to a fixed wave-shock incident angle $\beta$, and thus an inter-knot distance only  depending on the jet radius, as shown in Fig.~\ref{Fig::reflexion_chocs}.
For this scenario we consider the relativistic Mach number, defined as 
\begin{equation}\label{Eq:Mach}
{\cal M} = \frac{u}{u_s},
\end{equation}
with $u = \gamma v$ the proper speed of the fluid, $u_s = \gamma_s v_s$ the proper speed of the sound relative to the fluid, and $\gamma = 1 / \sqrt{1 - v^2}$ the associated Lorentz factor \citep{Konigl_1980}.

From this scenario, we obtain the relation
\begin{equation}
\tan{\beta} \, = \, \frac{r_1}{x_1 - k_1} \, = \, \frac{r_2}{x_2 - k_2} \, =\, {\rm cst}\,,
\end{equation}
leading to a direct proportionality between the jet radius and the inter-knot gaps,
\begin{equation}
r_n \, \propto \, \Delta k_n\,.
\end{equation}

\begin{figure*}[t!]
	\begin{minipage}[b]{0.33\linewidth}
   		\centering \includegraphics[width=6.2cm]{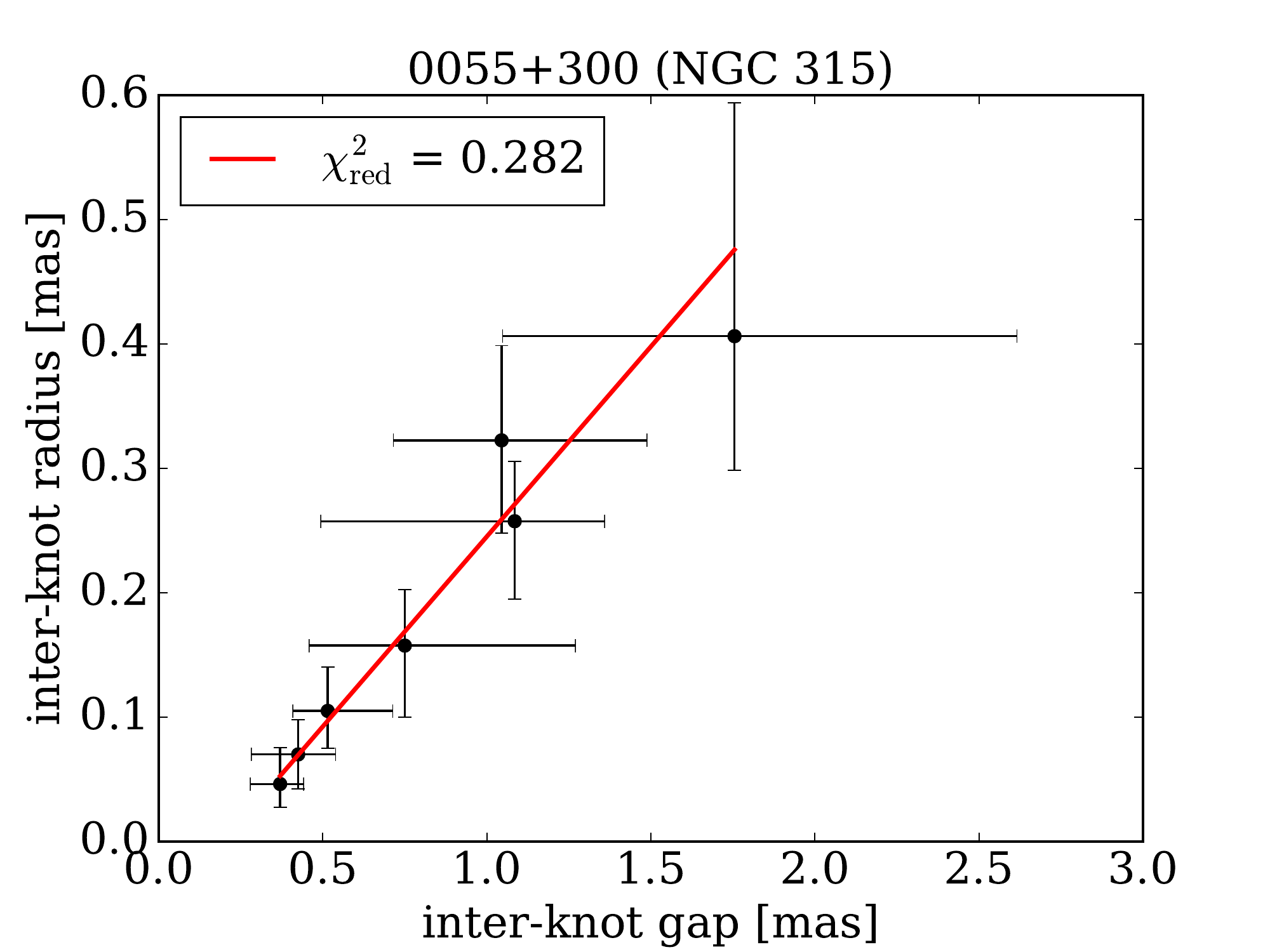}
	\end{minipage}\hfill
		\begin{minipage}[b]{0.33\linewidth}
      \centering \includegraphics[width=6.2cm]{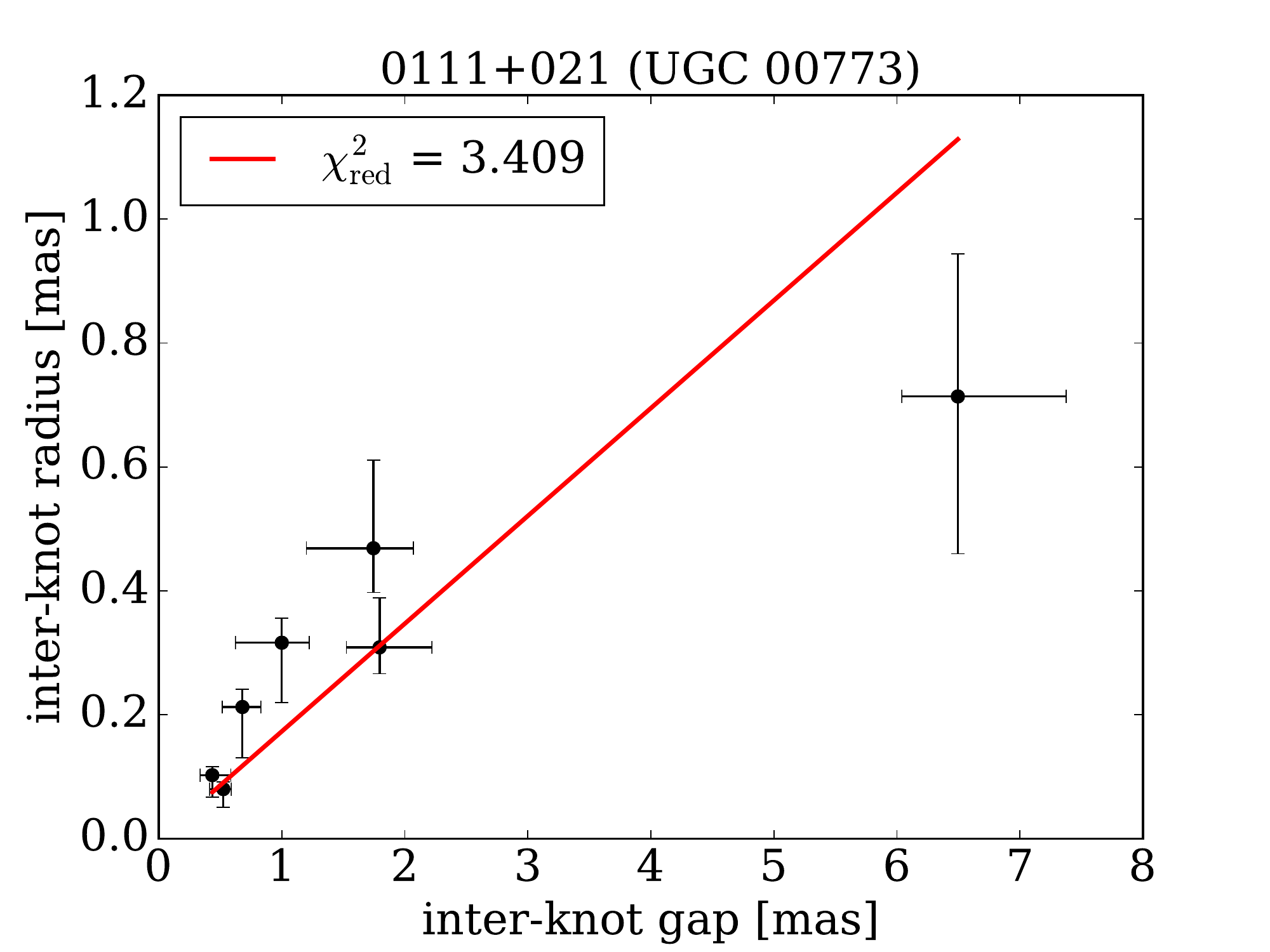}
	\end{minipage}\hfill
	\begin{minipage}[b]{0.33\linewidth}
      \centering \includegraphics[width=6.2cm]{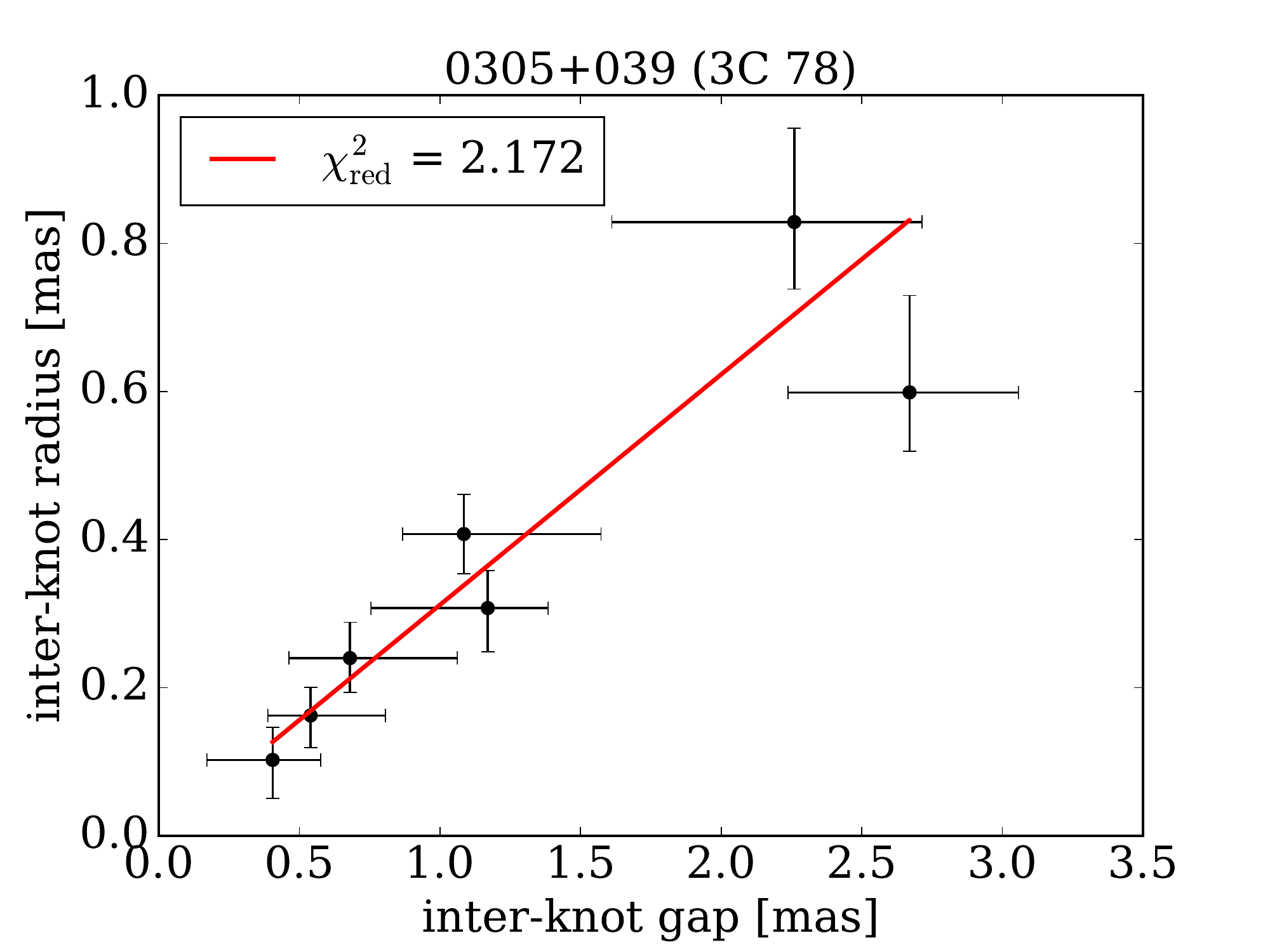}
	\end{minipage}\hfill
	\begin{minipage}[b]{0.33\linewidth}
      \centering \includegraphics[width=6.2cm]{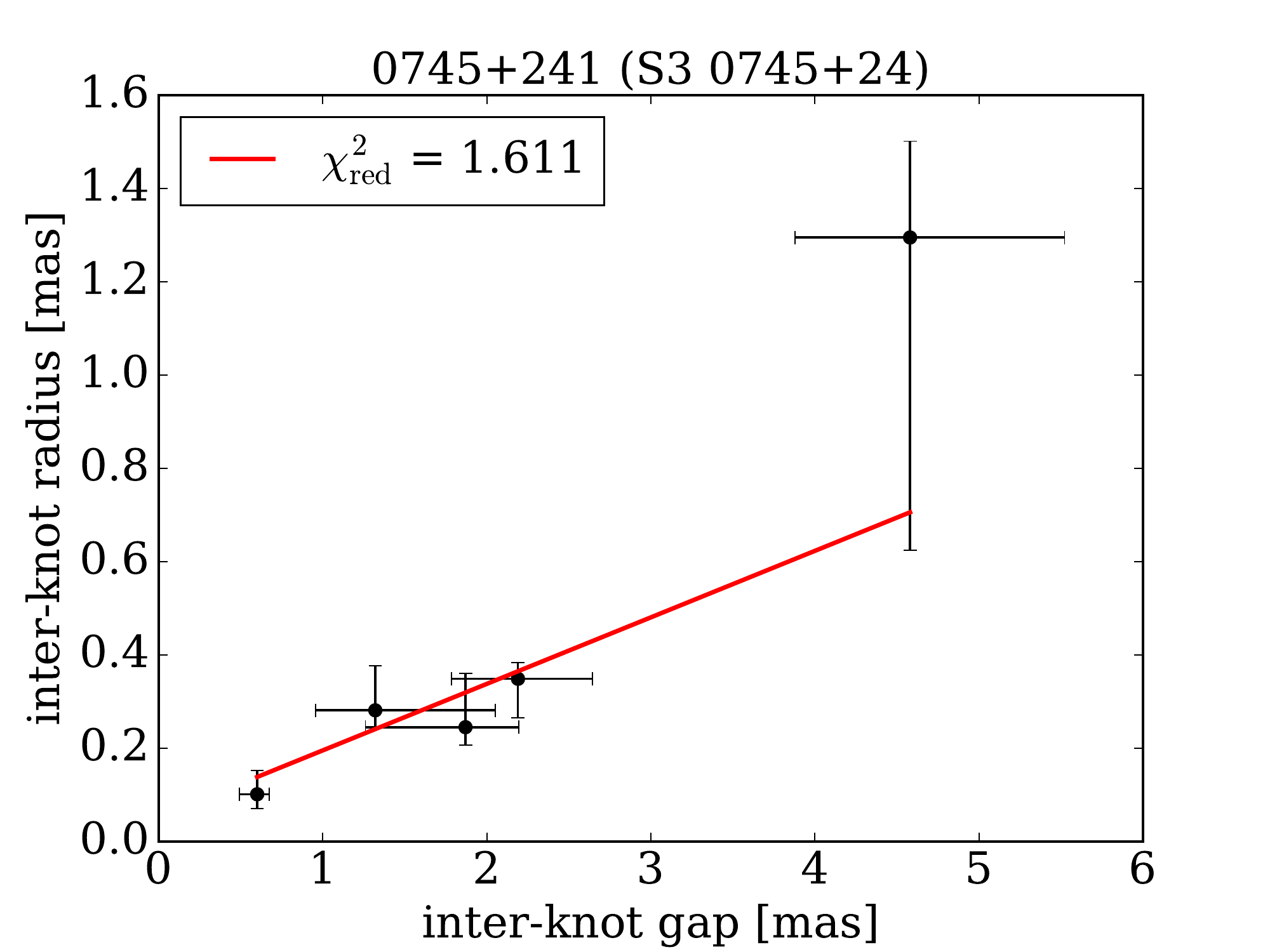}
	\end{minipage}\hfill
	\begin{minipage}[b]{0.33\linewidth}
   		\centering \includegraphics[width=6.2cm]{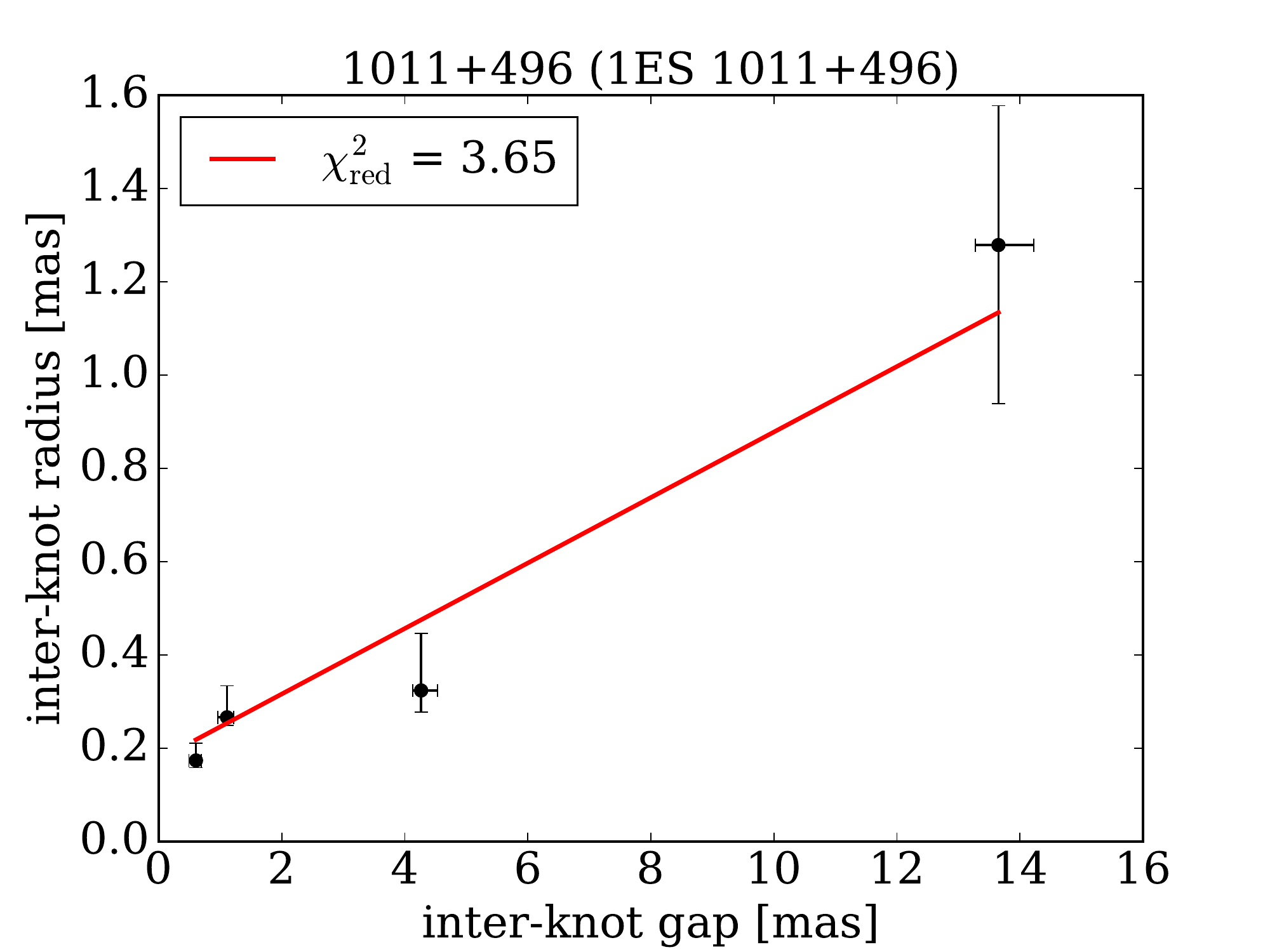}
	\end{minipage}\hfill
	\begin{minipage}[b]{0.33\linewidth}
      \centering \includegraphics[width=6.2cm]{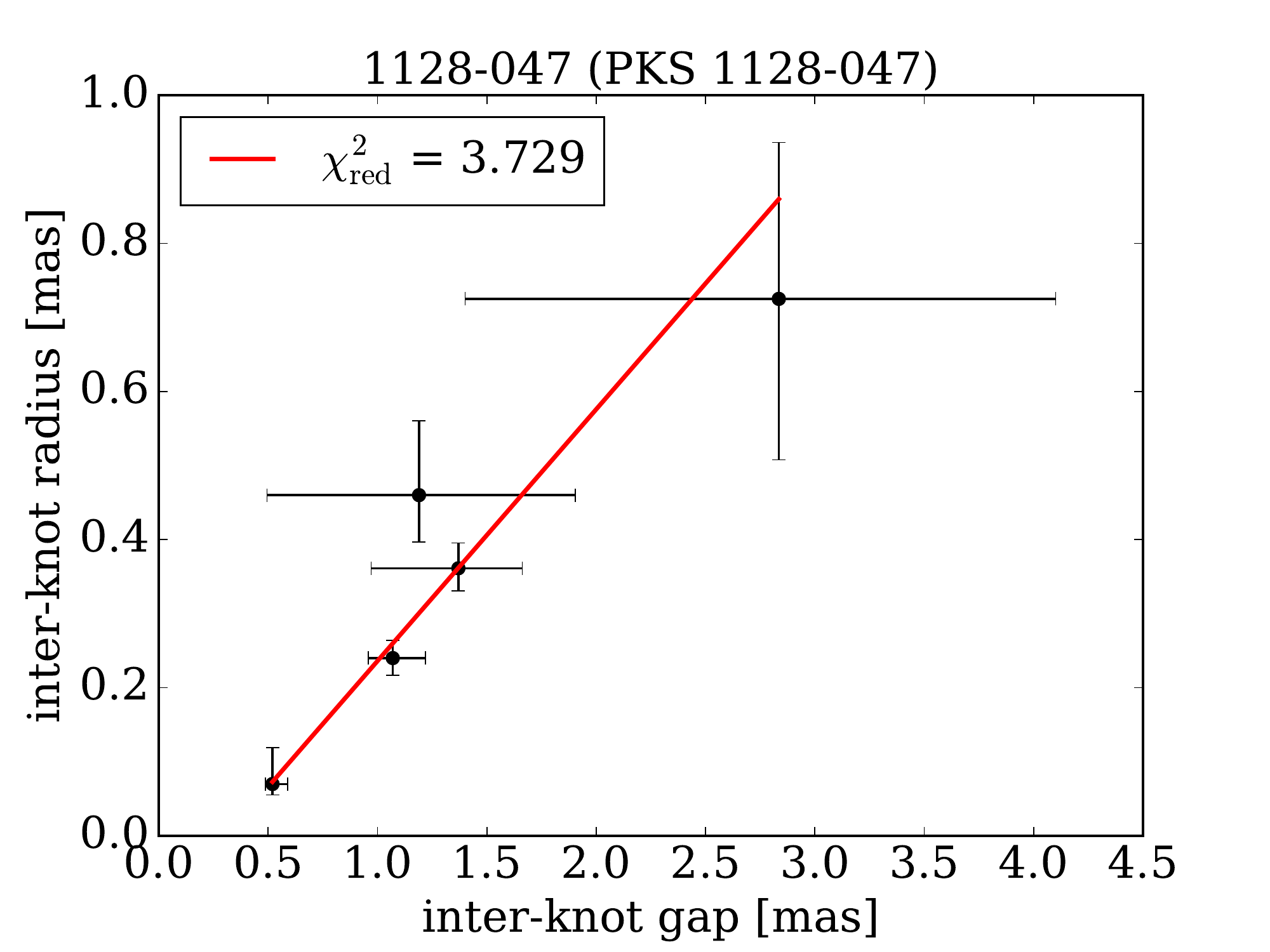}
	\end{minipage}\hfill	
	\begin{minipage}[b]{0.33\linewidth}
      \centering \includegraphics[width=6.2cm]{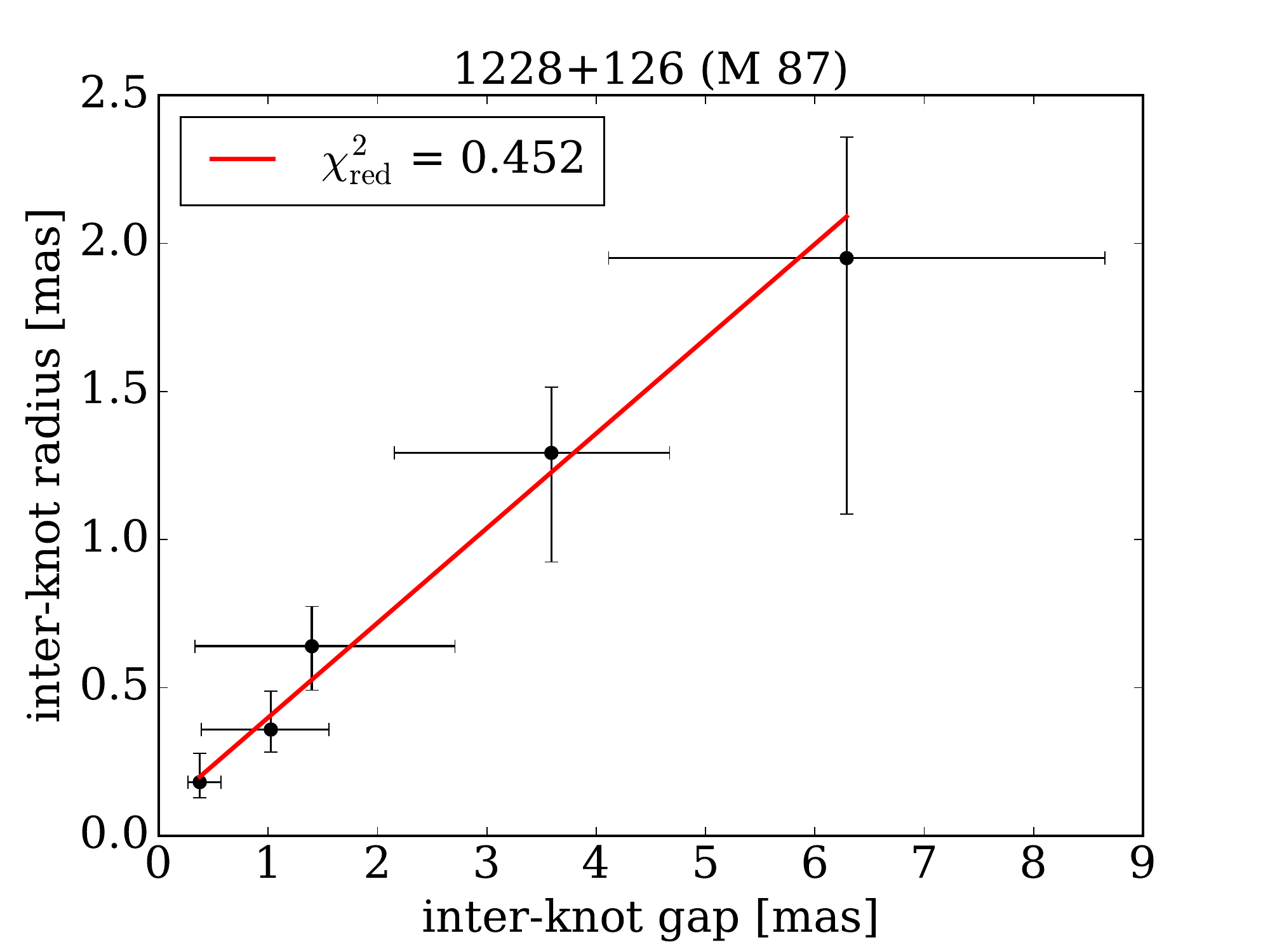}
	\end{minipage}\hfill
    \begin{minipage}[b]{0.33\linewidth}
      \centering \includegraphics[width=6.2cm]{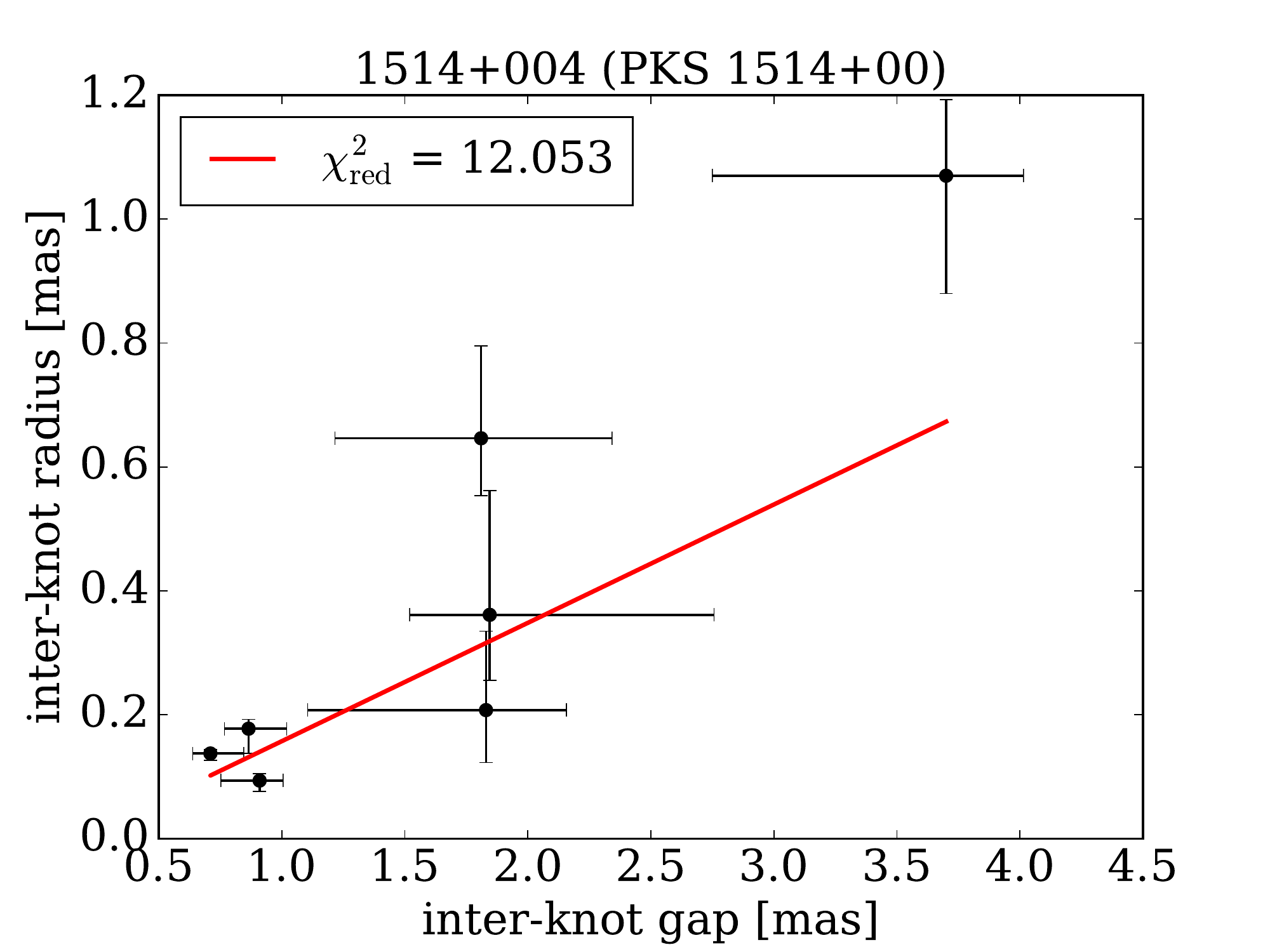}
	\end{minipage}\hfill
	\begin{minipage}[b]{0.33\linewidth}
      \includegraphics[width=6.2cm]{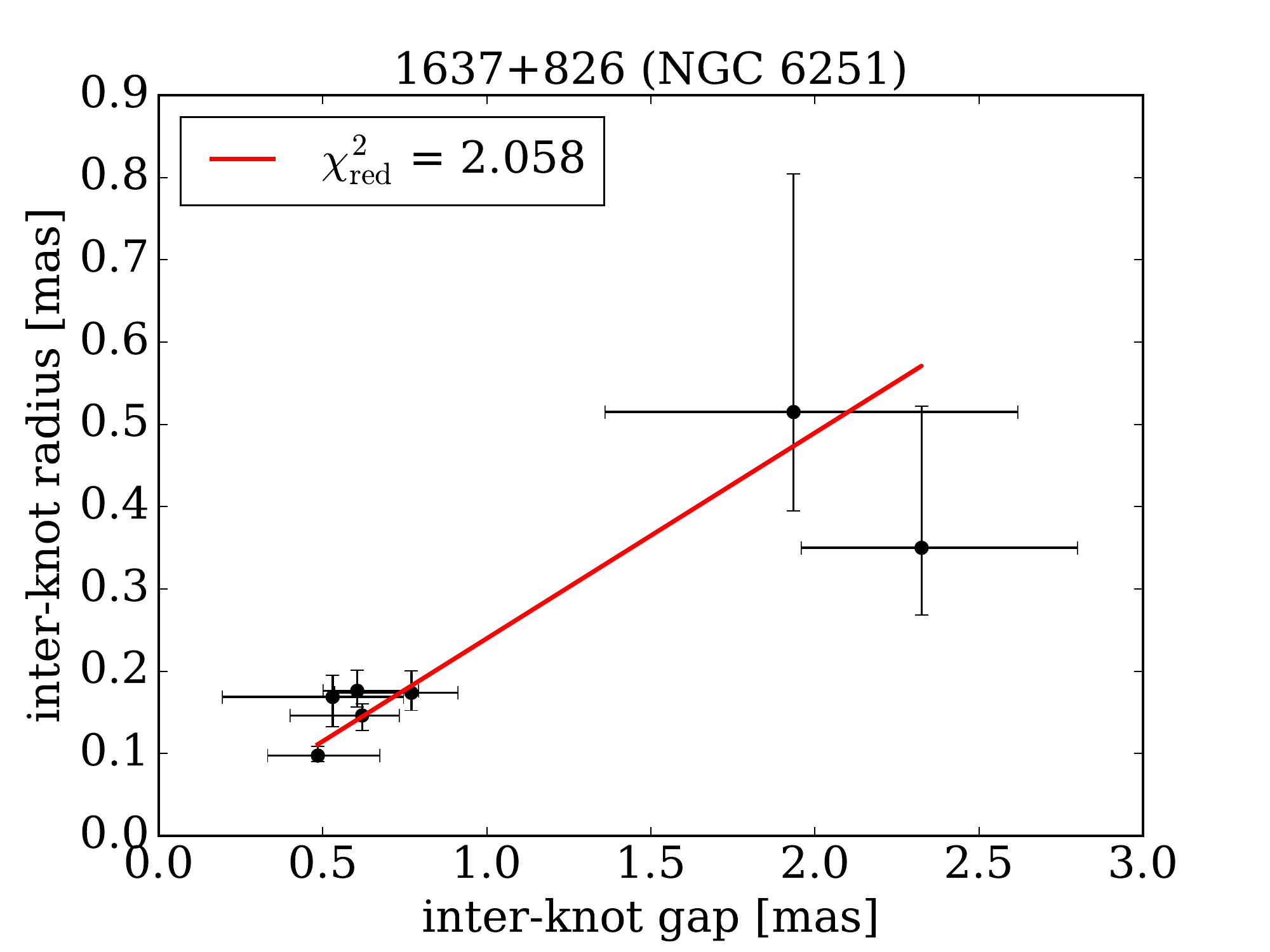}
	\end{minipage}\hfill
	\begin{minipage}[b]{0.33\linewidth}
      \includegraphics[width=6.2cm]{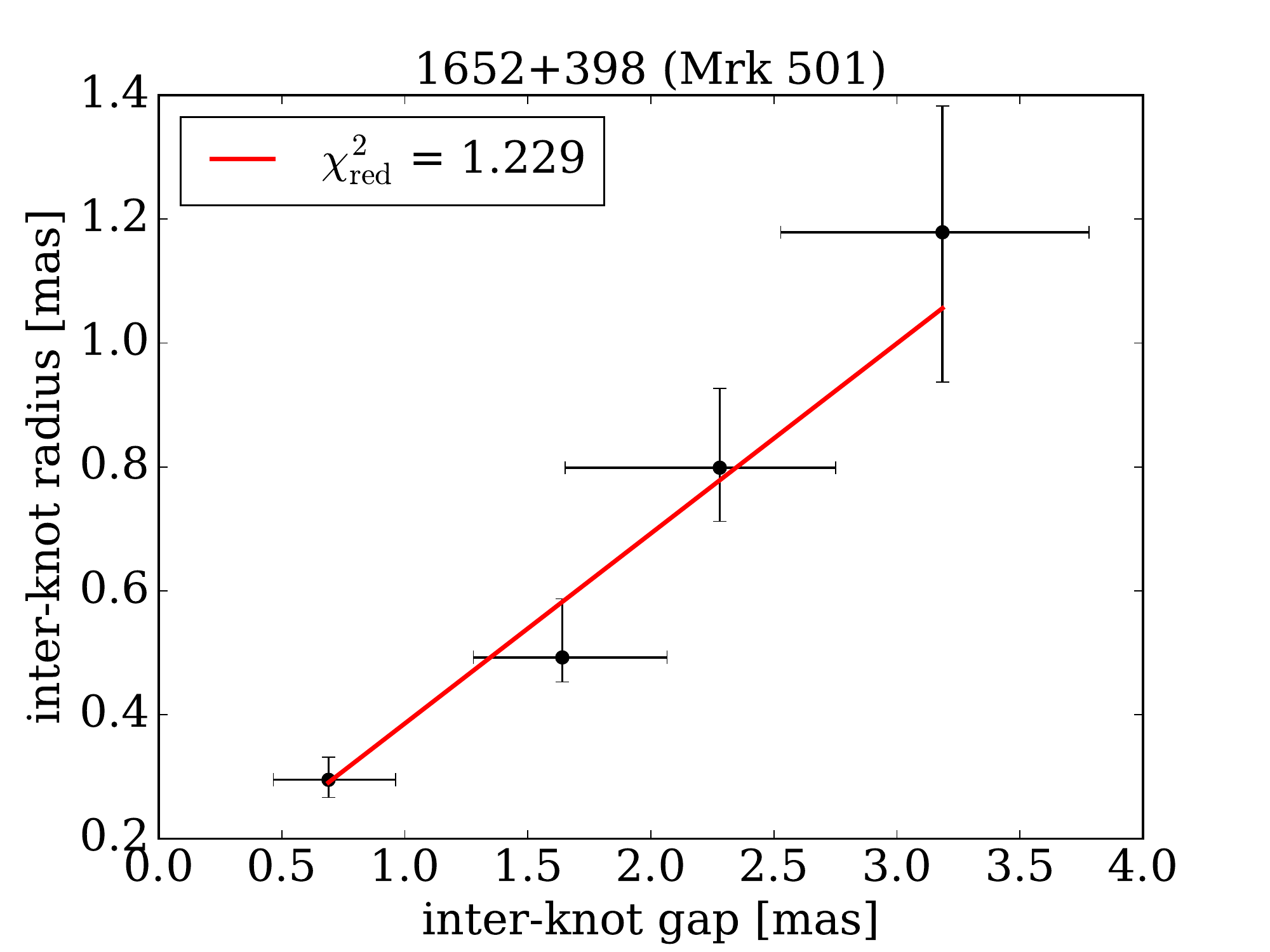}
	\end{minipage}\hfill
	\begin{minipage}[b]{0.33\linewidth}
      \centering \includegraphics[width=6.2cm]{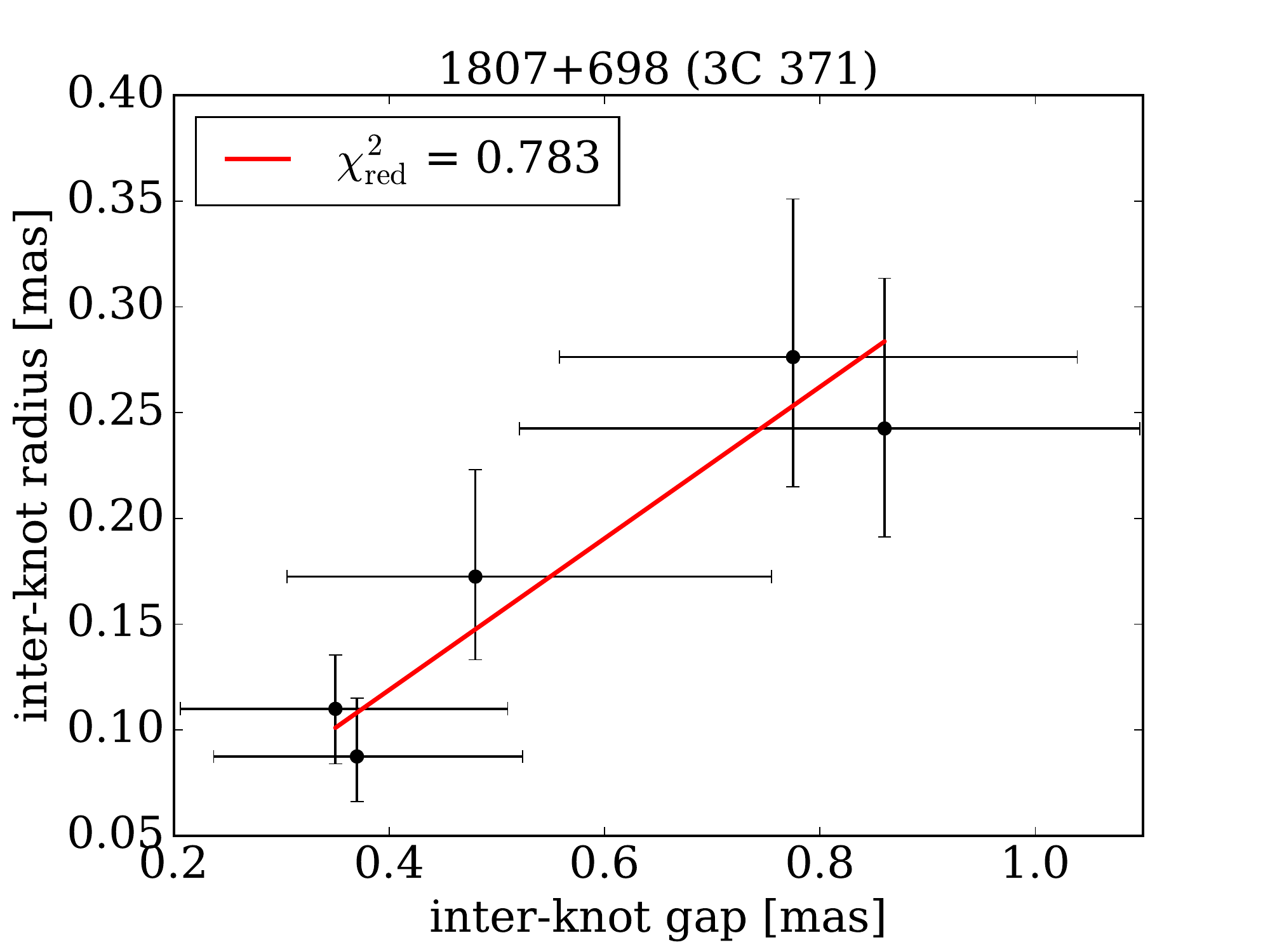}
	\end{minipage}\hfill
	\begin{minipage}[b]{0.33\linewidth}
      \centering \includegraphics[width=6.2cm]{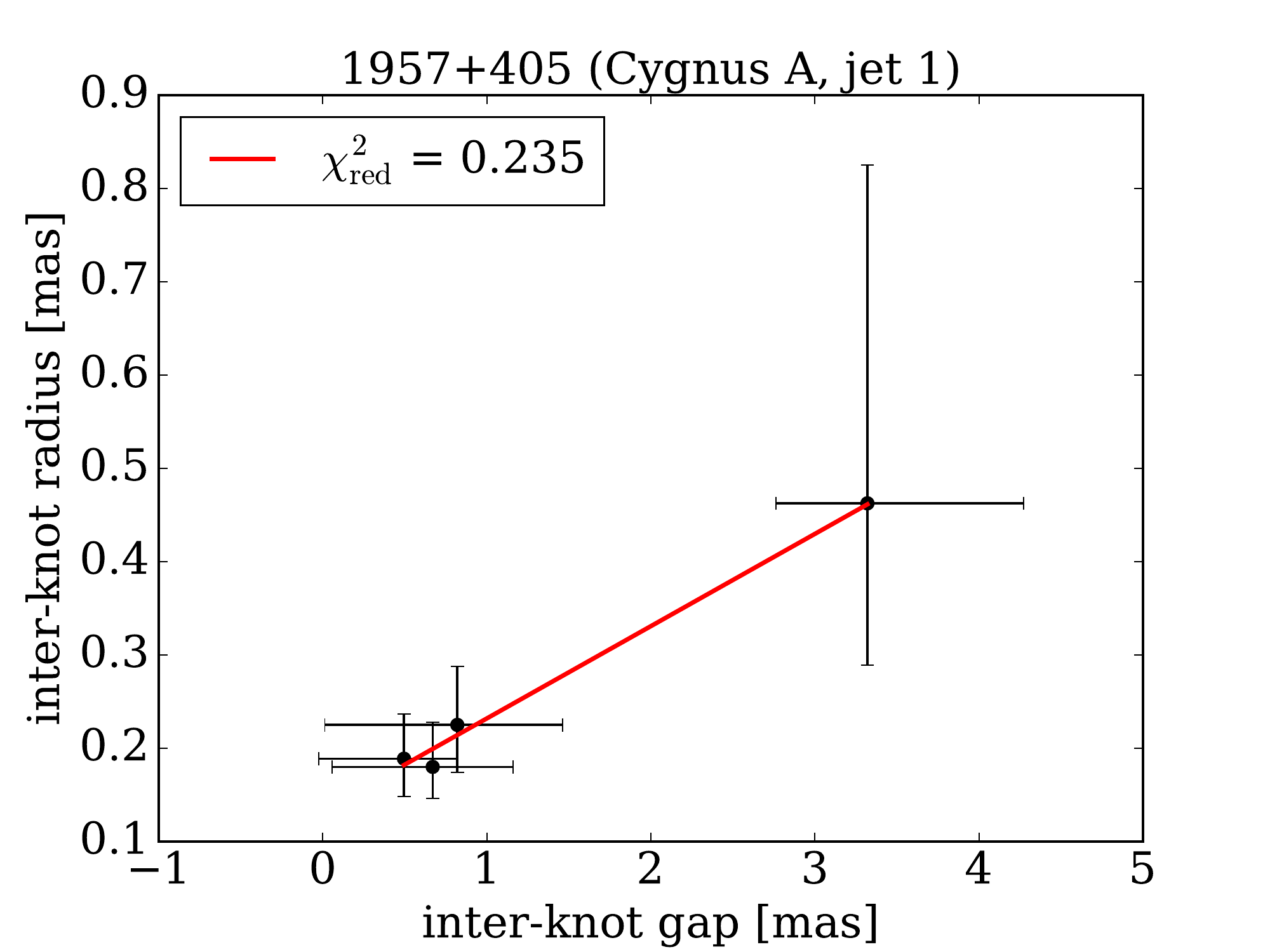}
	\end{minipage}\hfill
	\begin{minipage}[b]{0.33\linewidth}
      \centering \includegraphics[width=6.2cm]{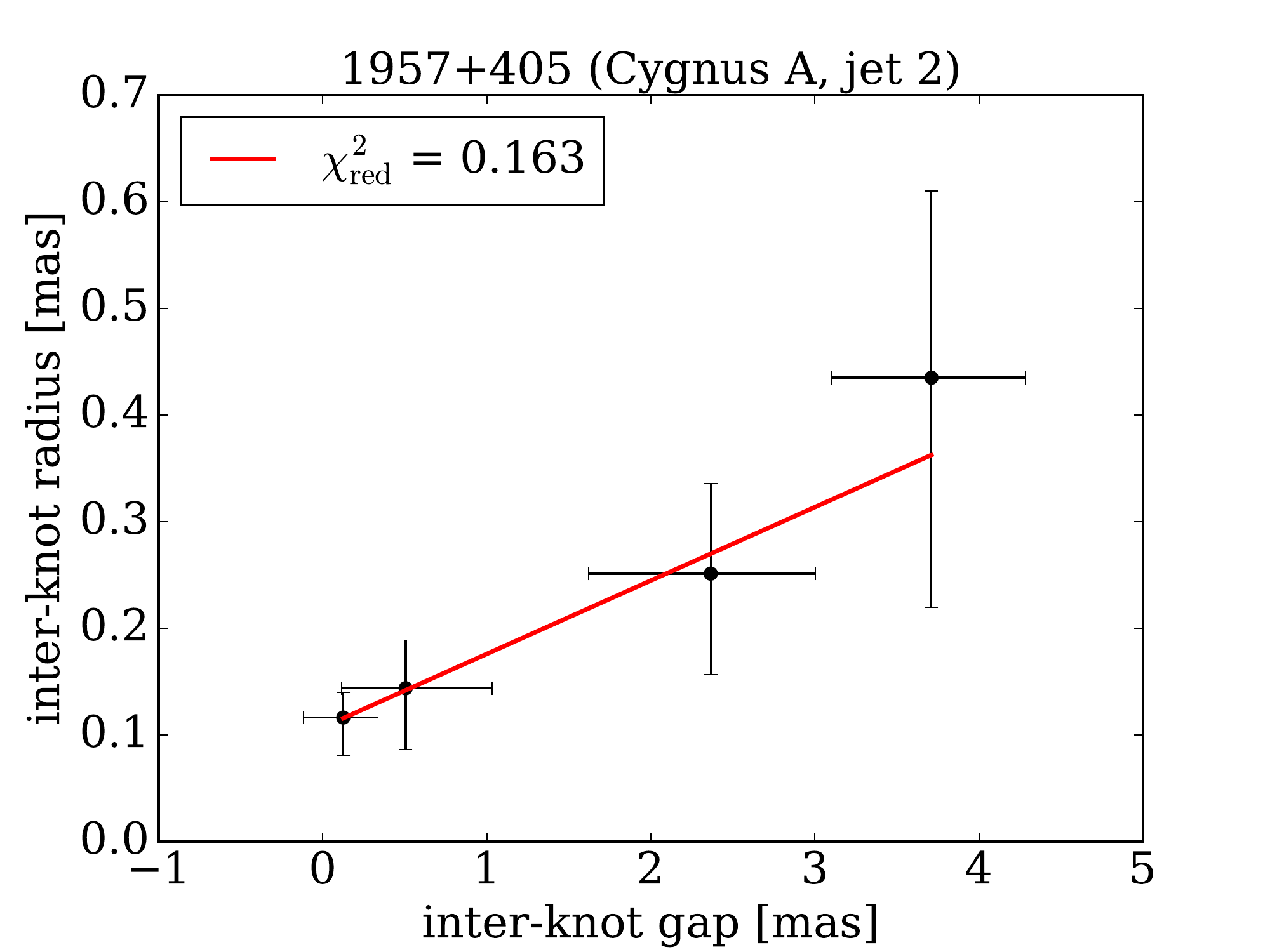}
	\end{minipage}\hfill
	\caption{Linear regression on the radio VLBI inner-jet radius associated with inter-knot gaps for all jets showing at least five slow moving ($V_{app} < 2$ c) or stationary knots referenced in \protect\cite{Listeretal13}. A linear evolution is expected when considering radio knots as successive re-collimation shocks.}
	\label{Fig:shock_scenario}
\end{figure*}
Thus, it is now possible to compare this proportionality relation considering a successive re-collimation shock scenario with the radio VLBI observations.
To achieve this, we select all jets showing at least five slow moving ($V_{app} < 2$ c) or stationary knots referenced in \cite{Listeretal13}, leading to a sample of 13 jets. 

The radius of the jet where the shock waves are reflected is defined as the half-FWHM of symmetrical 2D Gaussians fitted to the  radio flux of the knots. The knot sizes and inter-knot gaps are chosen at the median of the measures with uncertainties set at the 90th percentile around the median values. 
When multiple knots of various epochs have similar positions in the jet, we choose the one presenting the best temporal coincidence with the other monitored knots.

The proportionality relation is hence tested by performing a linear regression on these data. We can see in Fig.~\ref{Fig:shock_scenario} that the majority of the jets studied  indeed agree well with the prediction of the successive re-collimation shocks scenario.
Only the source PKS 1514+00 presents a significantly bad fit with the peculiarity of having multiple similar inter-knot gaps associated with various radius.
This could be associated with curved jets or complex propagations beyond the approximations used for this study.

The slopes of linear fits $a_{app}$ performed in Fig.~\ref{Fig:shock_scenario} allow us to constrain the physical parameters of the jets.
In order to do so, we need to use the intrinsic inter-knots gaps which can be deduced only knowing the jet angles with the line of sight. Thus, for this study we consider only the case of M87, which has a defined angle with the line of sight $\theta = 15^{+4}_{-5}$ deg \citep{Wang_2009}, and a previously measured apparent slope of $a_{app} = 0.32 \pm 0.09$. Hence, the jet Mach angle $\beta$ can be written as
\begin{equation}
\beta = \arctan \left( \frac{2 \, a_{app}}{\sin{\theta}} \right).
\end{equation}

The Mach number being related to the Mach angle as
\begin{equation}
\label{Eq:Mach_angle}
{\cal M} = \frac{1}{\sin{\beta}} ,
\end{equation}
we can deduce a Mach number ${\cal M} = 1.08_{-0.06}^{+0.15}$ for the inner jet of M87. 
 Equation~(\ref{Eq:Mach_angle}) is Lorentz invariant in the observer frame \citep{Konigl_1980}, and so there is no need to have a flow speed estimation to determine a Mach number value.
The Mach number is usually a free parameter, manually tuned for jet hydrodynamical simulations. With basic geometrical assumptions, this original method allows us to deduce a Mach number value associated with each astrophysical jet as soon as a viewing angle is defined and enough quasi-stationary VLBI knots are detected. Hence, the majority of the HD jet simulations performed in Sect.~\ref{s-model} assume a weakly supersonic flow for the inner jet, consistent with this value from M87.

\subsection{Probing a jet aperture change in intermediate blazars}
\label{Sec:Aperture of VLBI inner jets}

\begin{figure*}[t!]
\begin{center} 
\begin{minipage}[b]{0.33\linewidth}
 \centering \includegraphics[width=6.2cm]{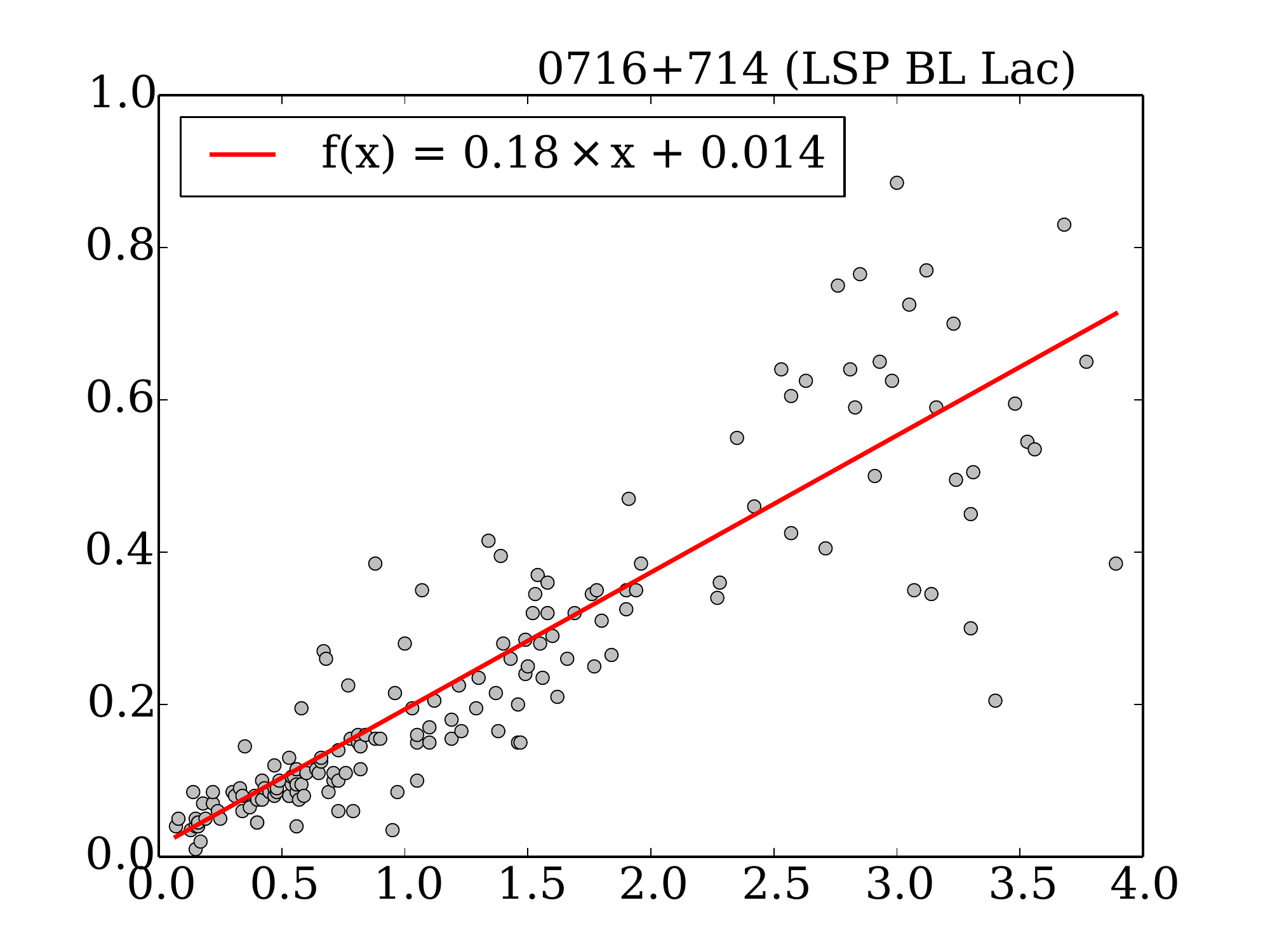}
\end{minipage}\hfill
\begin{minipage}[b]{0.33\linewidth}
 \centering \includegraphics[width=6.2cm]{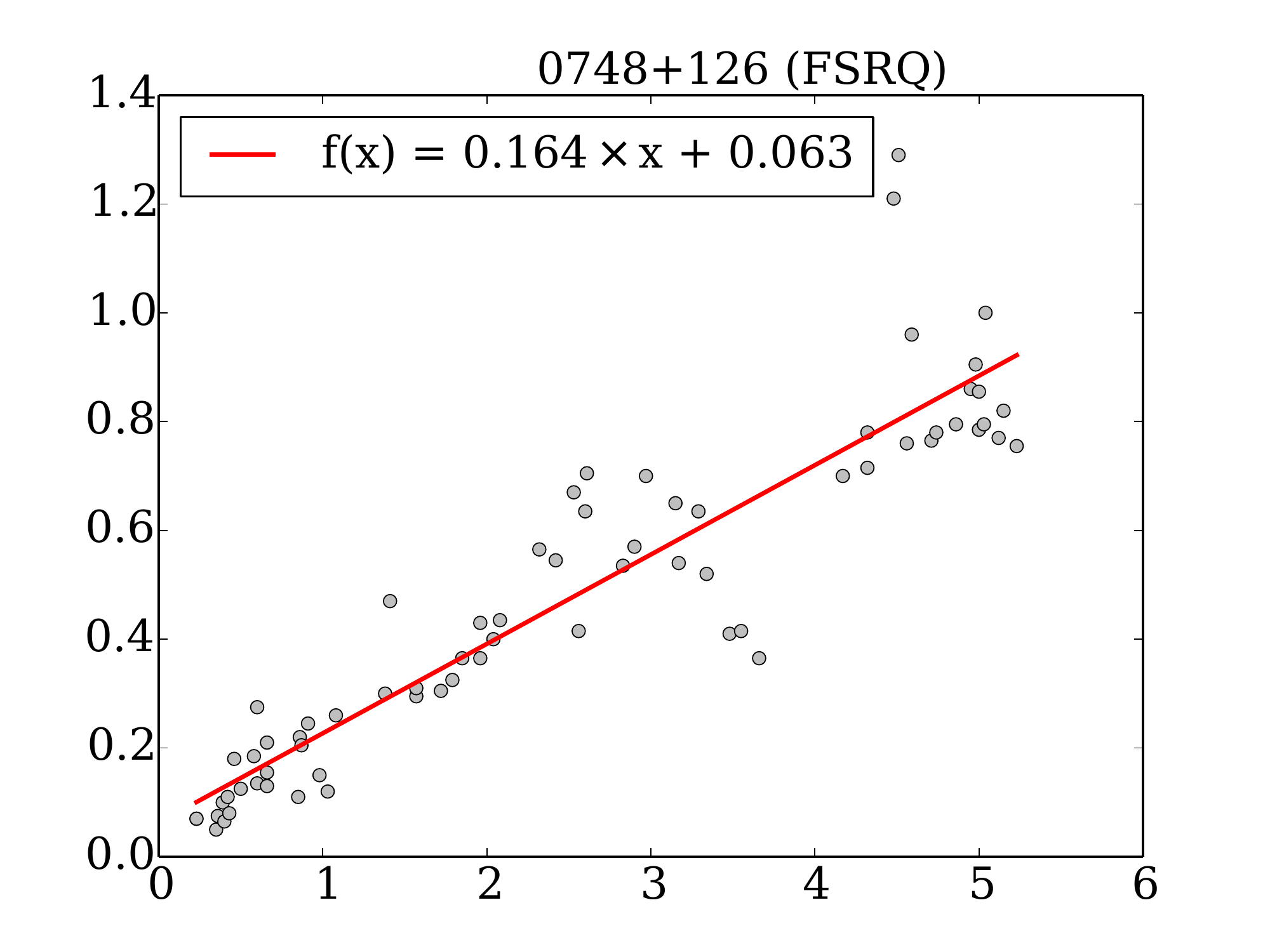}
\end{minipage}\hfill
\begin{minipage}[b]{0.33\linewidth}
 \centering \includegraphics[width=6.2cm]{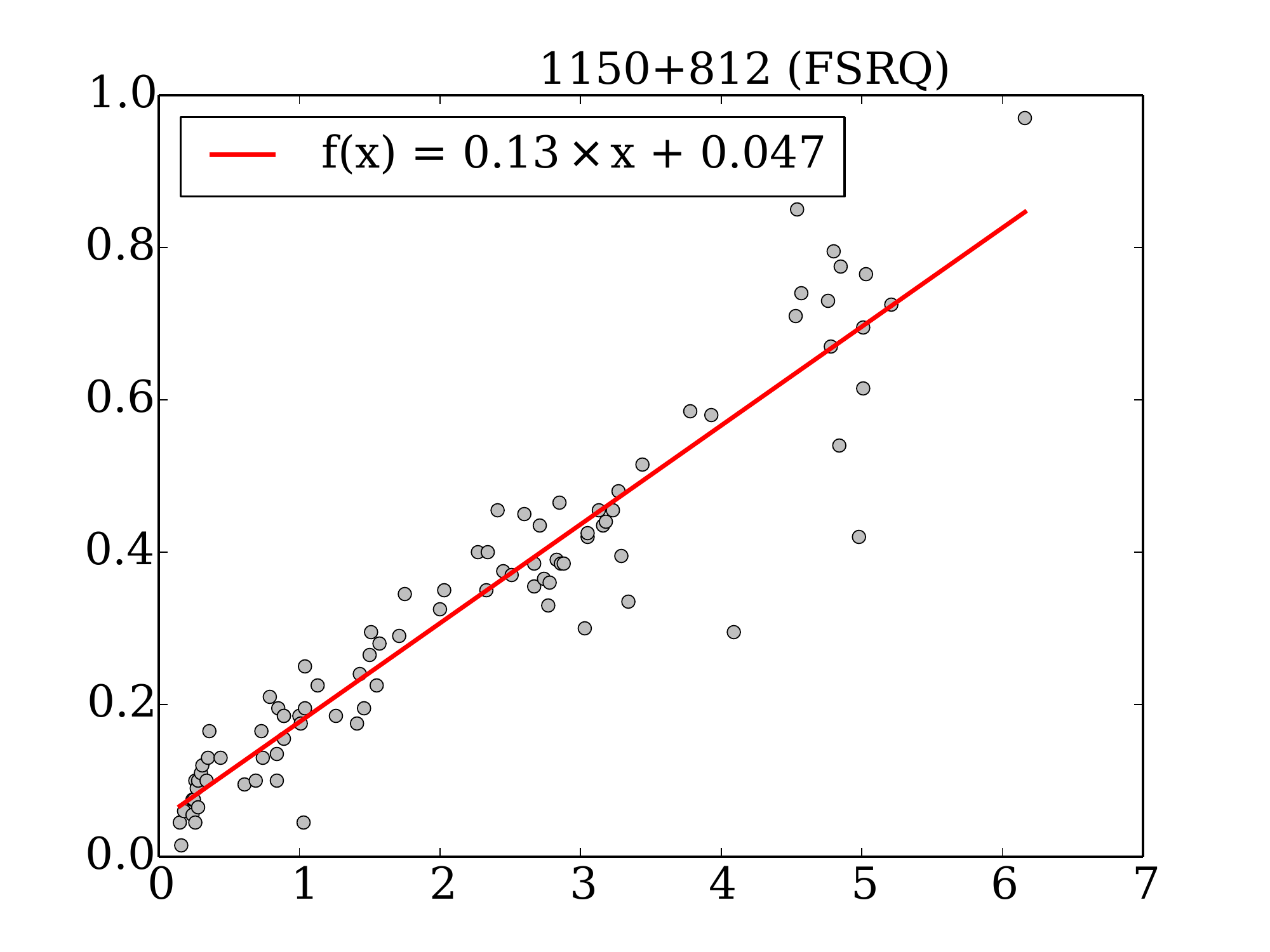}
\end{minipage}\hfill
\begin{minipage}[b]{0.33\linewidth}
 \centering \includegraphics[width=6.2cm]{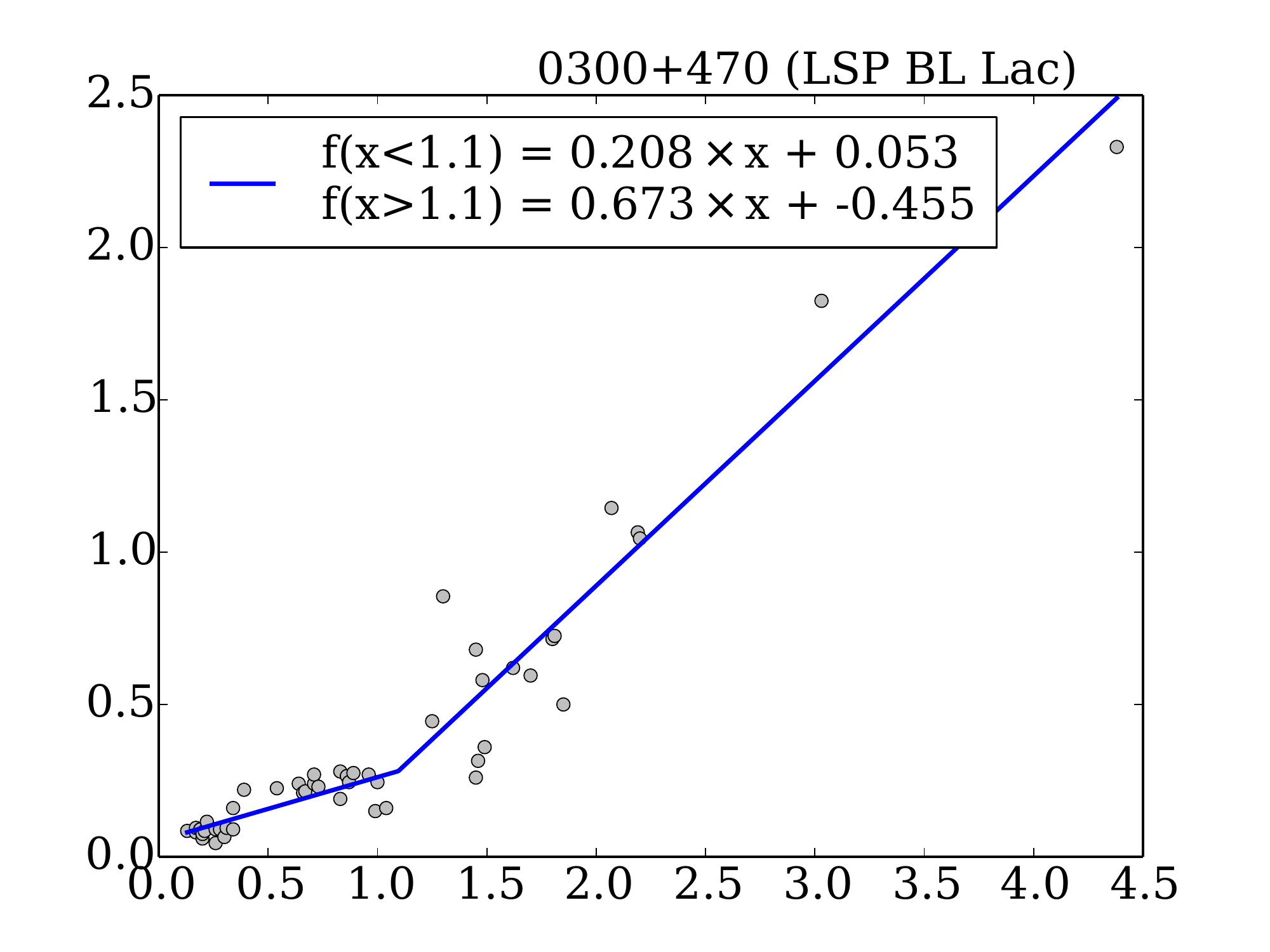}
\end{minipage}\hfill
\begin{minipage}[b]{0.33\linewidth}
 \centering \includegraphics[width=6.2cm]{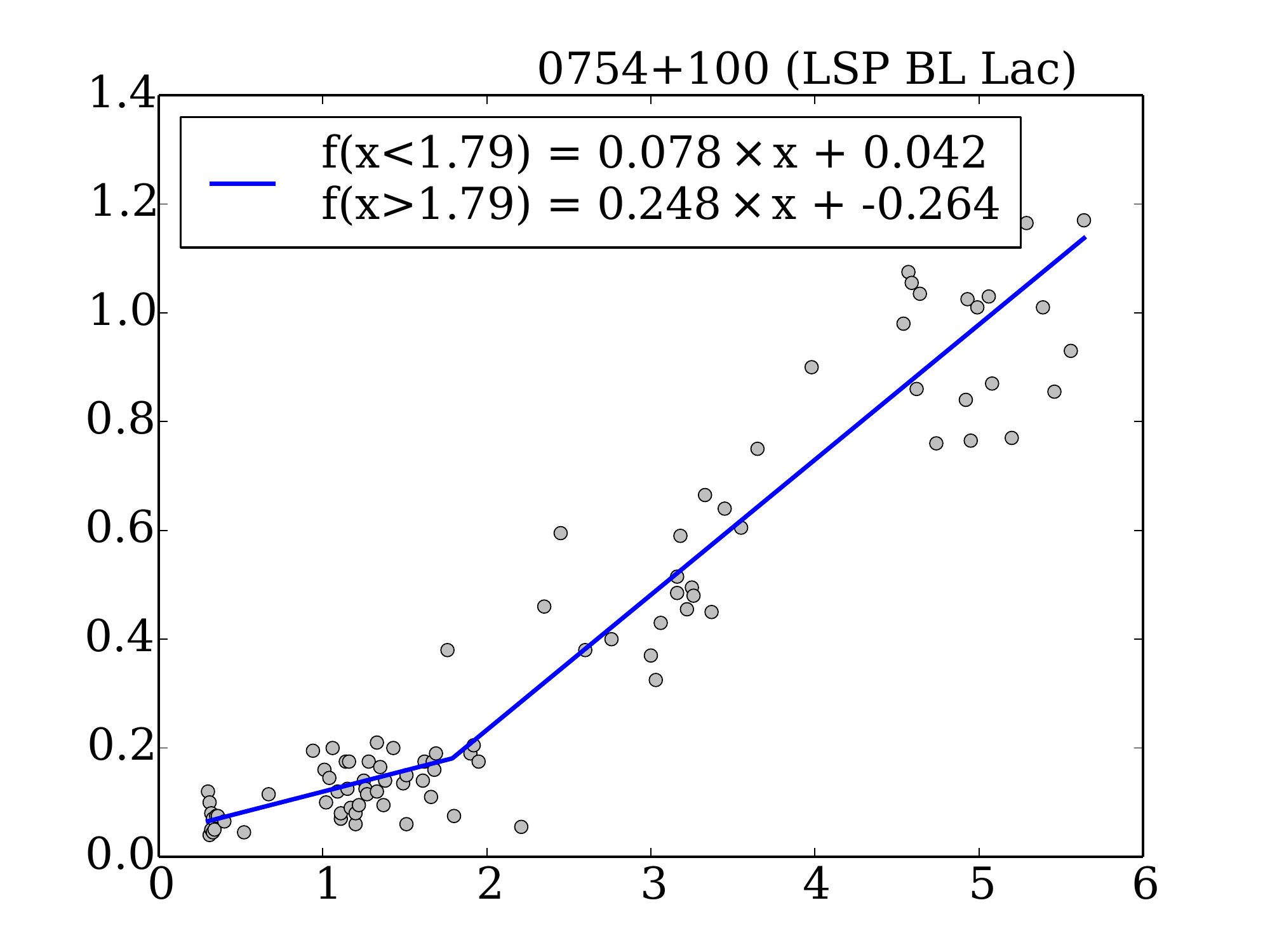}
\end{minipage}\hfill
\begin{minipage}[b]{0.33\linewidth}
 \centering \includegraphics[width=6.2cm]{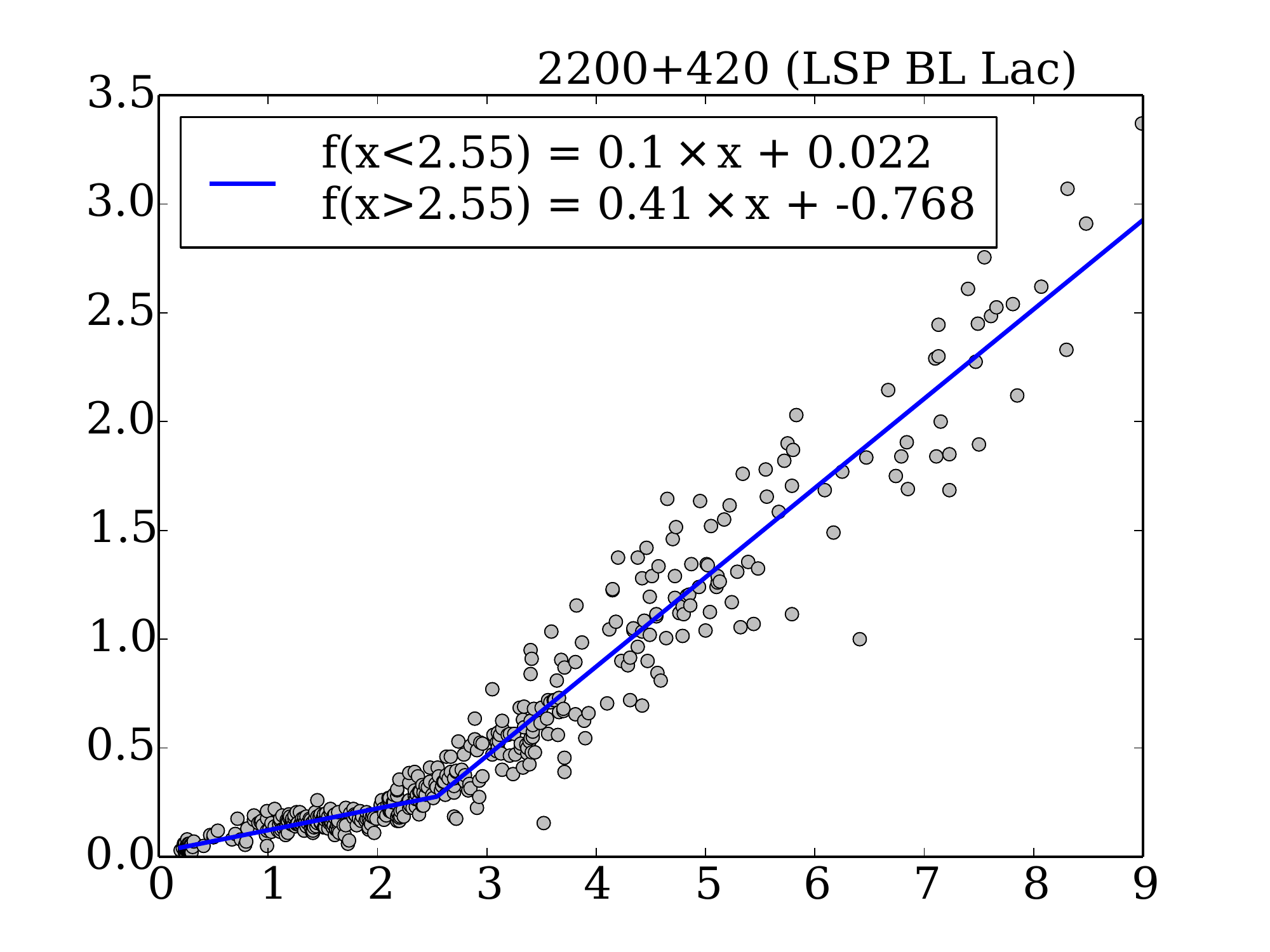}
\end{minipage}\hfill
 \put(-520,90){\rotatebox{90}{\makebox(0,0)[lb]{\large{Transverse size [mas]}}}}
 \put(-300,-10){\makebox(0,0)[lb]{\large{Core distance [mas]}}}
 \caption{Projected transverse size of the knots with their distance from the core for a representative sample of six sources. The solid lines represent the best fit, red for a linear function and blue for a broken linear function. The total sample of jets with a significant aperture increase is presented in Appendix \ref{Annexe::aperture_increase}.}
 \label{Fig:sample_apertures}
\end{center}
\end{figure*}

We call intermediate blazars the low- and intermediate-frequency synchrotron peaked BL Lacs (LBLs and IBLs), which are usually considered at the turnover point between the two main regimes of radio-loud AGN associated with high-frequency synchrotron peaked BL Lacs (HBLs) and flat spectrum radio quasars (FSRQs) \citep{Fossati_1998, Meyer_2011}. It has recently been shown  that in addition to peculiar multiwavelength spectral energy distributions, these sources also mostly have  a  hybrid behaviour of their radio VLBI jet kinematic with quasi-stationary knots close to the core and knots with relativistic motions from a definite zone in jets \citep{Hervet_2016}.
Even if this study does not find any significant global differences in their jet aperture,  a clear aperture increase in the jet of the LBL BL Lacertae was noticed. The fact that this peculiarity was also (and only) observed in the two other intermediate blazars 3C 66A and S5 1803+784 by \cite{Jorstadetal05} allows us to assume a common feature.

In order to check this link between the spectral classification and an increase in the jet aperture, we use the sample of $161$ blazars with defined redshift and apparent knot velocities of \cite{Hervet_2016}. We select then only the sources firmly identified as FSRQ, LBL, IBL, and HBL, reducing the sample to 156 blazars.
The aperture change can be detected by the measure of the knot size (half-FWHM defined in \cite{Listeretal13}) in function of the knots-core distance. For this purpose we use two fitting functions, a linear and a broken linear one.
The broken linear function is defined as
\begin{equation}\label{Eq:fit_fx}
f(x)\,=\, \left \{
\begin{array}{r}
 a_{1}\, x\, +\, b_{1}, ~x \, \leq \, x_{\rm b}\\
 a_{2}\, x\, +\, b_{2}, ~x \, \geq \, x_{\rm b}
 \end{array}
 \right .
 \,,
\end{equation}
with $x_{\rm b}$ the distance to the core where an aperture change occurs, and $a_{2} = a_{1} + (b_{1} - b_{2})/x_{\rm b}$.

First, we reject the sources presenting a determination coefficient $R^{2}\, <\, 0.1$ for both functions, meaning an impossible estimation of the aperture angle. Only two sources are rejected, reducing the sample to $154$ blazars.
Then, the linear broken function is favoured when its significance compared to the simple linear, determined by a F-test, reaches a minimum level of $95\%$.

The farthest knots from the core in VLBI jets can induce underestimated sizes. Indeed, considering adiabatic and radiative cooling of the jet along its propagation,  only the emission from the  centre of the last VLBI knots should be seen. Moreover, a small bend in a blazar jet will induce an apparent  decrease in its aperture due to projection effects.
This is why the end of the VLBI jets is not taken into account for the fitting method if it induces a decrease in the jet apertures.
Also, in order to obtain realistic apertures the fitting functions are constrained to present a positive radius at the core position,  and this results in a population of 36 sources that show a significant increase in jet apertures. We note that the method fails for the source 0355+508, which presents the strange behaviour of apparently quasi-stationary knots, but  important size fluctuations. Hence, we decided  not to consider this source, leading to jet aperture increase observed for 35/154 blazars. A representative sample of six sources with and without aperture increase is given in Fig.~\ref{Fig:sample_apertures}; the full sample of jets with a significant aperture increase and the associated broken-law parameters is given in Appendix \ref{Annexe::aperture_increase}.

While FSRQs and HBLs  respectively present this characteristic in $15\%$ (19/125) and $20\%$ (1/5) of their jets,  intermediate blazars are dominated by this jet aperture increase, which are significantly detected in $63\%$ (15/24) of the population. Due to their lower radio luminosity and less well-defined radio-knots, the HBL sample is very small with only five sources; this value of $20\%$ can thus  fluctuate widely for a much wider sample.

This trend observed in our sample suggests that the increase in jet aperture angle can be seen as a new property of intermediate blazar jets. Knowing that most of these sources  also  present a peculiar hybrid kinematic gives us  valuable information about the AGN jet structure and behaviour.

\section{Two-component jet model setting}\label{s-model}

\begin{table*}
\caption{Most relevant parameters for the two models investigated. In addition to these  values, the external medium has a normalized number density $\rho_{0} = 1$ and a normalized pressure $p_{0} \simeq 5\times 10^{-2}$ or  $p_{0} \simeq 1\times 10^{-3}$ (these two types of pressure are chosen according to the energy ratio between inner and outer jet components). The inner jet always presents  an initial Lorentz factor of $\gamma\,=\,10$, higher than that of the outer jet initialized at $\gamma\,=\,3$. The outer jet is assumed to be in pressure equilibrium with the external medium $\eta_{p,\,\rm out} =1 $, contrary to the inner jet which presents a larger pressure $\eta_{p,\,{\rm in}}=1.5$.}
\label{table:1}      
\centering                                      
\renewcommand{\footnoterule}{}
\begin{tabular}{c c c  c  c c c c c c  }
\hline
 &\multicolumn{1}{c}{external medium}&\multicolumn{2}{c}{inner jet}& \multicolumn{2}{c}{outer jet}&\multicolumn{3}{c}{Structured jet}\\
\hline
case& $p_0$&$\eta_{\rho\;,{\rm in}}$ & ${\cal M}_{\rm c, in}$& $\eta_{\rho\;,\rm out}$ &${\cal M}_{\rm c, out}$&$L_{\rm k,\, in}/L_{\rm k,\, total}$&$L_{\rm k,\, out}/L_{\rm k,\, total}$& Two-component jet\\
\hline
\hline
A&$5\times 10^{-2}$ & $4.5 \times 10^{-4}$ &$1.22 $&  &     &$ 1 $  & $0.0$& No\\
B&  $1\times 10^{-3}$ & $5 \times 10^{-1}$ &$4.34 $&$ 5\times 10^{-6} $ & $1.16 $&$0.95$ & $0.05$& Yes\\
C&  $5\times 10^{-2}$ & $5 \times 10^{-3}$ &$1.22 $&$ 5\times 10^{-1} $ & $16.34 $&$0.70$ & $0.30$& Yes\\
D&  $1\times 10^{-3}$ & $5 \times 10^{-6}$ &$1.22 $&$ 1\times 10^{-1} $ & $6$&$0.25$ & $0.75$& Yes\\
E&  $5\times 10^{-2}$ & $5 \times 10^{-3}$ &$1.22 $&$ 5\times 10 $ & $19.0 $&$5\times 10^{-3}$ & $0.995$& Yes\\
F&  $5\times 10^{-2}$ & $1 \times 10^{-3}$ &$0 $&$ 5\times 10^{-2} $ & $6.0 $&$0$ & $1$& Yes\\
\hline
\end{tabular}
\end{table*}
In the following, the speeds are normalized to light speed.

The observations show agreement between the knots and  the re-collimation shock scenario properties (Sect.~\ref{Sec:Link_Obs}). 
It gives us a strong basis to develop and check this scenario with special-relativistic-hydrodynamical simulations (SRHD), as presented in the following.

To study re-collimation shocks in transverse stratified jets, we adopt a two-component jet model with two uniform components. The model uses the basic characteristics of relativistic AGN jets, such as the total kinetic luminosity flux within an interval
 $L_{\rm k}\,=\left[10^{43}\,,\,10^{46}\right]{\rm ergs/s}$ \citep{Rawlings&Saunders91, Tavecchioetal04}, and the outer radius of the two-component jet $R_{\rm out}\,=\,R_{\rm jet} \sim 0.1 {\rm pc}$ at a parsec scale distance from the black hole.  For the less constrained
inner jet radius, we adopt the initial value $R_{\rm in}\,=\,R_{\rm jet}/3$.
As initial condition for simulations, we establish a cylindrical flow column along the jet axis with a radius $R_{\rm jet}$. Two types of jets are investigated in this paper,  uniform jets (the reference case)  and two-component jets. 
For structured jets, we have a discontinuity in the density, pressure, and velocity at the interface of the two components $R_{\rm in}$. The jet properties are related to the external medium density and pressure by

\begin{equation}\label{Eq:etarho}
\rho=\left\{
\begin{array}{ccr}
\rho_{ \rm 0}\;\; \eta_{\rho,\rm in }& & R \le R_{\rm in}\,,\\
\;\;\rho_{\rm 0}\;\; \eta_{\rho, \rm out}& & R_{\rm in}<R<R_{\rm jet} \,,
\end{array}
\right.
\label{Eq:rhoJet}
\end{equation}
and

\begin{equation}\label{Eq:etap}
p=\left\{
\begin{array}{ccr}
p_{ \rm 0} \;\;\eta_{p,\rm in }& & R \le R_{\rm in}\,,\\
\;\;p_{\rm 0} \;\;\eta_{p, \rm out}& & R_{\rm in}<R<R_{\rm jet} \,.
\end{array}
\right.
\label{Eq:pJet}
\end{equation}
where $\rho_{0}$ and $p_{0}$ are respectively the density and the pressure of the external medium, $\eta_{\rho,\rm in}$ and $\eta_{\rho,\rm out}$ are the inner and outer jet density ratio relative to the external medium density, and $\eta_{p,\rm in}$ and $\eta_{p,\rm out}$ are the inner and outer jet pressure ratio relative to the external medium pressure.

The kinetic energy flux  in each jet component is for the inner component,
\begin{equation}\label{Eq:Lkine}
 L_{\rm k,\, in} = \left(\gamma_{\rm in}\,h_{\rm in}-1\right)\,\gamma_{\rm in}\,\rho_{\rm in}\,R_{\rm in}^2\,\pi\,v_{\rm z,\, in}\,,
\end{equation}
and the outer component,
\begin{equation}
 L_{\rm k, out} = \left(\gamma_{\rm out}\,h_{\rm out}-1\right)\,\gamma_{\rm out}\,\rho_{\rm out}\,\left(R_{\rm out}^2-R_{\rm in}^2\right)\,\pi\,v_{\rm z, out}\,.
\end{equation}
Where $h$ is the specific enthalpy.

The inner jet is assumed to be faster than the outer jet ($\gamma_{\rm in} = 10$), whereas the outer jet has a Lorentz factor of $\gamma_{\rm out} = 3$. This outer jet Lorentz factor is the mean value derived at parsec scale for BL lac objects with the Very Long Baseline Array \citep{Girolettietal04b, Pineretal08}. The pressure ratios  $\left(\eta_{p,\rm in},\eta_{p,\rm out}\right)$ are set to $1.5$ and  $1.0,$ respectively. The density ratios $\left(\eta_{\rho,\rm in},\eta_{\rho,\rm out}\right)$  are chosen such that we can analyse a large fraction of the parameter space. Indeed, we aim to determine the influence of the kinetic energy flux and the Mach number distribution between the two components on the jet behaviours.


 We suppose that the outer jet  converts a large fraction of its thermal energy to kinetic energy near the launching region. However, we assume that the inner component conserves a fraction of its thermal energy. 
{ This is motivated by the fact that the outer jet is believed to be launched from the outer part of the accretion disc where the streamlines expand faster than the ones from the inner jet}. The inner jet is fast, and so we consider that it takes more time to reach pressure equilibrium with the outer jet component and with the external medium.

Five cases (A, B, C, D, E, and F) are investigated. The first case (A) presents a uniform jet with a Lorentz factor $\gamma=10$. All the others cases are two-component jet simulations; their order follows an increasing ratio of the outer/inner jets component kinetic powers. Hence,  case (B) has  the most powerful inner jet, carrying $95\%$ of the total kinetic power.  Cases (C) and (D) are set with inner and outer jet carrying relatively the same order of kinetic power. In  case (E), the jet is set with very powerful outer jet carrying $99,93\%$ of the total kinetic power; this is also the only case where the outer jet is denser than the external medium. In the last case (F), the inner jet is empty and all energy is carried by the outer jet.
All these cases are listed in Table~\ref{table:1}.

\subsection{Numerical method}\label{subs:NM}

The simulations are achieved with the special relativistic version of AMRVAC code \citep{Melianietal07a,vanderHolstetal08} using a Synge-type equation of state \citep{Melianietal04}. The simulations are performed in cylindrical coordinates using  HLLC flux formula \citep{Mignone&Bodo05} and  KOREN limiter \citep{Koren93}. The computational domain of this simulation is a 2D box $\left[0,r_{\rm max}\right]\times \left[0,z_{\rm max}\right]$. The box length  $\left(r_{\rm max}, z_{\rm max}\right)$  is chosen according to the distance between the internal shocks. In all the cases in this paper we chose  ($r_{\rm max} = 24 $ and $z_{\rm max} = 400 $). We take a resolution $80 \times 1312$ and we allow for three levels of refinement, reaching an effective resolution of $320\times 5248$.  We set fixed boundary conditions for the lower one in $z=0$ (at the jet inlet) and symmetrical axis for the cylindrical axis $r=0$,  and open boundaries for the external ones.

The simulations  run until  physical time  $t = 10000~R_{\rm jet}/{3c}$.

\section{Numerical simulation results}\label{Sec:NS_Results}

\begin{figure*}[t!]
\begin{center}
\resizebox{5cm}{10.8cm}{\includegraphics{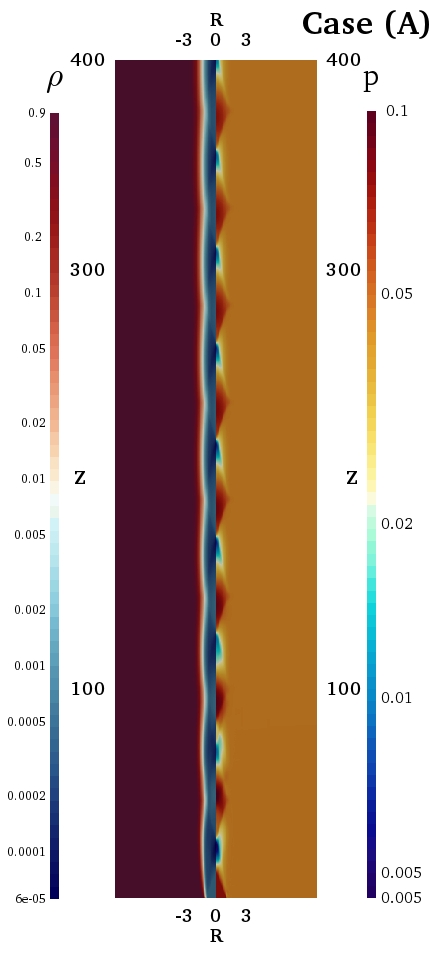}}
\resizebox{5cm}{10.8cm}{\includegraphics{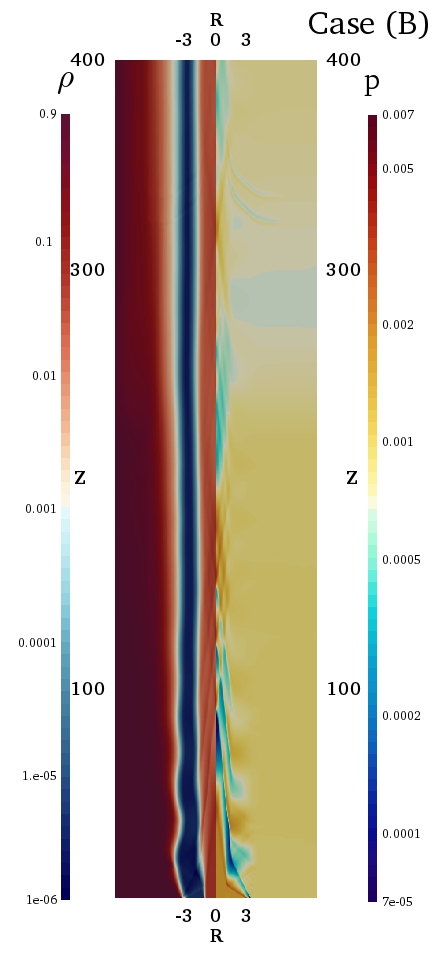}}
\resizebox{5cm}{10.8cm}{\includegraphics{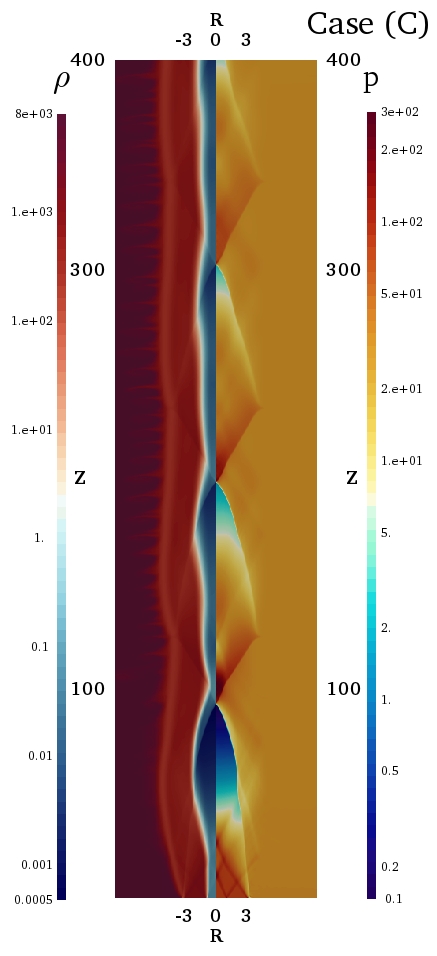}}

\resizebox{5cm}{10.8cm}{\includegraphics{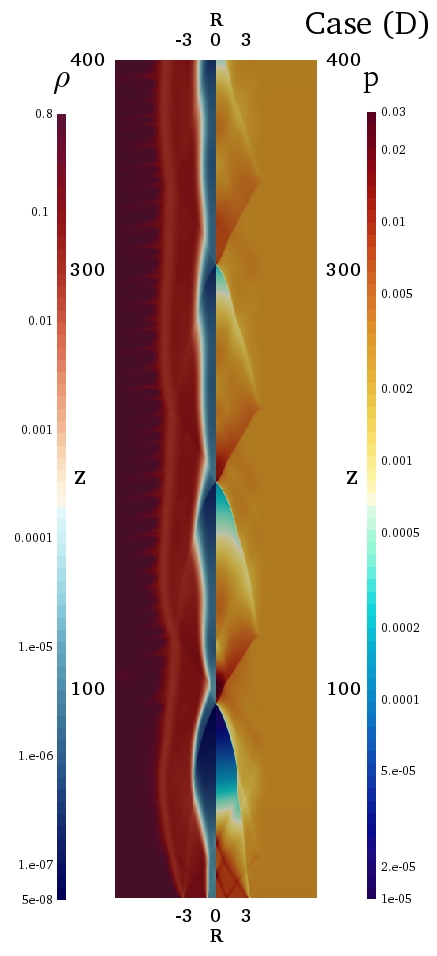}}
\resizebox{5cm}{10.8cm}{\includegraphics{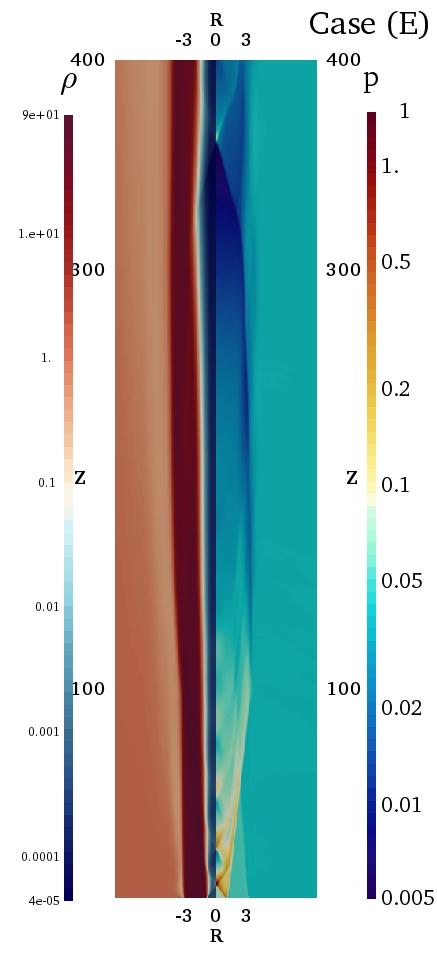}}
\resizebox{5cm}{10.8cm}{\includegraphics{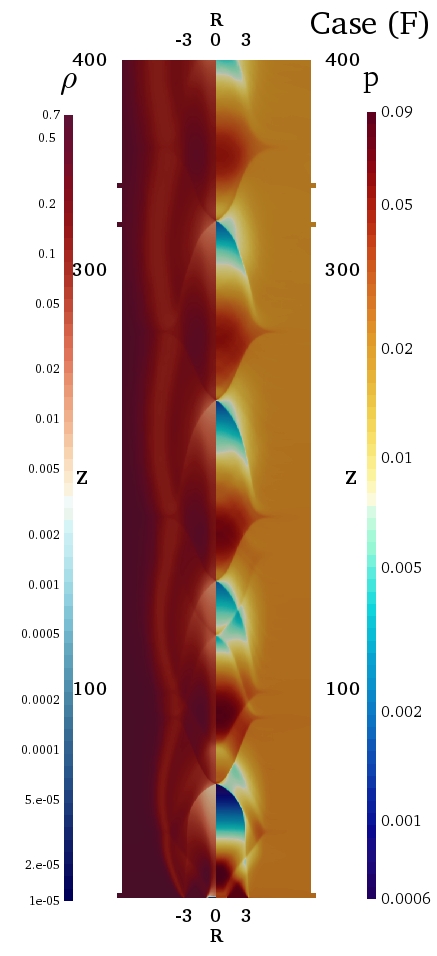}}
\caption{Two-dimensional view of all the simulated cases of the jets along the poloidal direction. In each figure, the density contour is drawn on the left side and the pressure contour on the right side. The jet figures are stretched in the radial direction and squeezed in the longitudinal direction.}
\label{Fig:Jet2C_2D_fromIToI}
\end{center}
\end{figure*}

\begin{figure*}
\begin{center}
\resizebox{6cm}{5cm}{\includegraphics{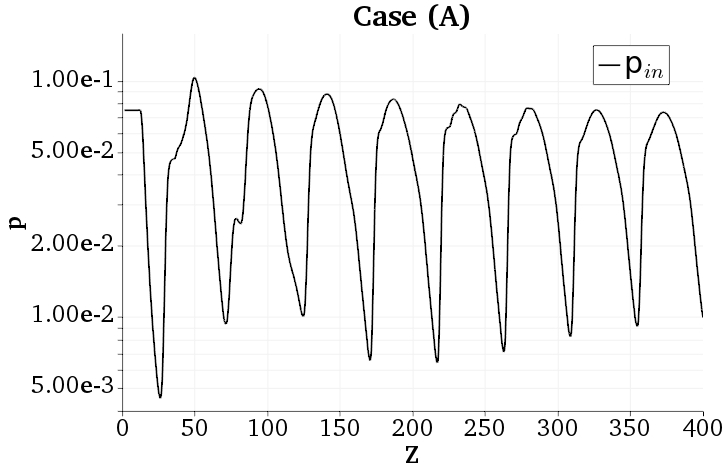}}
\resizebox{6cm}{5cm}{\includegraphics{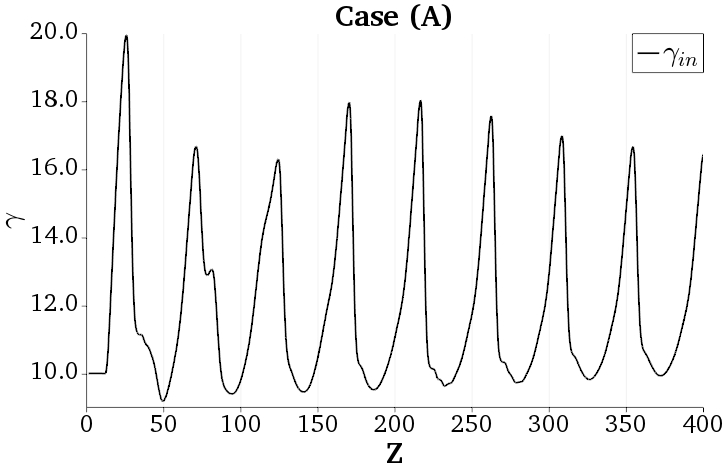}}
\caption{Case { (A)}. Variation along $R\,=\,0.01~R_{\rm jet}$ of the pressure ({left}) and the Lorentz factor ({right}). The drops in the Lorentz factor correspond to shocks crossing the line $R\,=\, 0.01~R_{\rm jet}$ and the increases in the Lorentz factor (decreases in the pressure) correspond to rarefaction waves crossing the line $R\,=\,0.01~R_{\rm jet}$.}
\label{Fig:Jet2C_2D_CaseA_1DCut}
\end{center}
\end{figure*}

\begin{figure*}
\begin{center}
\resizebox{6cm}{5cm}{\includegraphics{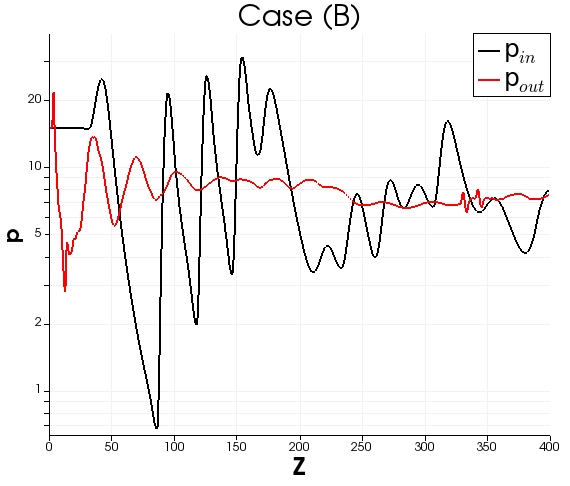}}
\resizebox{6cm}{5cm}{\includegraphics{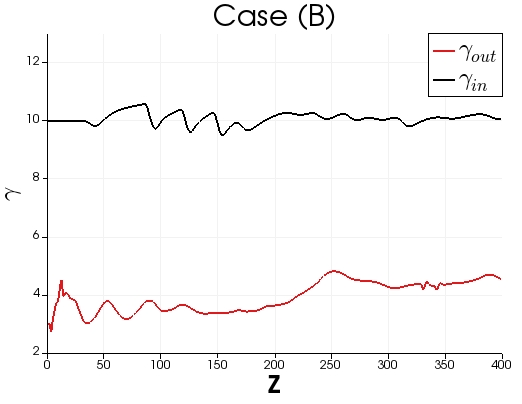}}
\caption{Case~(B). Variation along $R \,=\, 0.01~R_{\rm in}$ of the inner jet (black line),  and along $R\,=\,2\,R_{\rm in}$ of the external jet component (red line) of the pressure ({left}) and the Lorentz factor ({right}).}
\label{Fig:Jet2C_2D_CaseB_1DCut}
\end{center}
\end{figure*}

\subsection{Case (A), uniform jet}\label{Subsec:AandB}
 First, we discuss the results concerning the uniform jet with Lorentz factor  $\gamma\,=\,10$, denoted  case { (A)}. The properties of this case are shown in  Table~(\ref{table:1}), and the final results in 2D are shown in Fig.~\ref{Fig:Jet2C_2D_fromIToI}. Case {(A)} can be considered  the reference case.
The pressure difference at the interface between the jet inlet and the external medium generates two conical waves,  namely a shock wave and  a rarefaction wave (as in \citet{Mizunoetal15}).
 The shock wave is weak and propagates in the external medium. 
 It is rapidly reflected as it reaches equilibrium with external medium pressure. On the jet side, a conical rarefaction wave  propagates  toward the jet axis, converting the jet thermal energy to kinetic energy. 
 Since  the Bernoulli term $\gamma\,h$ is conserved through the rarefaction wave propagation  \citep{Aloy&Rezzolla06},  the jet  accelerates thermally to reach a Lorentz factor  $\gamma_{\rm max} \sim  21$ (Fig.~\ref{Fig:Jet2C_2D_CaseA_1DCut}).  
 After some time, the jet reaches a steady state, the mean value of the Lorentz factor increases, and then the distance between two successive internal shocks increases as well. 
 The jet Mach number increases and then the distance between two successive shocks  becomes $Z_{\rm osc}\sim 40$ (Fig.~\ref{Fig:Jet2C_2D_CaseA_1DCut}) instead of $Z_{\rm osc}\sim 25$, which should form according to the initial conditions. Indeed, the mean value of the jet Lorentz factor increases since the jet Lorentz factor varies between the initial value $\gamma\,\sim\,10 $ and $\gamma\,\sim\,20$  (Fig.~\ref{Fig:Jet2C_2D_CaseA_1DCut}).   Afterward, this shock structure remains stationary, and the relation between the distance between successive shock $Z_{\rm osc}$, Mach number ${\cal M}$ (Eq.~\ref{Eq:Mach}), and the jet radius $R_{\rm jet}$  is given by
\begin{equation}\label{Eq:Mach_oscilation}
Z_{\rm osc} \,=\,2.0\,R_{\rm jet}\,{\cal M}\,.
\end{equation}

The first rarefaction wave  accelerates the flow most efficiently, then the efficiency decreases due to a partial transmission into the external medium after each reflection (Fig.~\ref{Fig:Jet2C_2D_CaseA_1DCut}).
The large inertia of the external medium also prevents it from any alteration caused by the waves.
 Moreover, the thermal energy within the  jet decreases each time a rarefaction wave crosses the jet. The efficiency of these waves can be maintained if the pressure of the external medium decreases with distance and/or if there is a continued energy deposit within the jet.

We note that the initial structure with shock and rarefaction waves depends on the density ratio $\eta_{\rho}$ and the jet Mach number. Indeed, if the inner jet component has a larger Mach number, at the interface between the two component, a structure with two shock waves will arise.

\begin{figure*}
\begin{center}
\resizebox{6cm}{5cm}{{\includegraphics{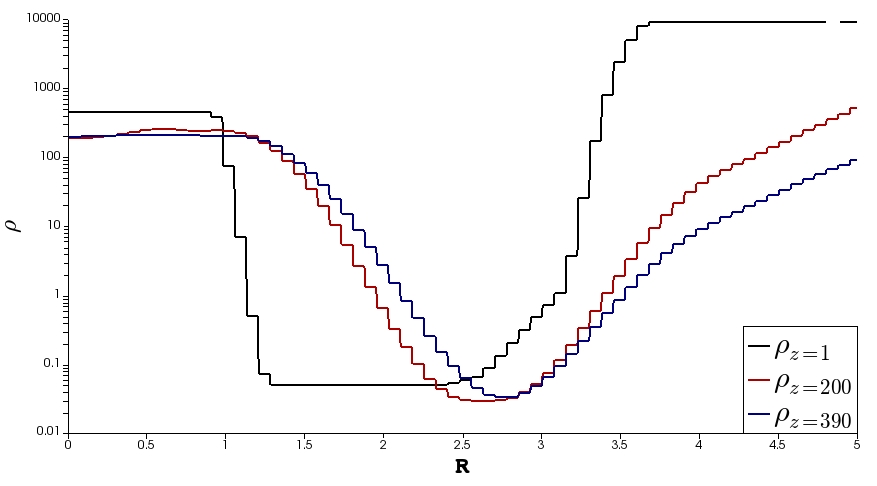}}}
\caption{Case {(B)}:  Transverse variation to the jet of the density along  line $Z=1\,R_{\rm in}$  ( black line), along line $Z=200\,R_{\rm in}$ (red line), and along line  $Z=390\,R_{\rm in}$ (blue line).}
\label{Fig:Jet2C_2D_CaseB_1DCutinZ}
\end{center}
\end{figure*}

\begin{figure*}
\begin{center}
\resizebox{6cm}{5cm}{\includegraphics{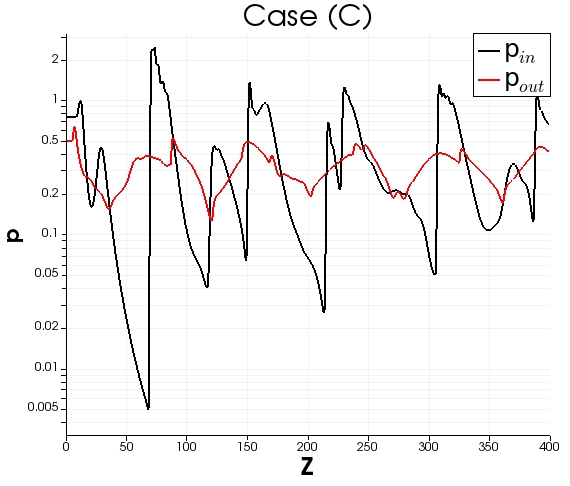}}
\resizebox{6cm}{5cm}{\includegraphics{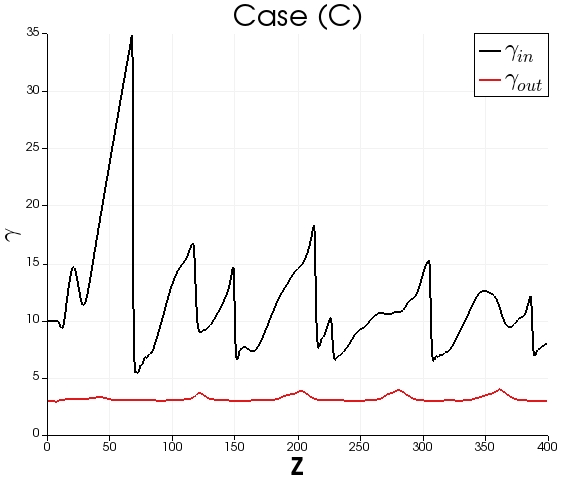}}
\caption{Case (C):  Variation along  line $R=0.01\,R_{\rm in}$ inside the inner jet ( black line) and along the line $R=2\,R_{\rm in}$ external jet component (red line) of    the pressure ({left})  and  the Lorentz factor ({right}).}
\label{Fig:Jet2C_2D_CaseC_1DCut}
\end{center}
\end{figure*}

\subsection{Case (B)}
We now set  a structured jet according to  case ({B}) (see  Table~\ref{table:1}). This jet, presented in 2D in Fig.~\ref{Fig:Jet2C_2D_fromIToI},  is characterized by a powerful inner component surrounded by a low-energy outer component. The inner component carries $95\%$ of the jet's total kinetic energy and has a large Mach number ${\cal M}\,=\,4.34$.
In this configuration, the rarefaction waves  accelerate the flow less efficiently with a maximum Lorentz factor limited to  $\gamma_{\rm in} \sim  10.5$ (see Fig.~\ref{Fig:Jet2C_2D_CaseB_1DCut}). 
The difference in the kinetic energy flow and then in the Mach number between the two components induces an energy transfer from the inner to the outer jet. This energy transfer is achieved by the shock and rarefaction waves.

Moreover, the inner jet compresses the outer jet. Indeed, the radius of the inner jet increases (Fig.~\ref{Fig:Jet2C_2D_CaseB_1DCutinZ}) at large distance and Lorentz factor (Fig.~\ref{Fig:Jet2C_2D_CaseB_1DCut}) (it reaches $\gamma\,\sim\, 4.5$) and density (Fig.~\ref{Fig:Jet2C_2D_fromIToI}) of the outer jet  increase.

 Regarding the re-collimation shocks within the inner component jet, they are weak and damped within a short distance from the core at $Z \sim 200 R_{\rm in}$. Then, they become weak compression waves at greater distances.
 
It is important to note that the low-energy outer jet behaves as a  jet sheet that damps waves propagating in jet spin, as was shown in the case of the Kelvin-Helmholtz instability in the context of a stratified jet \citep{HanaszSol98}.

\subsection{Case (C)}
The structured jet defined as case (C) (2D view in Fig.~\ref{Fig:Jet2C_2D_fromIToI}), is characterized by an inner component carrying $70\% $ of the jet total kinetic energy ($30\% $ for the outer component). The Mach number difference between the two components is relatively high,   a factor greater than 13 in favour of the outer component (see Table~\ref{table:1}). This high Mach number ratio is chosen with the aim of  increasing the efficiency of energy transfer from the inner to the outer jet component. Moreover, this change results from the increase in the outer jet inertia, which corresponds to the sound speed decrease.

At the jet inlet, along the inner-outer component interface, a shock-rarefaction wave structure arises. 
 The shock propagates in the inner jet and the rarefaction wave propagates in the outer jet. At the edge of the outer jet is another conical rarefaction wave that propagates toward the jet axis (Fig.~\ref{Fig:Jet2C_2D_fromIToI}, case (C)). When this wave reaches the inner component, it is partially reflected by the outward propagating conical shock wave from the inner jet.  At short distances, this wave-wave interaction induces oscillations at the inner-outer jet interface.  Furthermore,  at $Z\,>\,300 R_{\rm in}$ the waves from the inner-outer component synchronize and become stationary. The interference between these waves results from  the wavelength difference between the inner  and outer waves. The wavelengths are given by the jet Mach number and its radius (Eq.~\ref{Eq:Mach_oscilation}).

 In the inner jet, the two first shock waves are steady and the distance between them is  $Z_{\rm osc} \,\sim\,10\, R_{\rm in}$.  Between these shocks and the jet axis a rarefaction region arises. Thus, the resulting nozzle accelerates the flow to Lorentz factor $\gamma_{\rm in} \,\sim\, 35$ (Fig.~\ref{Fig:Jet2C_2D_CaseC_1DCut}, right). This structure of two successive shocks in the inner jet followed by the rarefaction region is reproduced along the jet, but the maximum Lorentz factor remains smaller than $\gamma_{\rm in} \,< \, 16 $. Indeed, as in  case {(A)}, the thermal energy carried at jet inlet is converted into kinetic energy and another small fraction of the energy  is partially transferred to the external medium  each time the waves reach the outer jet edge.
 Furthermore, these waves induce pressure equilibrium between the two jet components  (Fig.~\ref{Fig:Jet2C_2D_CaseC_1DCut}).

\begin{figure*}
\begin{center}
\resizebox{6cm}{5cm}{\includegraphics{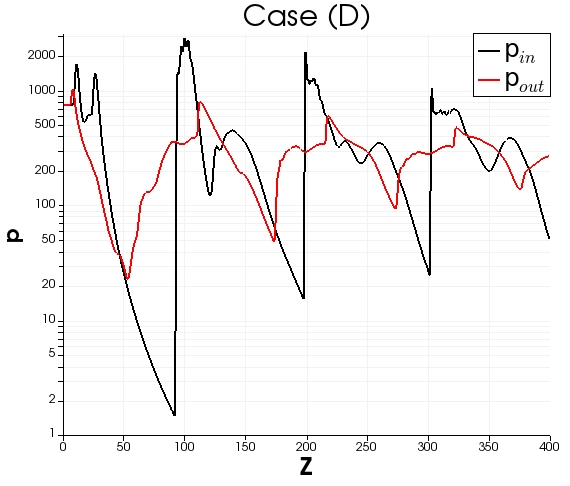}}
\resizebox{6cm}{5cm}{\includegraphics{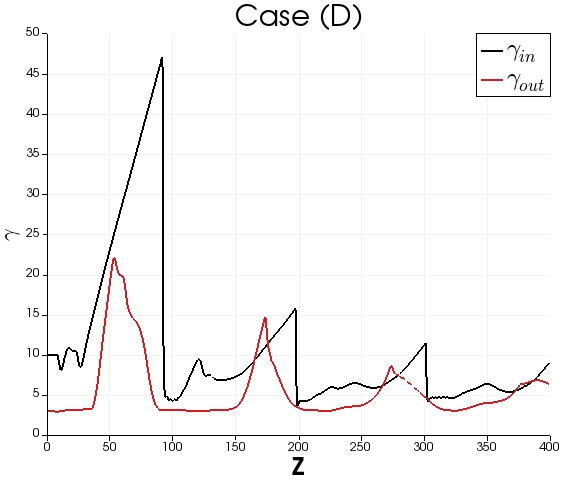}}
\caption{Case {(D)}:  Variation along  line $R=0.01\,R_{\rm in}$ inside the inner jet ( black line) and along the line $R=2\,R_{\rm in}$ external jet component (red line) of the pressure ({left}) and the Lorentz factor ({right}).}
\label{Fig:Jet2C_2D_CaseD_1DCut}
\end{center}
\end{figure*}

\subsection{Case (D)}
 In  case ({D}), the jet is set according to Table~\ref{table:1}, and shown in Fig. \ref{Fig:Jet2C_2D_fromIToI}. The inner jet contributes to $25\%$ of the jet's total kinetic energy flux.  In this configuration, at the inner-outer jet interface, a weak inward-outward shock structure forms. 
 At the outer component edge  an inward propagating rarefaction wave arises. 
 When the strong inward rarefaction wave from the outer component edge reaches the inner jet, it pushes all the shock waves within the inner component toward the jet axis. This wave is then reflected outward when it reaches the jet axis. When it reaches the jet edge, the jet radius increases. Consequently, the efficiency of the thermal acceleration of the jet is enhanced.  This process produces a powerful acceleration of the flow with a maximal Lorentz factor of the inner jet $\gamma_{\rm in}\,\sim\,50$ and of the outer jet $\gamma_{\rm out}\,\sim\,30$ (Fig. \ref{Fig:Jet2C_2D_CaseD_1DCut}, right). However, after the first strong jet acceleration, the Lorentz factor drops to $\gamma\,\sim\,10$ for both components.

\begin{figure*}
\begin{center}
\resizebox{6cm}{5cm}{\includegraphics{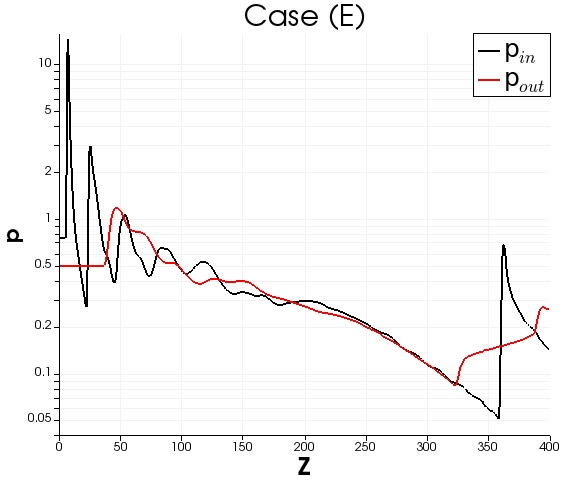}}
\resizebox{6cm}{5cm}{\includegraphics{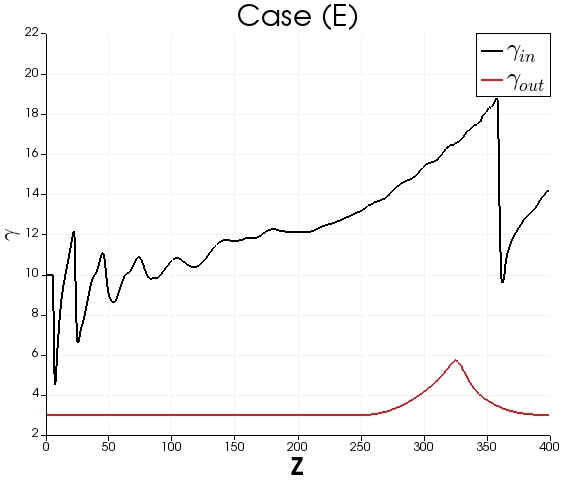}}
\caption{Case {(E)}:  Variation along  line $R=0.01\,R_{\rm in}$ inside the inner jet ( black line) and along  the line $R=2\,R_{\rm in}$ external jet component (red line) of    the pressure ({left})  and  the Lorentz factor ({right}).}
\label{Fig:Jet2C_2D_CaseE_1DCut}
\end{center}
\end{figure*}

\subsection{Case (E)}

  Case {(E)} is set according to Table~\ref{table:1}, and the final state is shown in Figure~\ref{Fig:Jet2C_2D_fromIToI}. It presents a powerful outer jet with an inertia that is much larger than that of the inner jet, $\gamma_{\rm out}^2\;h_{\rm out}\,=\,24 \gamma_{\rm in}^2\;h_{\rm in}$.
 This difference induces the formation of a conical compression-compression wave structure at the interface between the two components. The one propagating within the inner jet has a  characteristic wavelength of $Z_{\rm in, osc}\,\approx 24\, R_{\rm in}$  and the one propagating in the outer jet has $Z_{\rm out, osc}\,\approx\,220\, R_{\rm in}$. 
 This large difference decreases the interference between the two shock waves.

The compression wave that arises in the inner jet edge reaches the jet axis within a  distance $ 12\, R_{\rm in}$ (Fig.~\ref{Fig:Jet2C_2D_CaseE_1DCut}). 
As it reaches the interface between  the two components, a fraction is reflected toward the jet axis and another fraction is transmitted to the outer jet. The angle between the transmitted wave and the jet axis is smaller than the angle between the incoming wave and the jet axis. This wave diffraction is caused by a Mach number jump.  The outer component has a larger Mach number ${\cal M}_{\rm out} \,\sim \,19$ and the inner component has only $M_{\rm in} \,\sim \,1.2$ (Table~\ref{table:1}).

\begin{figure}[t!]
\begin{center}
\resizebox{6cm}{5cm}{\includegraphics{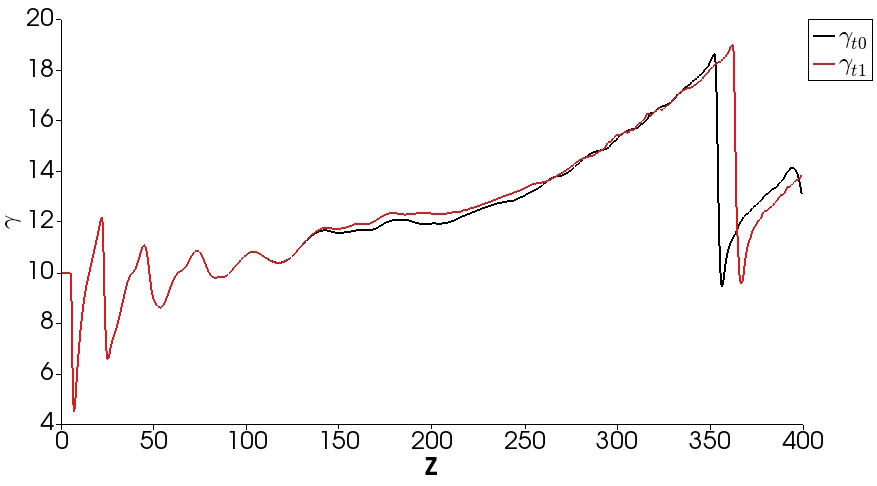}}
\caption{Case {(E)}:  Lorentz factor along the inner jet at time $t_{0}~=~7500 $ (black) and $t_{1}\,=\,12000$ (red).}
\label{Fig:Jet2C_2D_CaseL_1DCut_LF}
\end{center}
\end{figure}

  The compression wave strength decreases with distance since the waves are transmitted to the outer jet.  After six reflections, the large fraction of the inner compression waves are transferred to the outer component jet, and the rest is converted by the rarefaction wave to kinetic energy.  Consequently, the compression-rarefaction waves within  the inner jet are damped (Fig. \ref{Fig:Jet2C_2D_CaseE_1DCut}, left).  When the transmitted waves reach the edge of the outer jet they are reflected. This compression wave induces the expansion of the outer jet. When this wave reaches the inner jet, it becomes a rarefaction wave and accelerates the flow to reach $\gamma\,\sim\,18$. Until   time $t= 2000 {\rm R_{\rm in}/c}$, this wave structure is not stationary. For longer times, it  evolves toward steady state, when the   shock wave moves across the outer boundary.  However, this wave decelerates with time, and at distance $Z\,=\,350\,R_{\rm in}$  its speed is subsonic $0.0025\,c$ (Fig.\ref{Fig:Jet2C_2D_CaseL_1DCut_LF}). This wave induces a rarefaction region. The outer component pressure decreases, and thus its ability to confine the inner jet diminishes. The inner jet component expansion accelerates the jet more efficiently thanks to the thermal Laval nozzle mechanism.

\subsection{Case (F)}
%
 Case {(F)} is set according to Table~\ref{table:1} with an empty spine, and all the energy flux is carried by the outer jet. In this case, the shock waves push the outer jet component to expand toward the axis. Moreover, the shock waves reflect outward and reshape the jet, making a conical shape. Near the jet inlet a two-shock structure arises;  at large distances as the outer jet fills the inner region, only one shock wave structure remains (Fig.~\ref{Fig:Jet2C_2D_fromIToI}, case ({F})).
\section{Classification of two-component jets}\label{Sec:Discussion}

Our transverse structured jet model shows that energy distribution between the inner and the outer jet could be the key to  jet classification. Indeed, the energy distribution has a significant influence on the formation and  the state of internal shocks and on the local jet acceleration.   The model we propose here shows how the jet structure affects the development of the stationary and the non-stationary shocks observed in AGN jets. The jet structures modify the configuration of the internal shocks and the accelerations in the rarefaction regions. This structure can result from the jet launching mechanism, and is discussed  in more detail in Sect.~\ref{Sec:ImproveClassAGN}.

 We can classify the relativistic structured jets according to the transverse energy distribution between the two components as follows.  This distribution affects the transverse variation Mach number since the density between the two components is related  to the energy flux within each component.

The jets with low-energy outer jet (case {B}) show weak shocks. Moreover, the rarefaction waves are inefficient at accelerating the jet. The outer jet plays the role of a shear layer isolating the inner jet from the external medium, as in \cite{Porth&Komissarov15}. Moreover, the low inertia of the outer component allows it to absorb waves. This makes a more efficient energy transfer from the inner to the outer jet. The Lorentz factor of the outer component increases with distance.
Overall, the inner jet's Lorentz factor remains near the initial values.

The jets with near-equal energy distribution between the two components (cases  C and D) show two shock wave structures with different wavelengths.

In case ({D}), the  Lorentz factor  reaches locally  $\gamma\,\sim\,30$ and even  $\gamma\,\sim\,50$.  This acceleration is the result of the energy transfer from the outer to the inner jet by the inward propagating rarefaction waves that rise at the edge of the outer jet. The large difference in the Mach number between the hot inner jet and cold outer jet  increases the efficiency of the energy transfer from outer to inner jet.

 The jet with large energy carried by the outer jet, such as in case (E) (Fig.~\ref{Fig:Jet2C_2D_fromIToI}), could be representative of a jet with steady knots near the core and moving features  at large distances like those observed in some sources.
In the region with the steady shocks, the jet radius remains relatively constant, but downstream this radius increases with distance.
The jet expansion at large distance is the result of the large inertia of outer jet that propagates in a rarefied external medium.


The last case (F) shows that jets with empty spines could evolve to   conical shapes under the influence of the internal shocks.

These simulations show that the transverse structure in relativistic jets could be responsible for the diversity in knots  observed in radio sources. In the following section we  develop how this diversity can be related to multiple radio-loud AGN properties.

\section{Improving the blazar classification}\label{Sec:ImproveClassAGN}

%

\begin{figure*}[t!]
\begin{center}
	\includegraphics[width= 14cm]{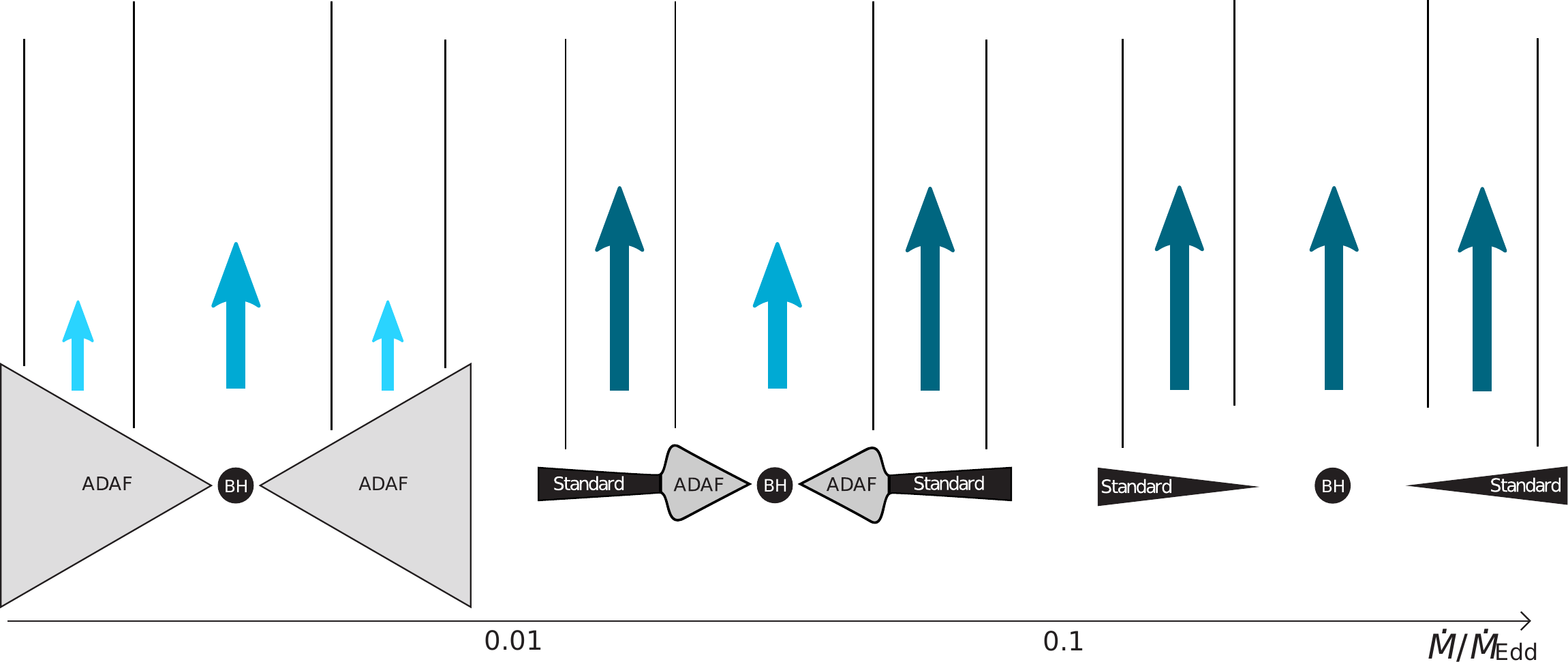}
	\put(-350,150){\color[rgb]{0,0,0}\makebox(1,0)[lb]{\smash{\large{\textbf{HBL}}}}}
	\put(-222,150){\color[rgb]{0,0,0}\makebox(1,0)[lb]{\smash{\large{\textbf{IBL/LBL}}}}}
	\put(-78,150){\color[rgb]{0,0,0}\makebox(1,0)[lb]{\smash{\large{\textbf{FSRQ}}}}}
	\caption{Conventional model of accretion regime following the critical accretion efficiency based on \citep{Odaetal09}, associated with the various AGN jet structures discussed here. Blue arrows represent the kinetic power carried by internal and external jets.}
	\label{Fig:illustration_accretion}
\end{center}
\end{figure*}

The 2D RHD simulations performed in Sect.~\ref{Sec:NS_Results} allow us for the first time to reproduce a wide variety of AGN jets depending on the energetic distribution between the inner and the outer jets. In these current conditions, the simulations reproduce standing and moving knots, which correspond to a large fraction of the observed AGN jets properties. 
However, the high relativistic motions of radio knots observed in some VLBI jets is  not described by these simulations. We assume that these fast motions result from strong instabilities at the launching region and/or at the last quasi-steady knot able to break this re-collimation shock structure and drag them close to the speed of the inner flow. This perturbative approach,  not taken into account in our models, will be investigated in future studies.
Even not considering these fast motions, it appears that various properties linking blazar types to these simulated structured jets can be highlighted.

\textbf{HBLs:}
HBLs are blazars with  less powerful jets, but  relative to their power, they also show the strongest emission at high and very high energies. This indicates a more efficient particle acceleration process than in other blazars. This strong acceleration can be linked to the multiple successive stationary shocks \citep{Meli_2013} as suggested by the VLBI observations. The relatively low spectral signature of an extended jet, added to the radio polarimetric measurements \citep{Kharb_2008A,Kharb_2008B,Gabuzda_2014}, favours an energetic dominance of their inner jet.
These sources are consequently naturally close to  simulations {(A) } and {(B)} in which the outer jet is absent or weak, and the inner jet creates multiple successive stationary re-collimation shocks.

\textbf{FSRQs:}
FSRQs are the most powerful blazars; they have  a strong spectral signature of an extended jet in addition to a stronger compact emission zone emitting at high energies. The inverse-Compton interaction of the high-energy particles on the BLR radiation, the occasional very powerful flares, as well as the non-observation of stationary knots far from the radio core suggest a powerful particle acceleration zone very close to the core, which can be associated with a strong shock.
We find these features in simulations {(C)} and {(D)}, testifying to a kinetic power of the same order as that  between internal and external jets, and also an extremely powerful shock close to the jet's base with a shock strength that quickly decreases with distance.

\textbf{IBLs/LBLs:}
These intermediate blazars have the peculiarity of presenting an extended jet emission relatively strong compared to the high-energy zone \citep[e.g.][]{Hervet_2015}.
Also, they often present  stationary knots close to the core and fast moving knots downstream  \citep{Hervet_2016}, and are mainly associated with an increase in the jet aperture angle, as discussed in Sect.~\ref{Sec:Aperture of VLBI inner jets}.
Case {(E)}, strongly dominated by its external jet, represents remarkably well these characteristics, such as stationary knots close to the core, a strong shock downstream, and an increase in the jet aperture.

Thus, these links appear as a first step for a significant achievement in the improvement of blazar unification schemes where the properties are not only defined by the power of their jets, but mainly by the energetic equilibrium between inner and outer jets components.
This also suggests that the various jet properties are more likely defined by intrinsic AGN core properties such as accretion and ejection mechanisms rather than the direct influence of an external medium.
This idea leads us to propose a new interpretation of the link of the various main known accretion regimes with the transverse jet structures and thus with AGN classification.

The AGN population is usually divided into two main accretion regimes: standard discs (optically thick, geometrically thin),  and advection dominated accretion flow (ADAF) or inefficient radiative accretion flow (IRAF)  \citep{Narayan&Li94}. 
These regimes can be differentiated following the accretion efficiency, considering an energy input of the supermassive black hole fixed at the Eddington limit (subscript ${\rm Edd}$ in the following formulas),
$\eta\,=\,\dot{M}\,/\,\dot{M}_{\rm Edd}$, where $\dot{M}$ is the mass accretion rate.
 
The ADAF regime corresponds to a low accretion efficiency of $\eta\,\lesssim\,0.01$ with a disc luminosity $L_d = L_{\rm Edd}\,(\dot{M}\,/\,\dot{M}_{\rm Edd})^2$, while the standard disc corresponds to a high accretion regime $\eta\,\gtrsim\,0.01$ with a disc luminosity as $L_d = L_{\rm Edd}\,(\dot{M}\,/\,\dot{M}_{\rm Edd})$ \citep{Esin_1997}. The usual dichotomy scenario of blazars binds ADAF with FR~I jets and BL~Lacs on the  one hand, and standard discs with FR~II jets and FSRQs on the other hand \citep{Ghisellini&Tavecchio08}.
 The standard disc we describe here corresponds to the jet emitting disc (JED) described in \cite{Ferreiraetal06}. In JED, the magnetic field pressure is near equipartition with the thermal pressure and contributes strongly to  jet launching.
It is favoured that the transition between standard discs and ADAFs passes via a hybrid regime: ADAF close to the black hole and Standard for larger radius  \citep{Esin_1997,Narayan&McClintock08,Odaetal09}.
It appears that LBLs or IBLs can be associated with this peculiar regime, as shown for the blazar AP Librae by \cite{Hervet_2015}.
Thus, a scenario linking accretion regimes and jet structures can be developed.

The {ADAF} regime, associated with the less powerful jets, is not favoured to generate a strong disc wind, leading to jets widely dominated by one component as seen in {HBLs} and simulation cases {(A)} and {(B)}.

The intermediate accretion regime assumes a relatively weak internal jet linked to the {ADAF} centre of the disc, but also a strong wind generated by the external standard disc regime. This corresponds to the upper description of {IBLs} and {LBLs} and simulation case {(E)}, in which the outer jet is relatively strong compared to the inner jet.

 Finally, with high accretion rate the  standard disc extends down to the innermost, stable orbit. The launching mechanism for the inner and outer jet should have the same efficiency. Consequently, this leads to a global powerful jet with inner and outer components likely with similar powers. These types of jets correspond to FSRQs and they can be linked to the simulation cases ({C}) and ({D}).
A scheme of this unification scenario is presented in Figure~(\ref{Fig:illustration_accretion}).

\section{Conclusion}
The investigation of the VLBI knotty structure of numerous AGN jets, their association with multiple re-collimation shocks, and the differentiation of blazar inner jet apertures provide us with a strong basis to develop a hydrodynamical scenario of structured jets.
We perform for the first time, fully 2D special relativistic hydrodynamics simulations of  overpressured two-component jets. 
The comparisons of simulation results with observations suggest a new blazar classification by linking the proprieties of jet knots to the accretion disc characteristics. According to the accretion rate and to the disc configuration (ADAF/standard disc), the energy distribution between the inner and outer jets changes. 
 The increase in the outer jet energy flux contribution induces a damping in the stationary knot near the core. When the contribution  of the outer jet prevails,  a moving knot rises at a large distance from the core. 
 In our new blazar classification, the relative contribution of the outer jets increases in the following order: { HBL}, { FSRQ}, { IBL/LBL}.

In a following paper we will present the influence of the magnetic field on the re-collimation shocks and flow re-acceleration in stratified jets.

\begin{acknowledgements}
Part of this work was supported by the PNHE. This work acknowledges financial support from the UnivEarthS Labex program at Sorbonne Paris Cit\'e (ANR-10-LABX-0023 and ANR-11-IDEX-0005-02). O.H. thanks the U.S. National Science Foundation for support under grant PHY-1307311 and the Observatoire de Paris for financial support with ATER position.
All the computations made use of the High Performance Computing OCCIGEN and JADE at CINES within the DARI project c2015046842. 
This research has made use of data from the MOJAVE database that is maintained by the MOJAVE team \citep{Lister_2009}.
\end{acknowledgements}

\input{TwoCjet2D_HD_lib.tex}

\onecolumn
\begin{appendix}
\section{Jets with a significant aperture increase}
\label{Annexe::aperture_increase}

\begin{figure*}[h]
\begin{center}
\begin{minipage}[b]{\linewidth}
\begin{minipage}[b]{0.33\linewidth}
 \centering \includegraphics[width=6.2cm]{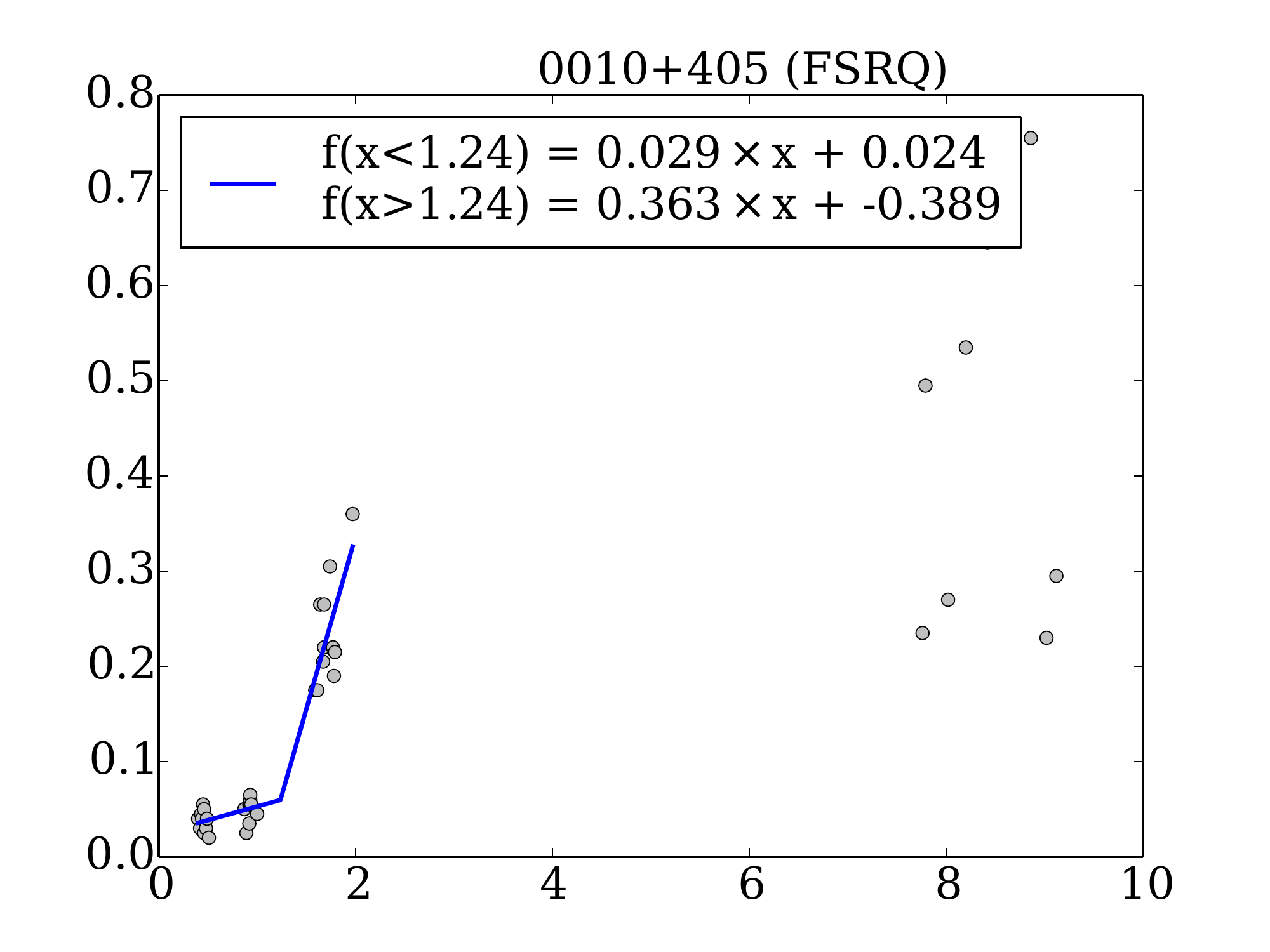}
\end{minipage}\hfill
\begin{minipage}[b]{0.33\linewidth}
 \centering \includegraphics[width=6.2cm]{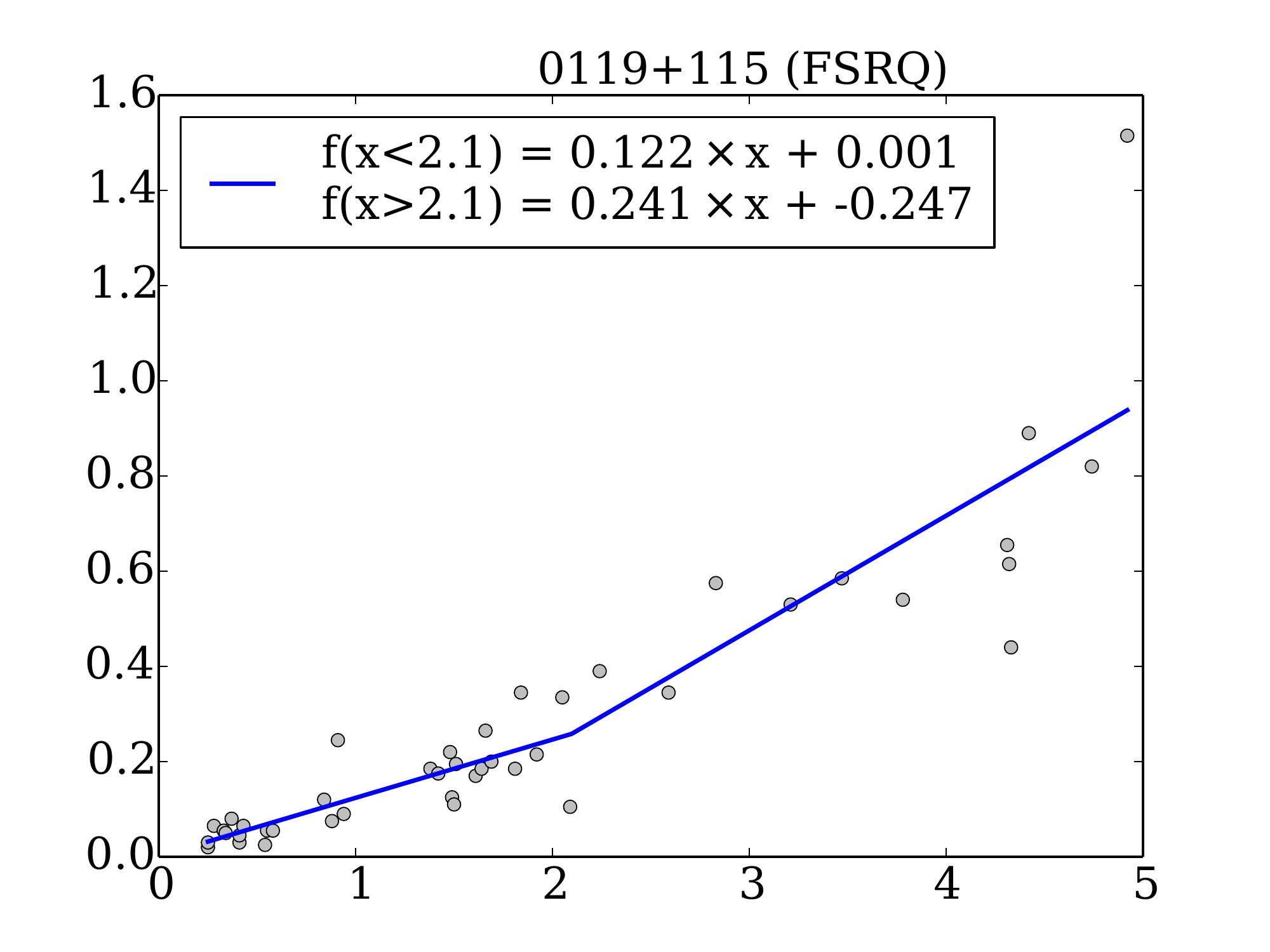}
\end{minipage}\hfill
\put(-520,90){\rotatebox{90}{\makebox(0,0)[lb]{\large{Transverse size [mas]}}}}
\begin{minipage}[b]{0.33\linewidth}
 \centering \includegraphics[width=6.2cm]{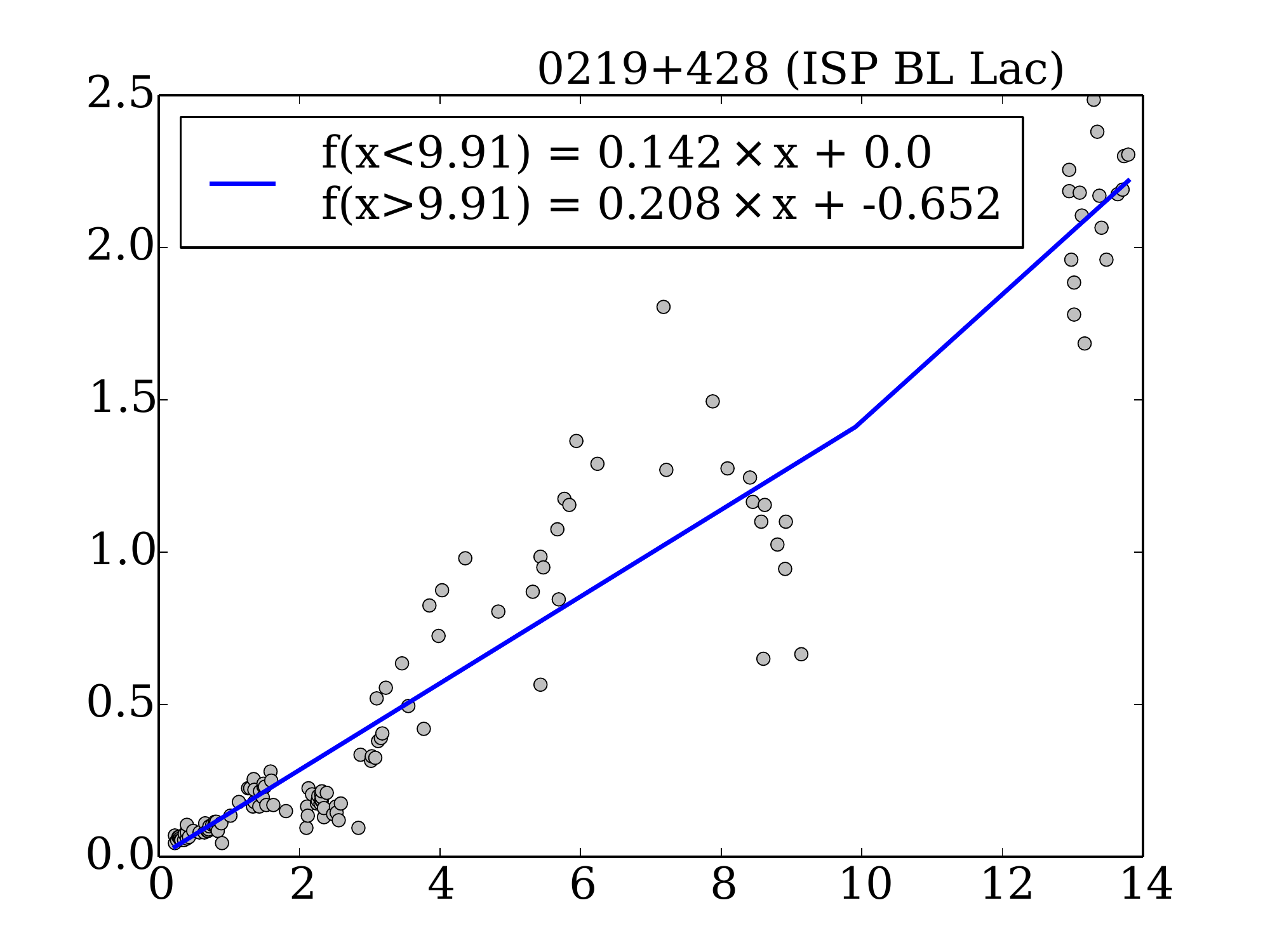}
\end{minipage}\hfill
\begin{minipage}[b]{0.33\linewidth}
 \centering \includegraphics[width=6.2cm]{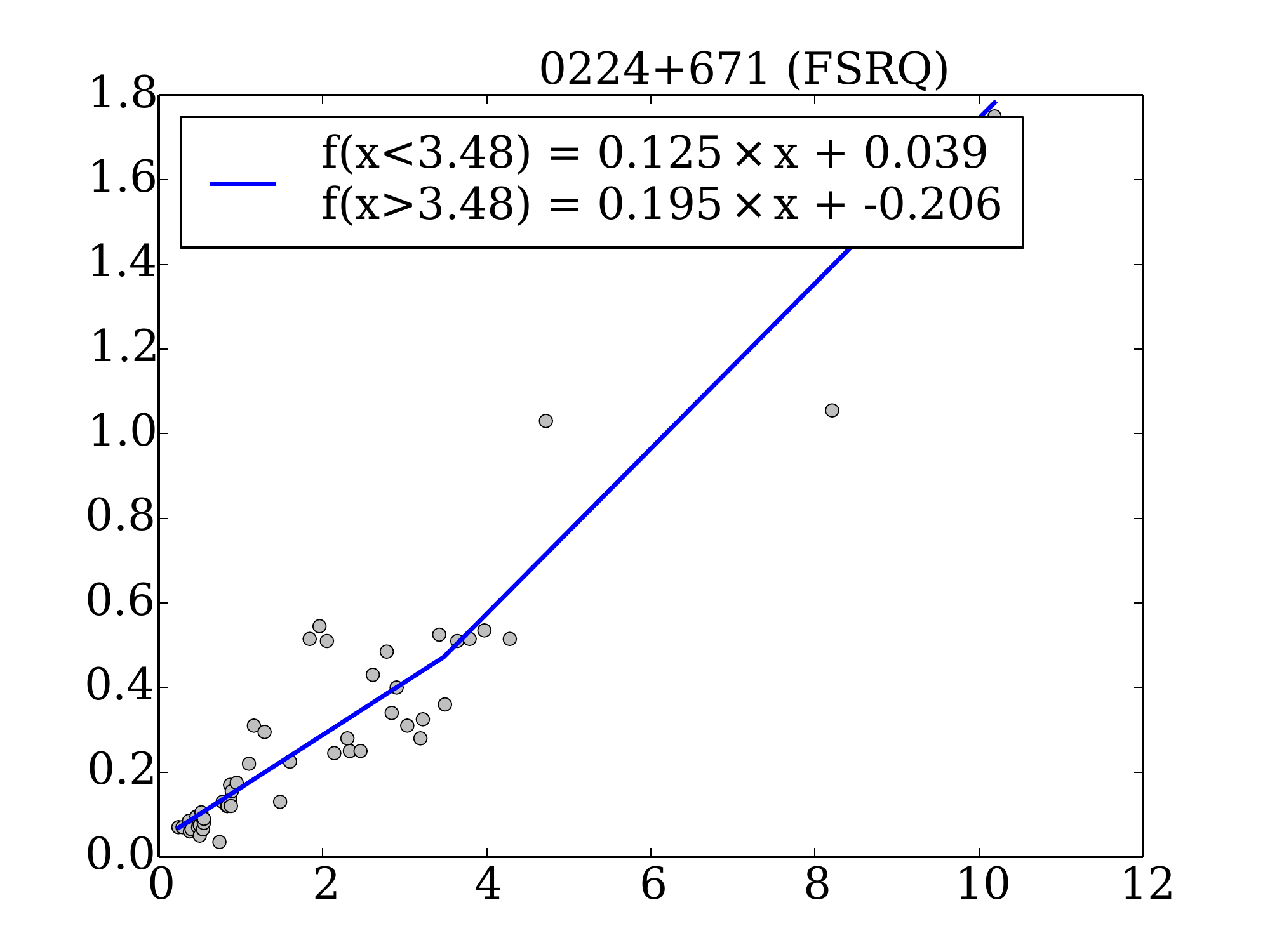}
\end{minipage}\hfill
\begin{minipage}[b]{0.33\linewidth}
 \centering \includegraphics[width=6.2cm]{Figures/Ouverture_jets_VLBI/jet_interne_0300+470.pdf}
\end{minipage}\hfill
\begin{minipage}[b]{0.33\linewidth}
 \centering \includegraphics[width=6.2cm]{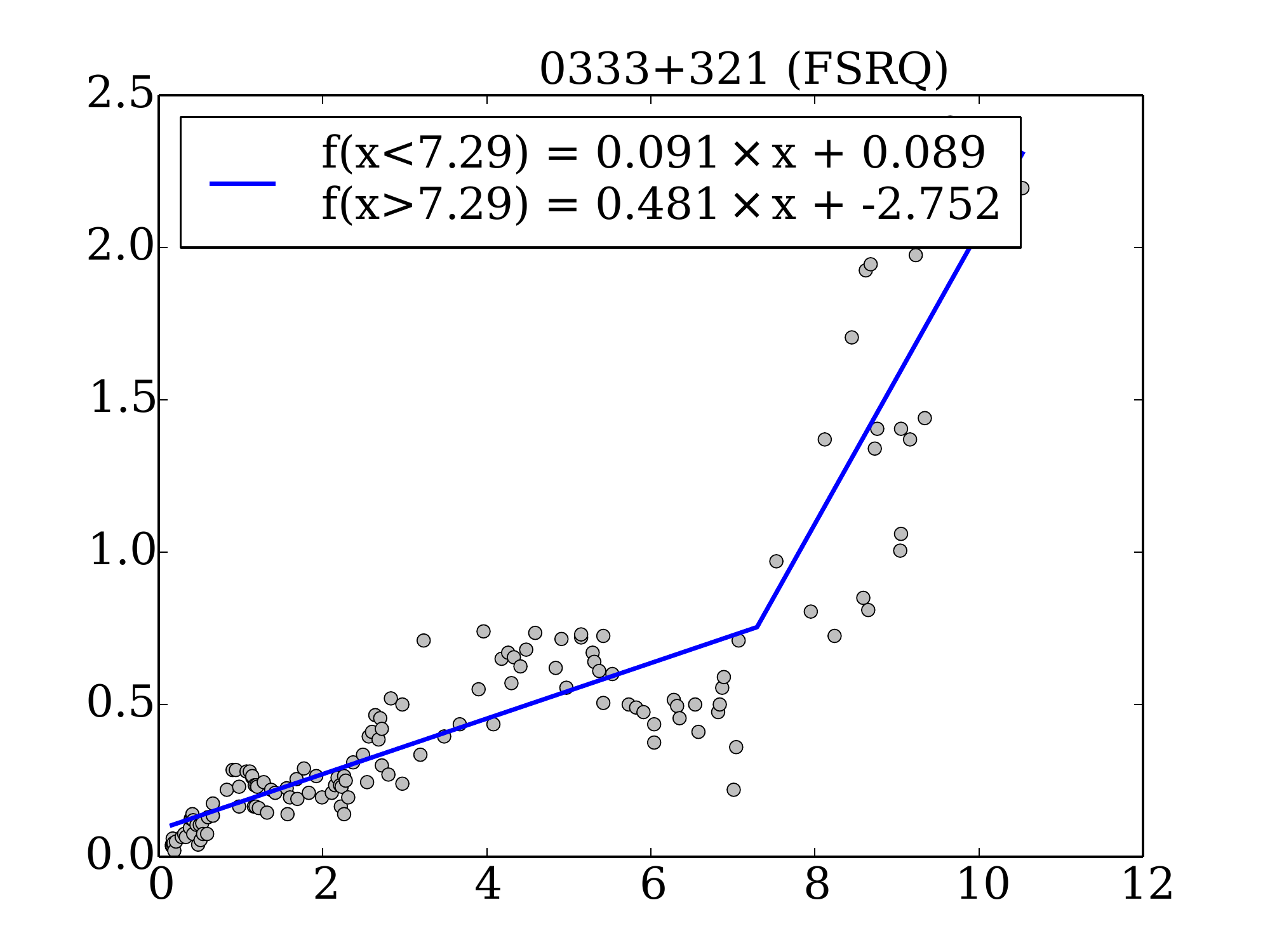}
\end{minipage}\hfill
\begin{minipage}[b]{0.33\linewidth}
 \centering \includegraphics[width=6.2cm]{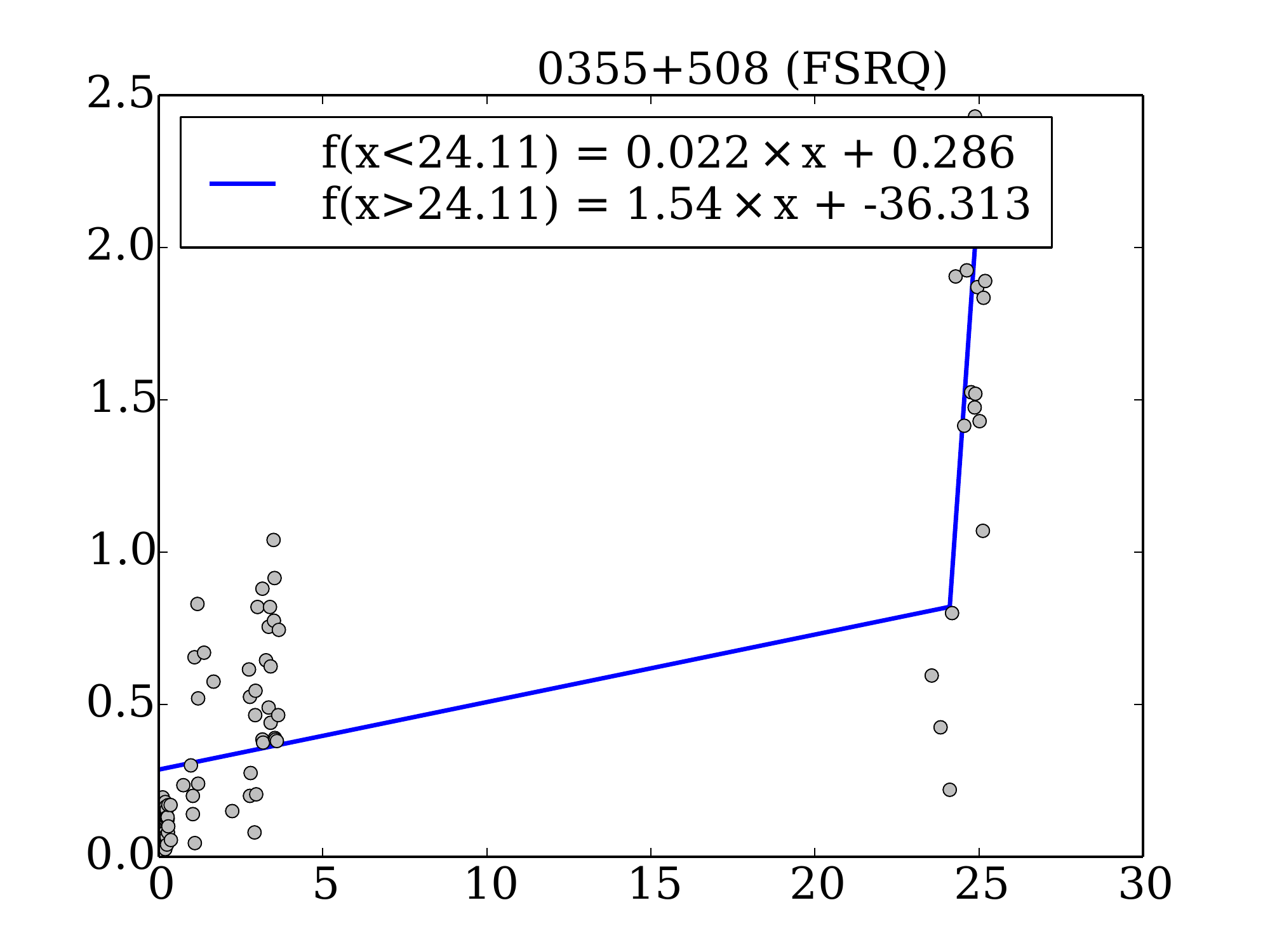}
\end{minipage}\hfill
\begin{minipage}[b]{0.33\linewidth}
 \centering \includegraphics[width=6.2cm]{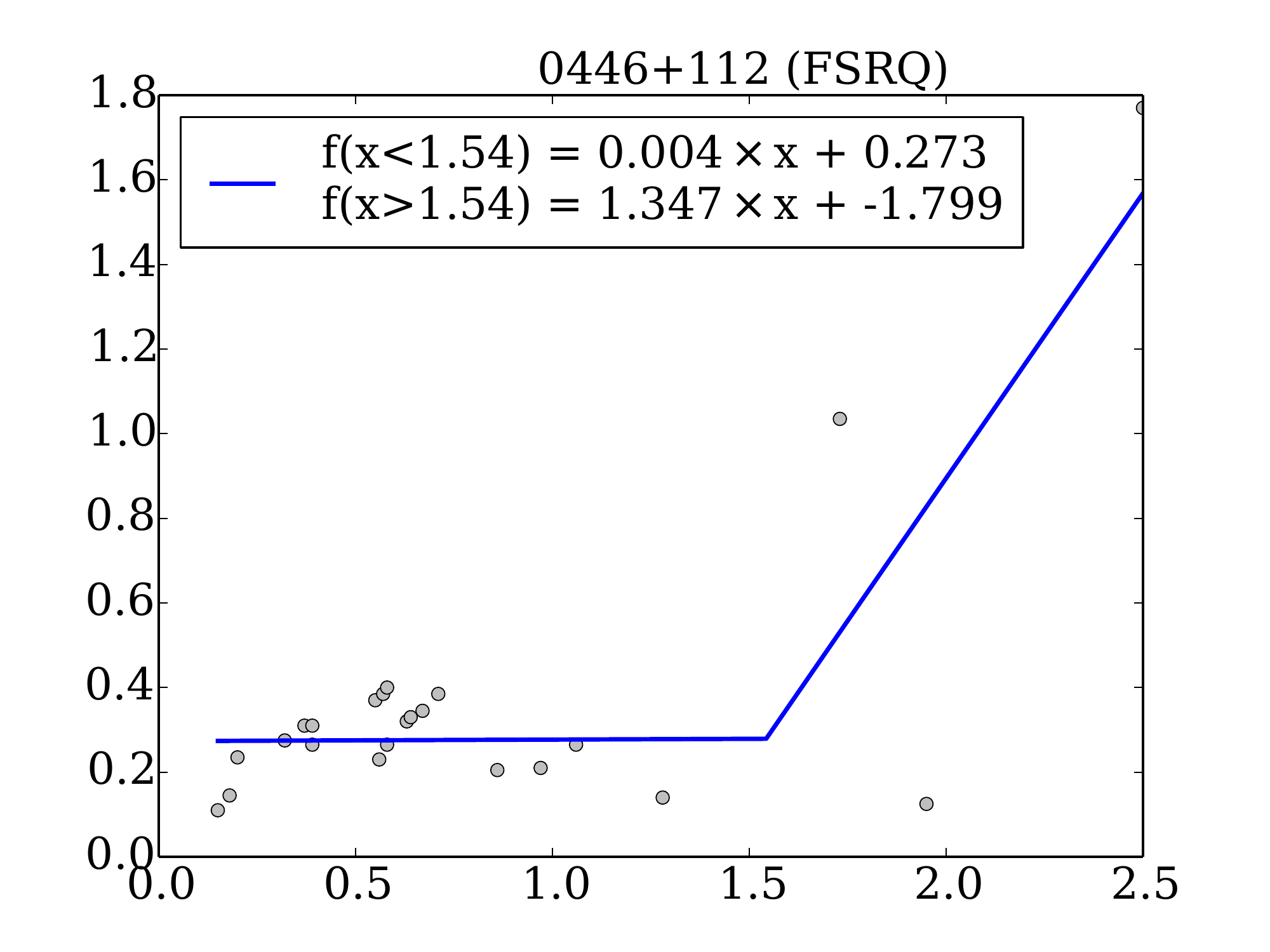}
\end{minipage}\hfill
\begin{minipage}[b]{0.33\linewidth}
 \centering \includegraphics[width=6.2cm]{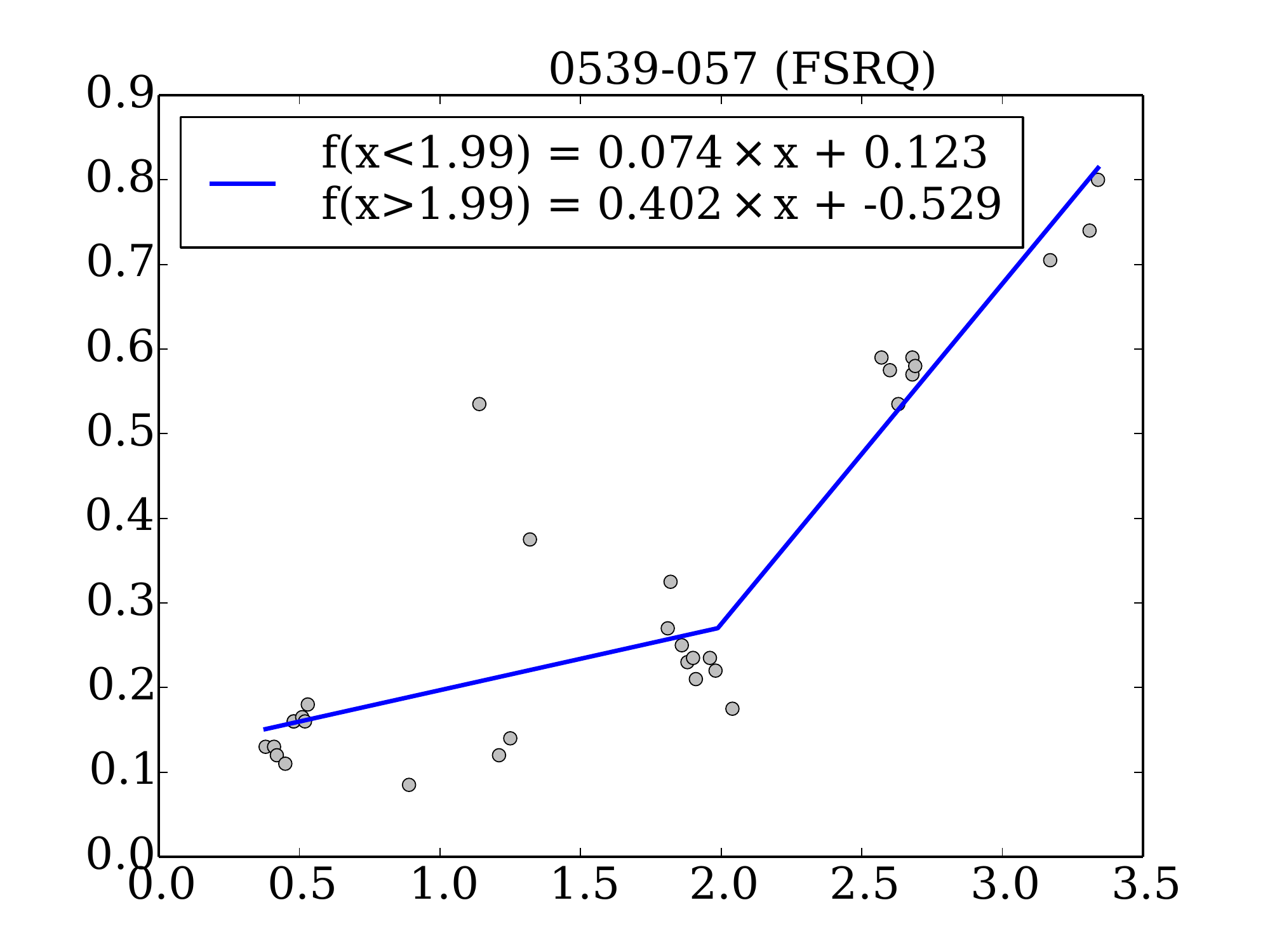}
\end{minipage}\hfill
\begin{minipage}[b]{0.33\linewidth}
 \centering \includegraphics[width=6.2cm]{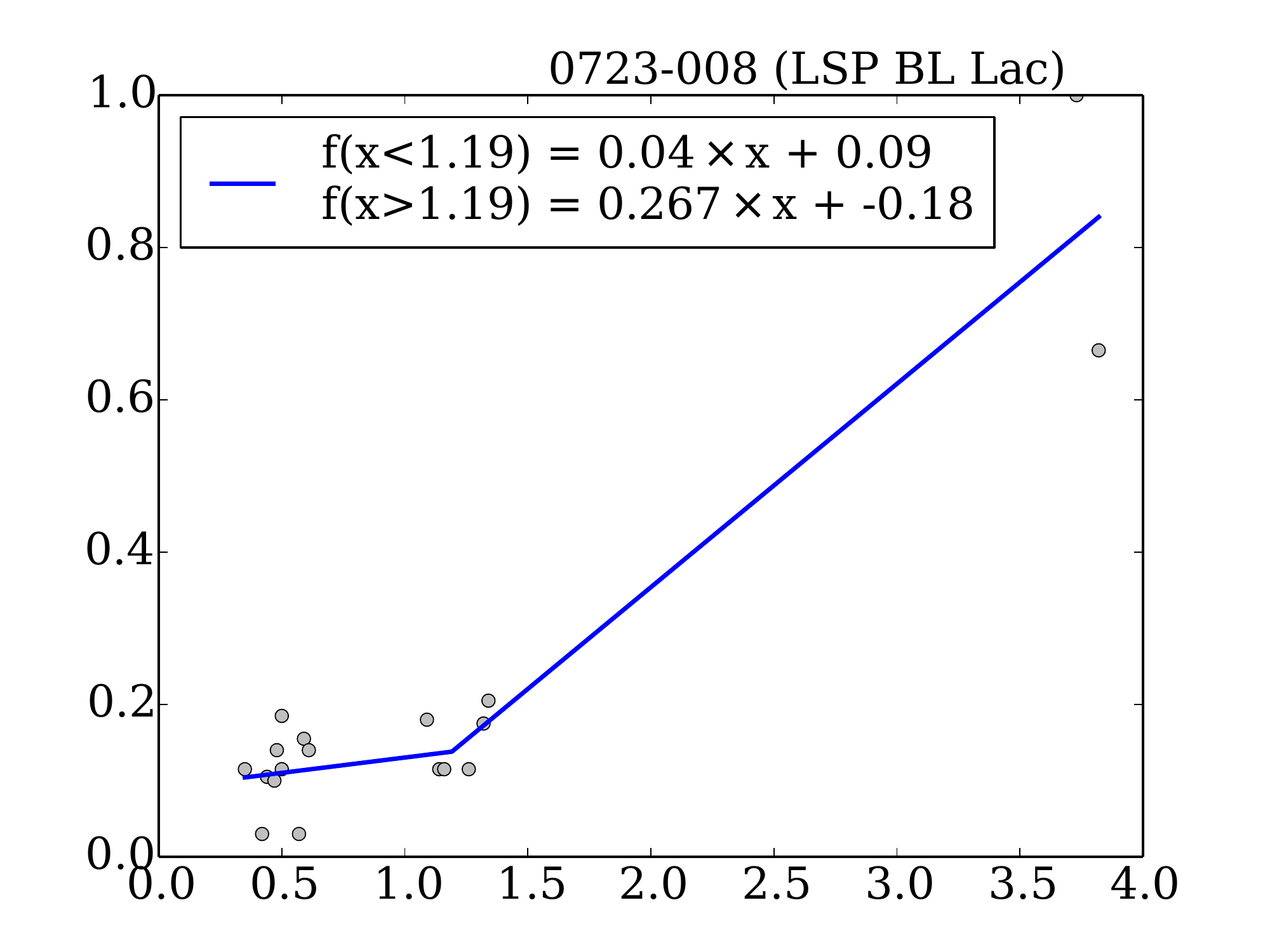}
\end{minipage}\hfill
\begin{minipage}[b]{0.33\linewidth}
 \centering \includegraphics[width=6.2cm]{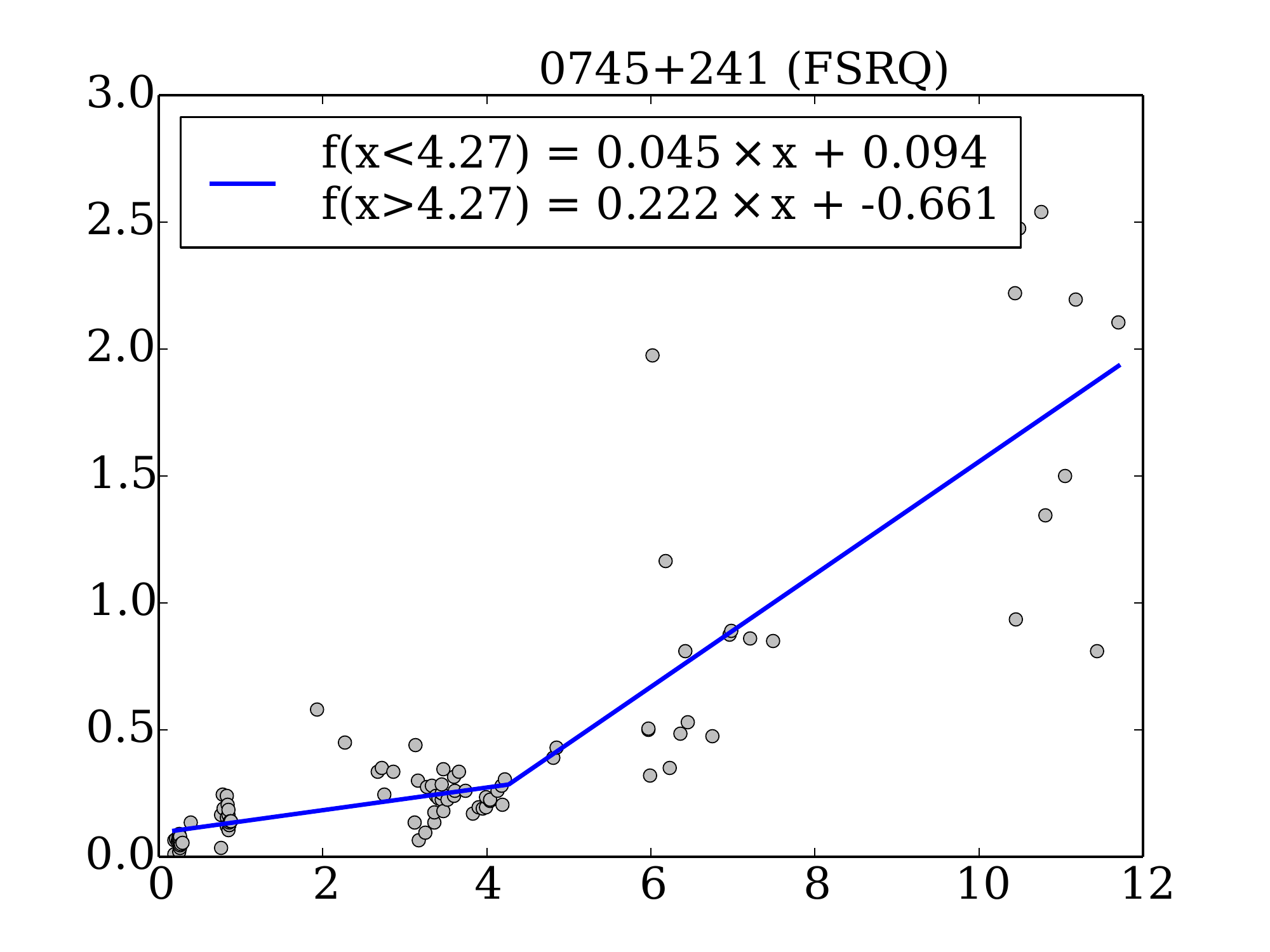}
\end{minipage}\hfill
\begin{minipage}[b]{0.33\linewidth}
 \centering \includegraphics[width=6.2cm]{Figures/Ouverture_jets_VLBI/jet_interne_0754+100.pdf}
\end{minipage}\hfill
\end{minipage}
 \put(-300,-10){\makebox(0,0)[lb]{\large{Core distance [mas]}}}
 \put(-520,230){\rotatebox{90}{\makebox(0,0)[lb]{\large{Transverse size [mas]}}}}
\end{center}
\end{figure*}

\begin{figure*}[h]
\begin{center}
\begin{minipage}[b]{\linewidth}
\begin{minipage}[b]{0.33\linewidth}
 \centering \includegraphics[width=6.2cm]{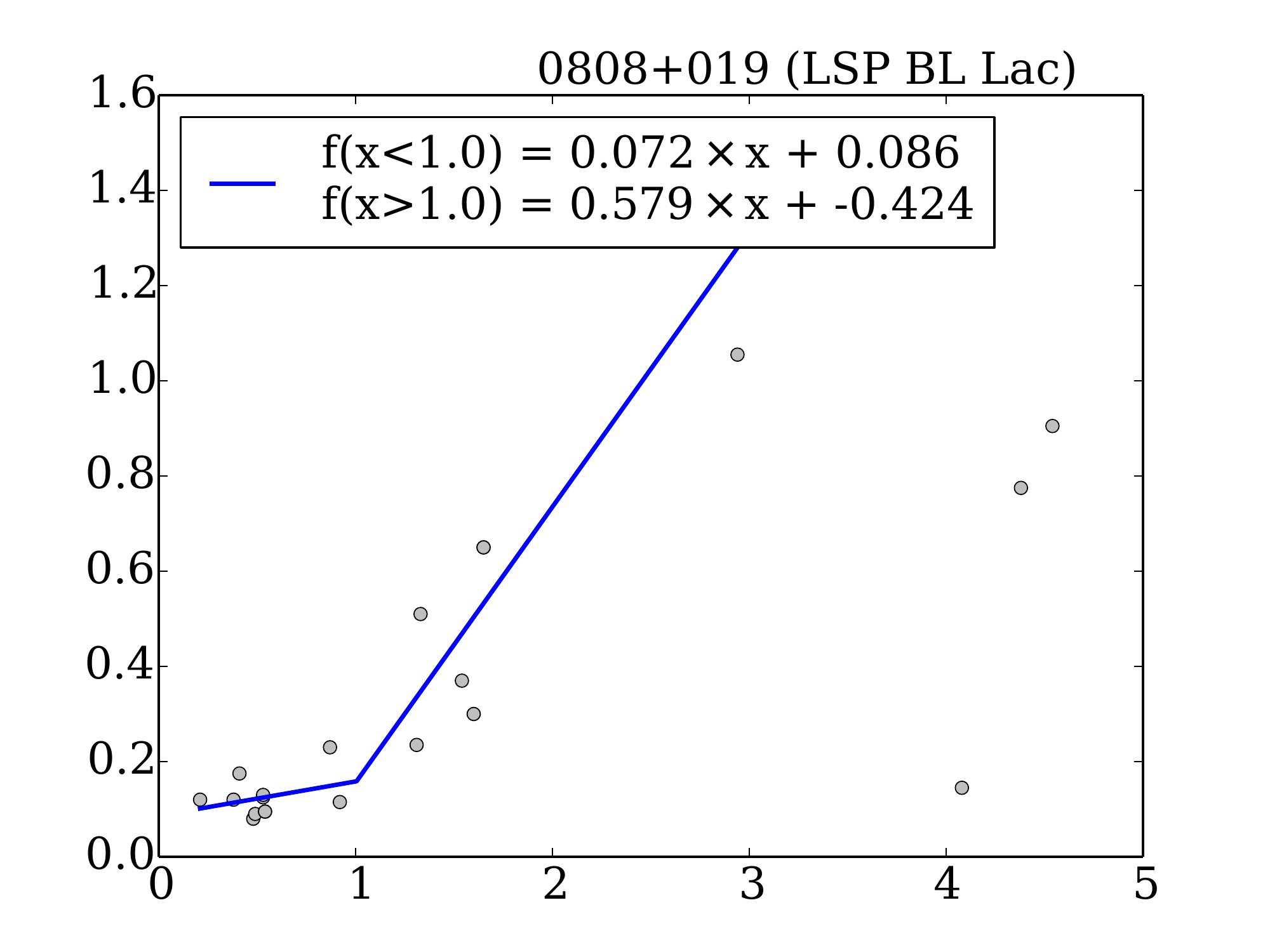}
\end{minipage}\hfill
\begin{minipage}[b]{0.33\linewidth}
 \centering \includegraphics[width=6.2cm]{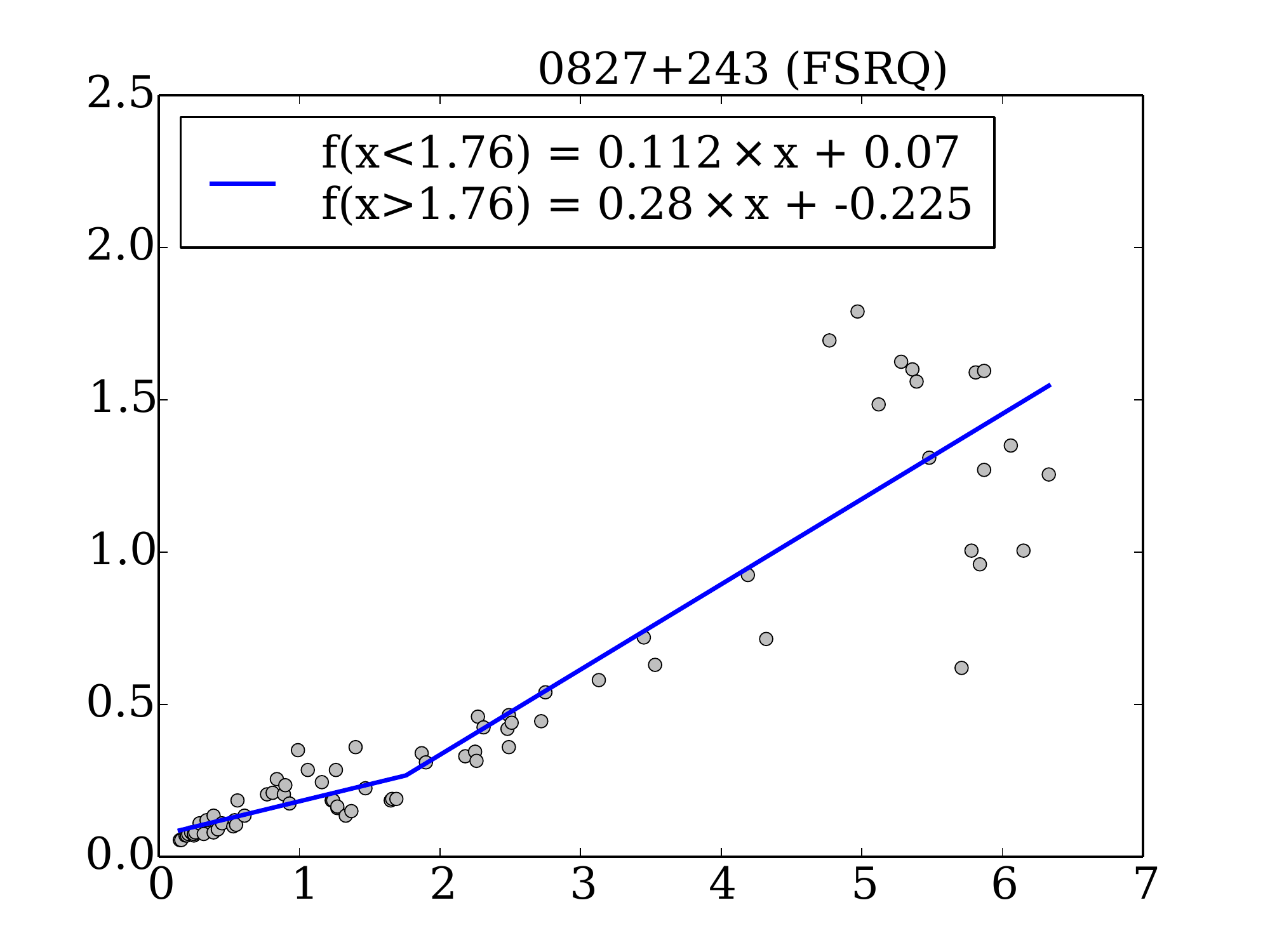}
\end{minipage}\hfill
\begin{minipage}[b]{0.33\linewidth}
 \centering \includegraphics[width=6.2cm]{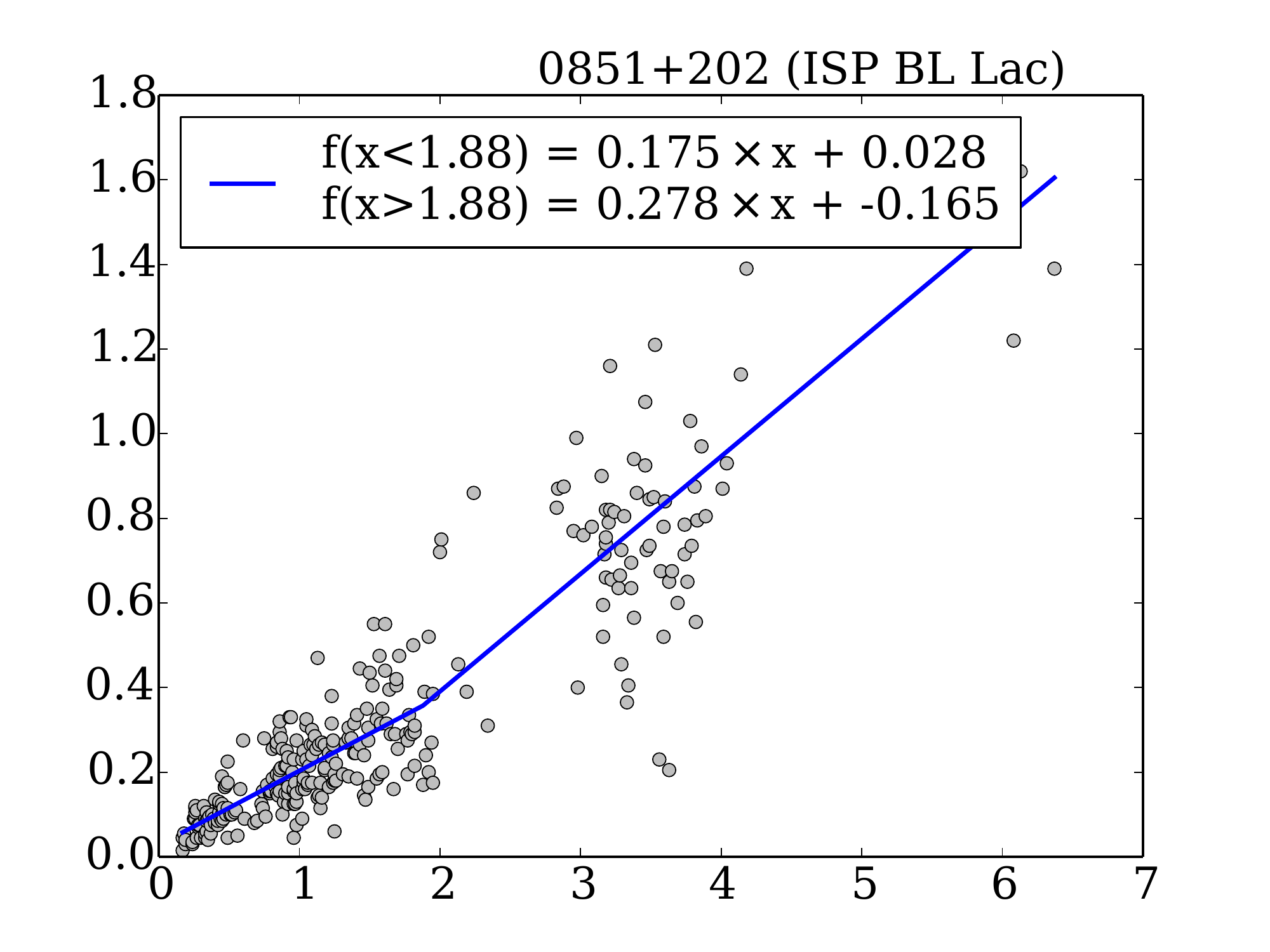}
\end{minipage}\hfill
\begin{minipage}[b]{0.33\linewidth}
 \centering \includegraphics[width=6.2cm]{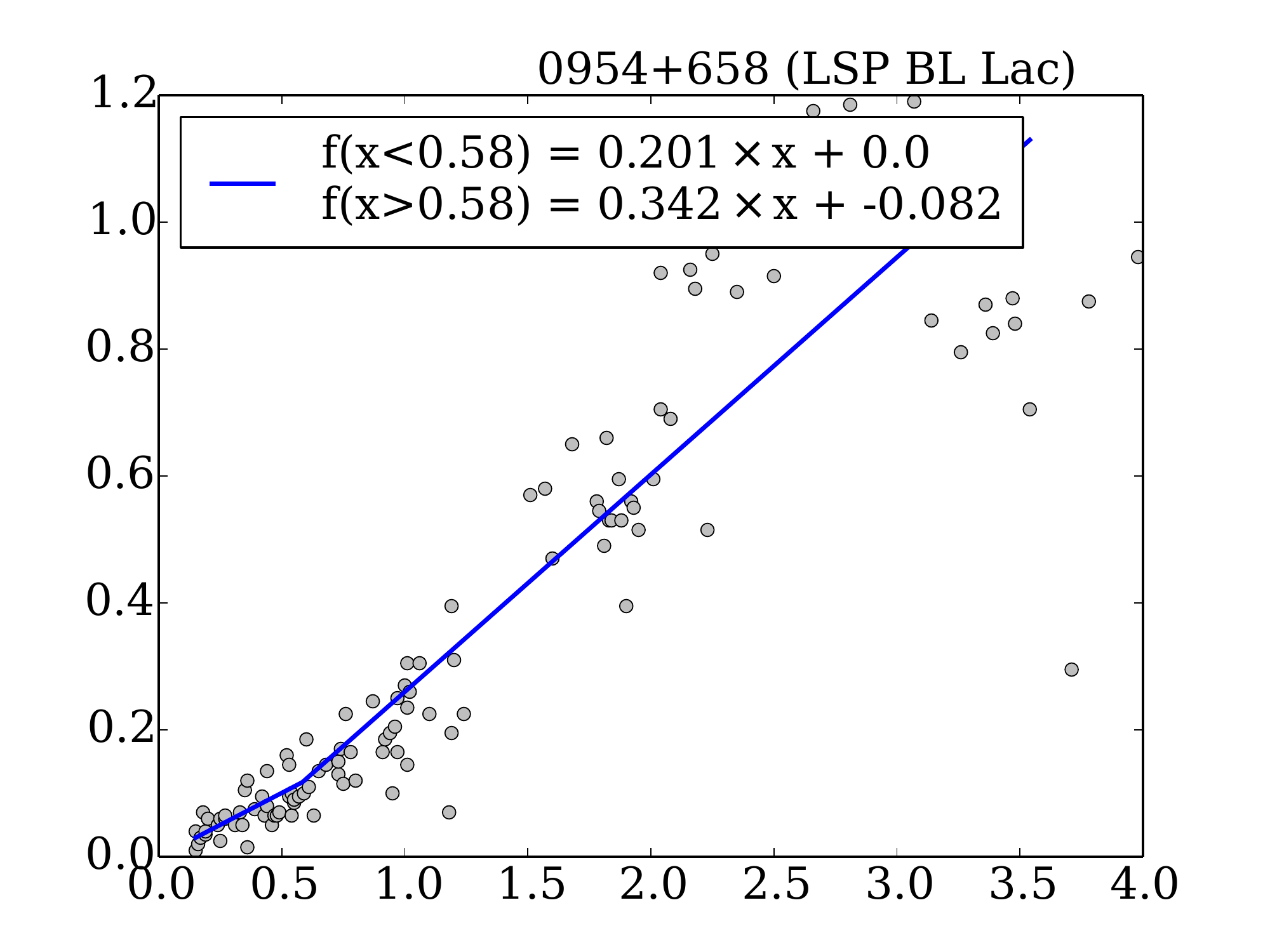}
\end{minipage}\hfill
\begin{minipage}[b]{0.33\linewidth}
 \centering \includegraphics[width=6.2cm]{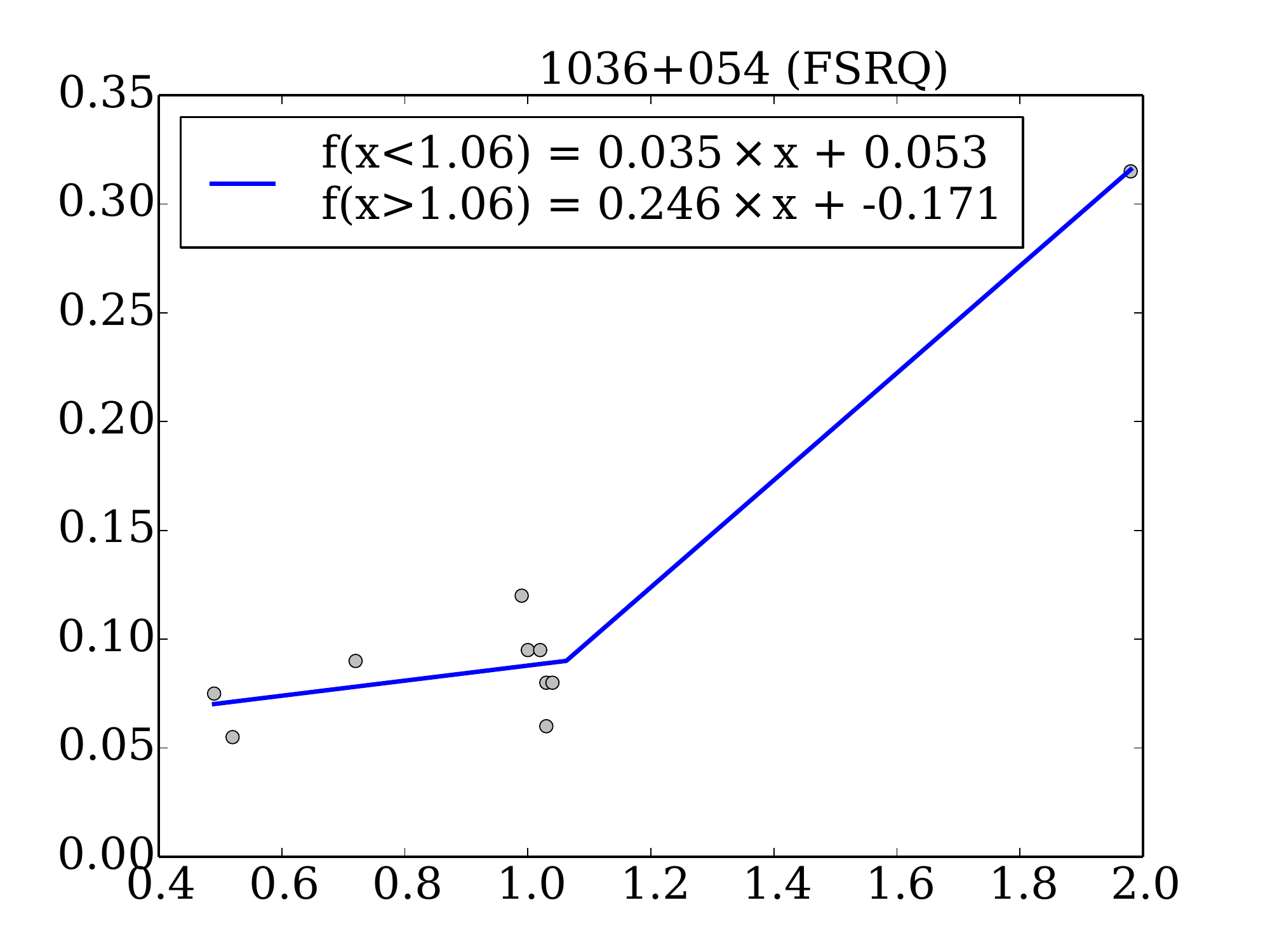}
\end{minipage}\hfill
\begin{minipage}[b]{0.33\linewidth}
 \centering \includegraphics[width=6.2cm]{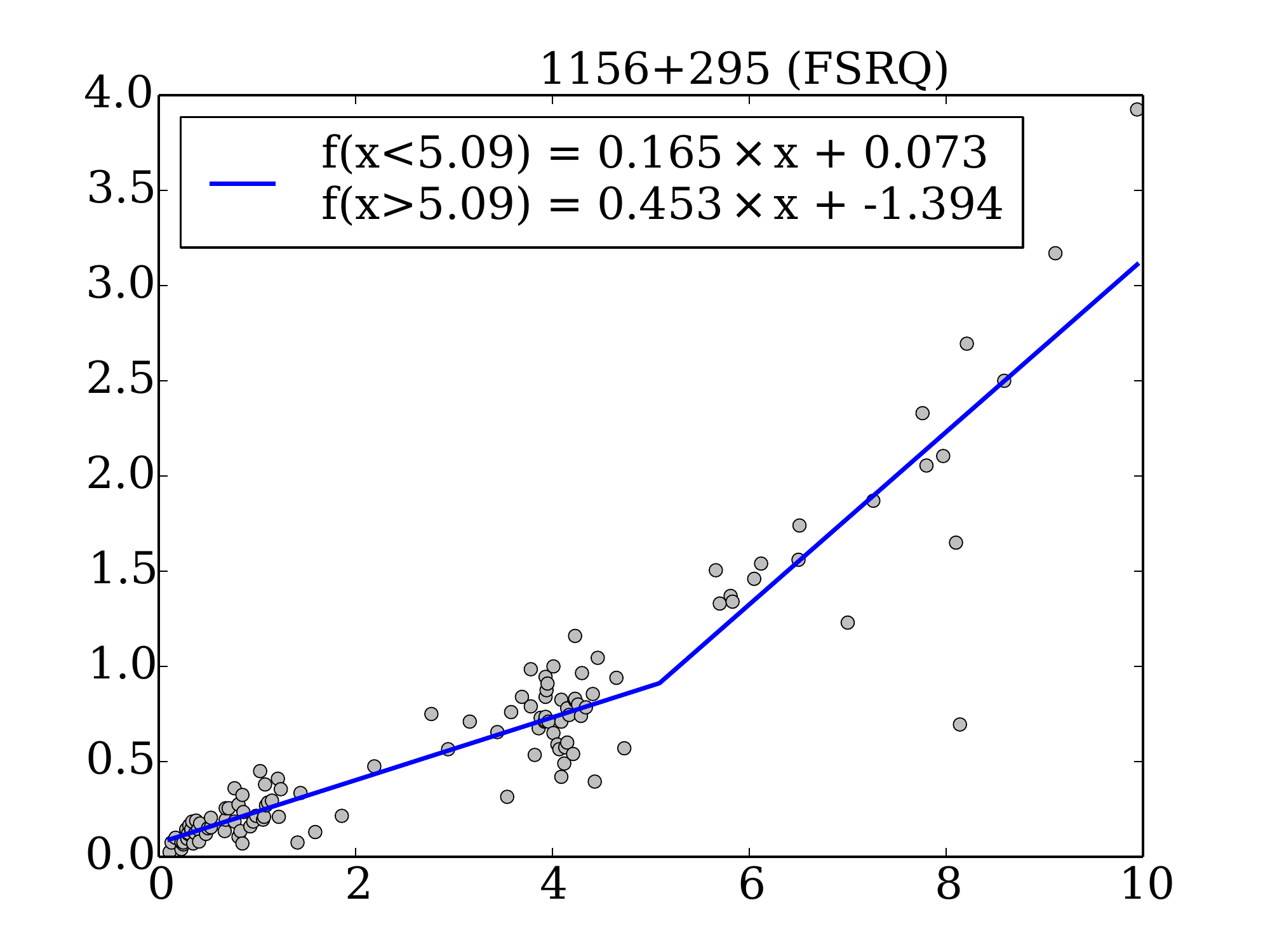}
\end{minipage}\hfill
\begin{minipage}[b]{0.33\linewidth}
 \centering \includegraphics[width=6.2cm]{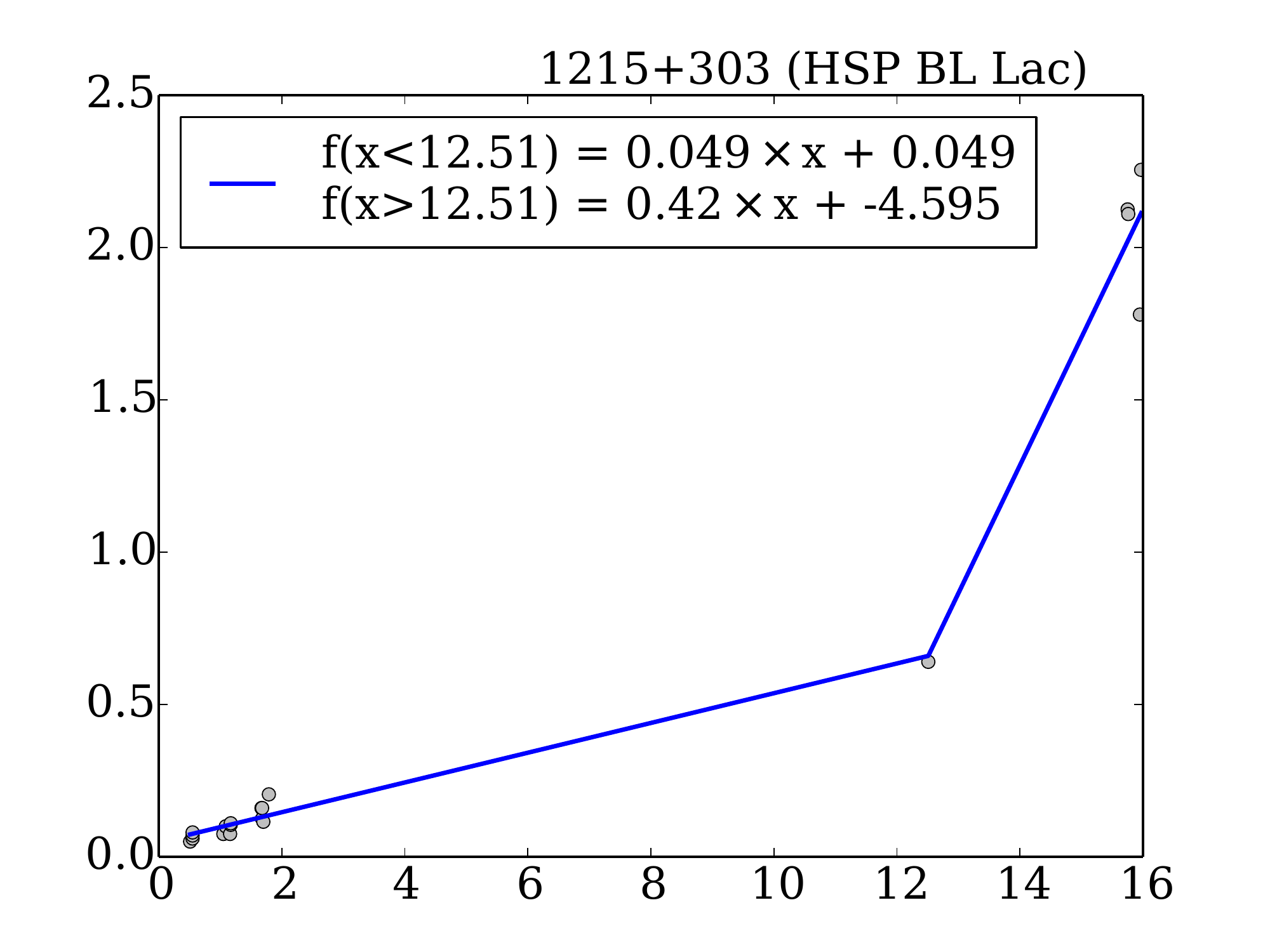}
\end{minipage}\hfill
\begin{minipage}[b]{0.33\linewidth}
 \centering \includegraphics[width=6.2cm]{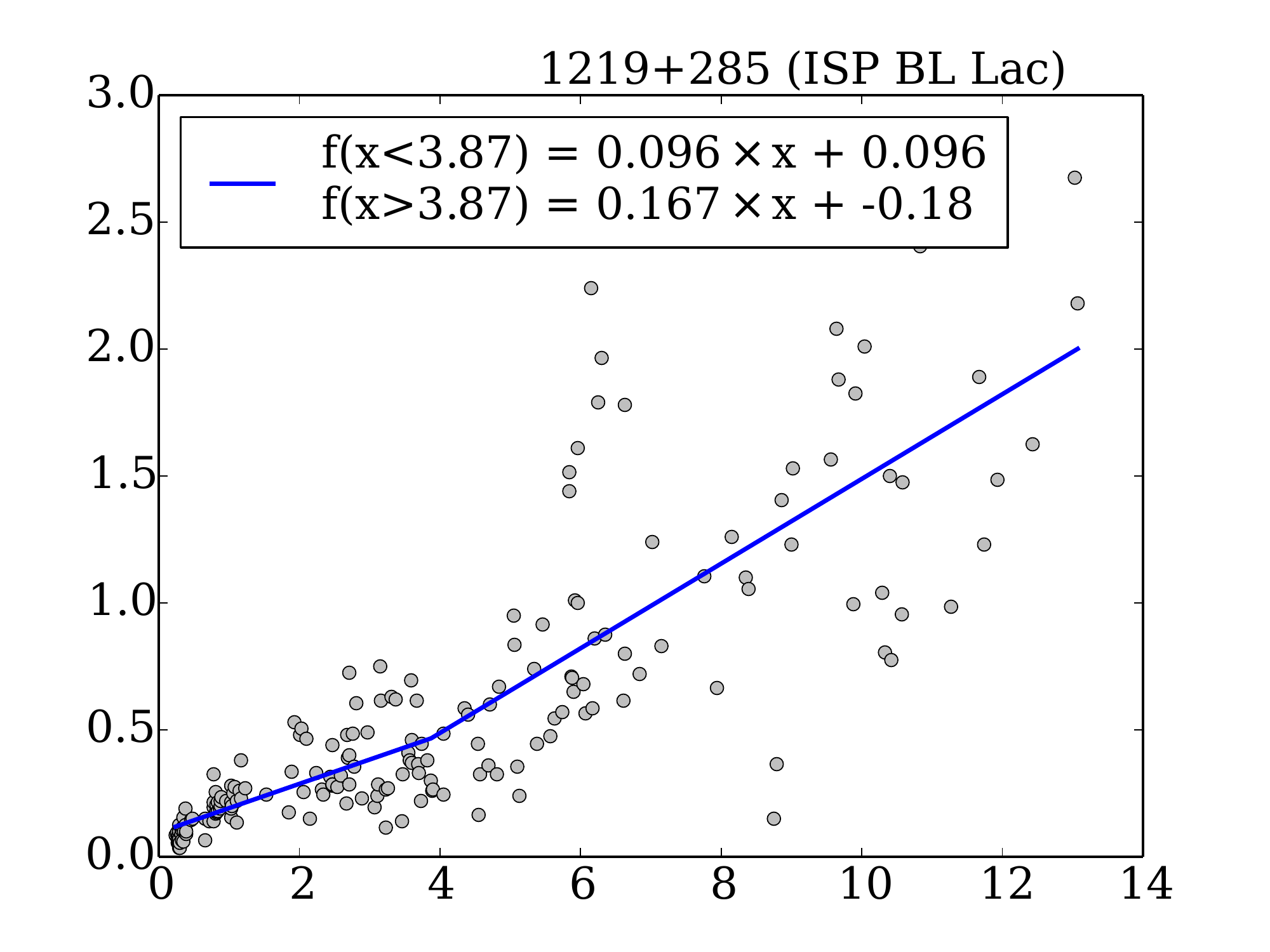}
\end{minipage}\hfill
\begin{minipage}[b]{0.33\linewidth}
 \centering \includegraphics[width=6.2cm]{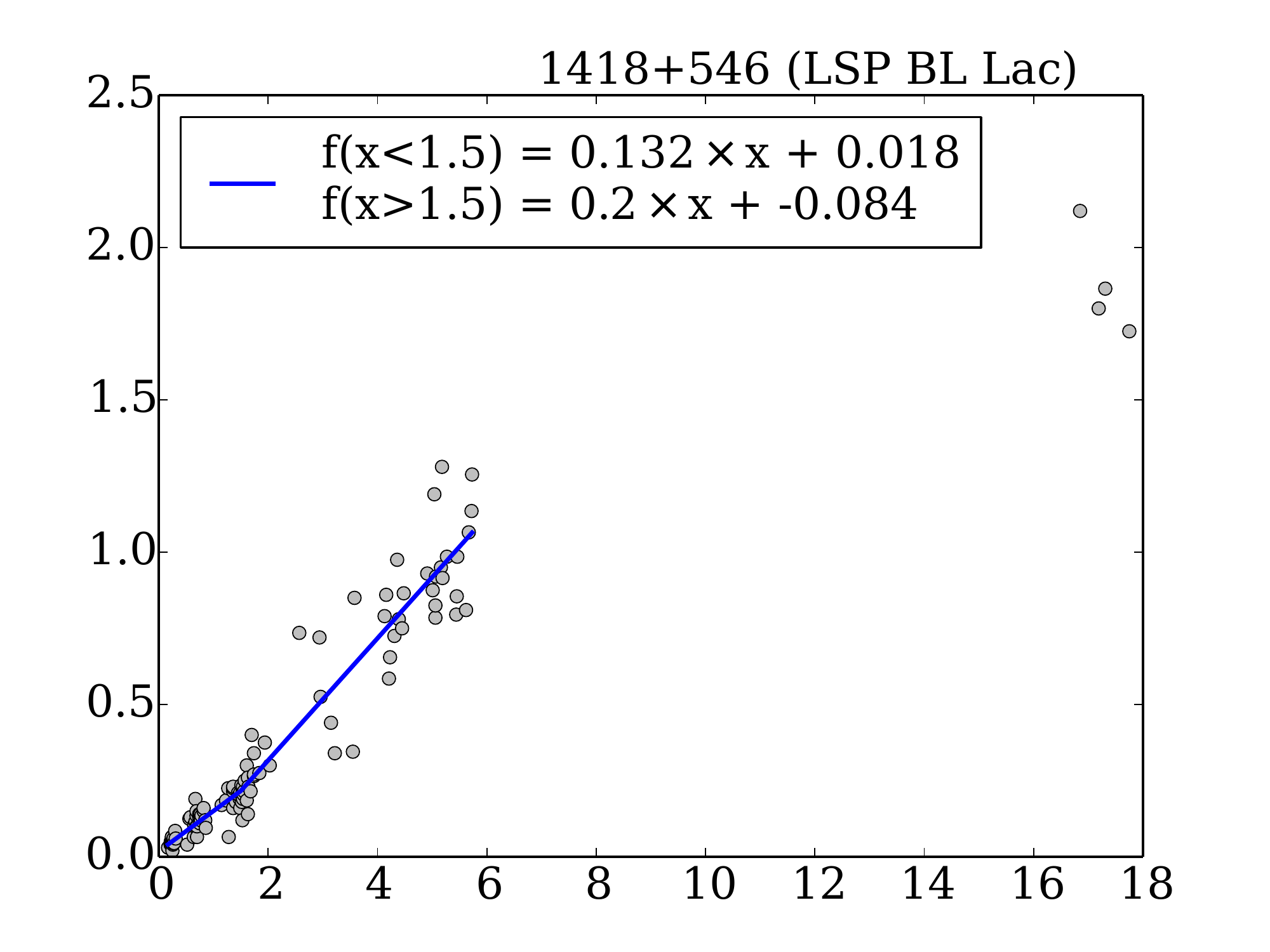}
\end{minipage}\hfill
\begin{minipage}[b]{0.33\linewidth}
 \centering \includegraphics[width=6.2cm]{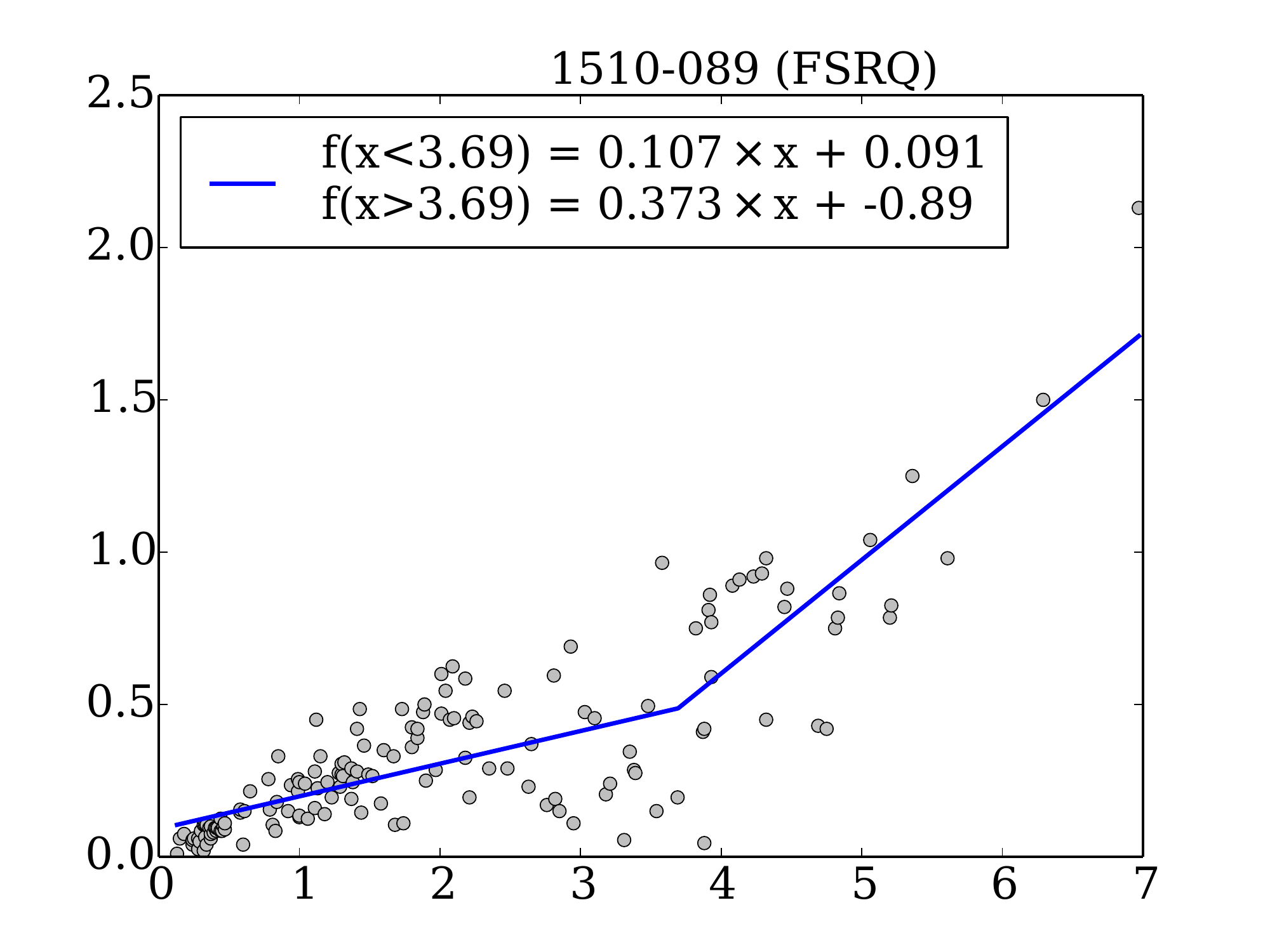}
\end{minipage}\hfill
\begin{minipage}[b]{0.33\linewidth}
 \centering \includegraphics[width=6.2cm]{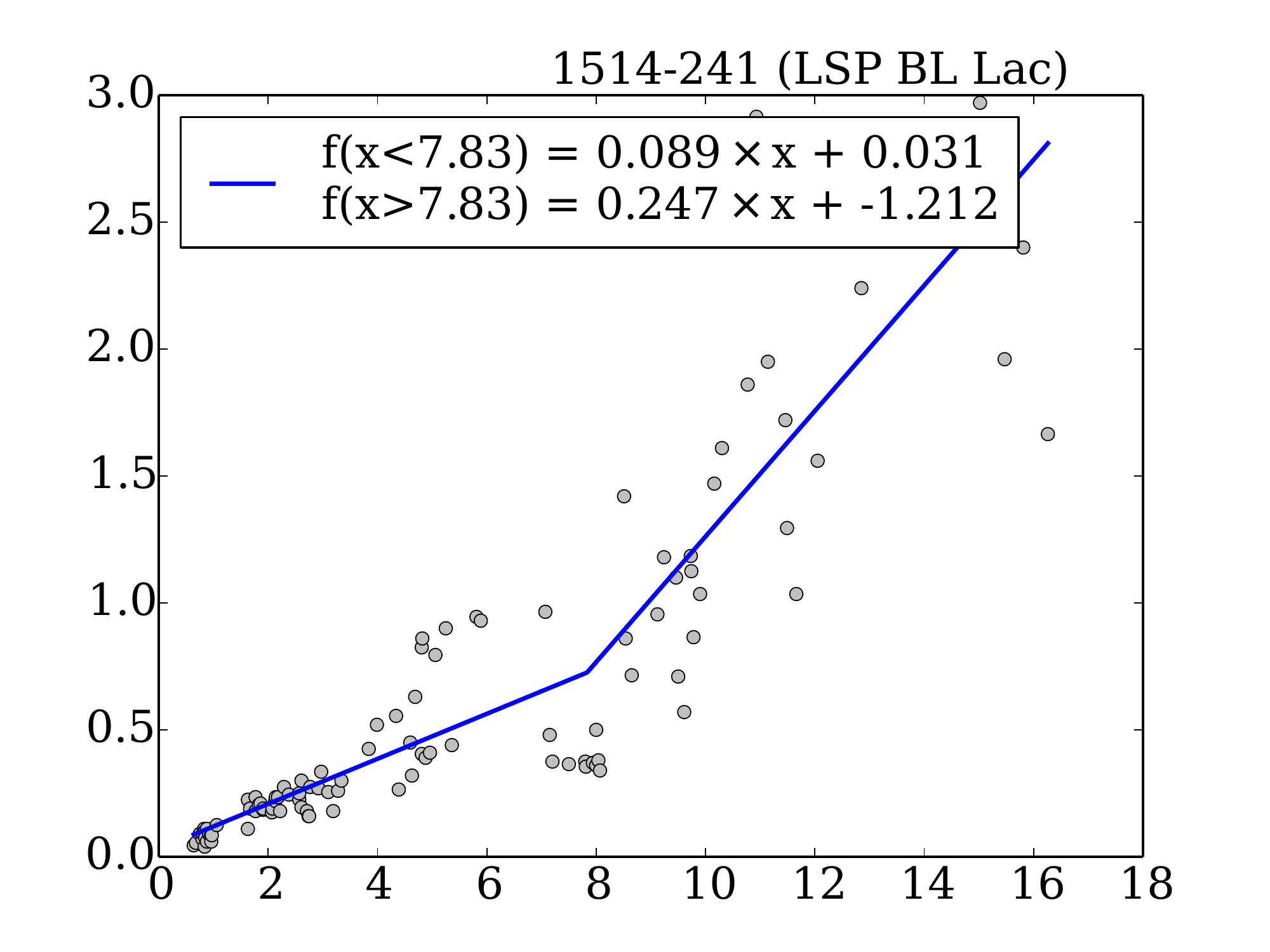}
\end{minipage}\hfill
\begin{minipage}[b]{0.33\linewidth}
 \centering \includegraphics[width=6.2cm]{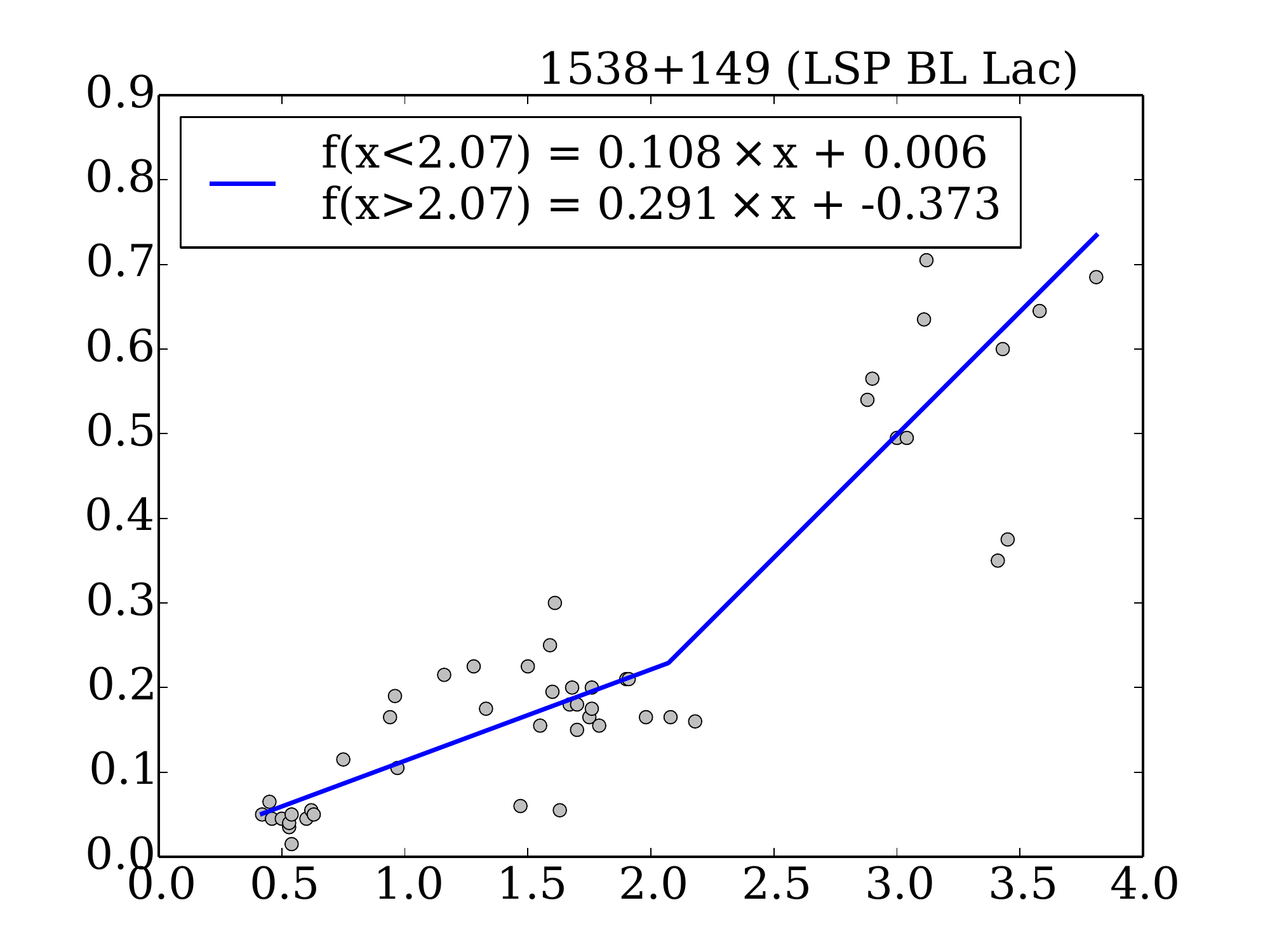}
\end{minipage}\hfill
\begin{minipage}[b]{0.33\linewidth}
 \centering \includegraphics[width=6.2cm]{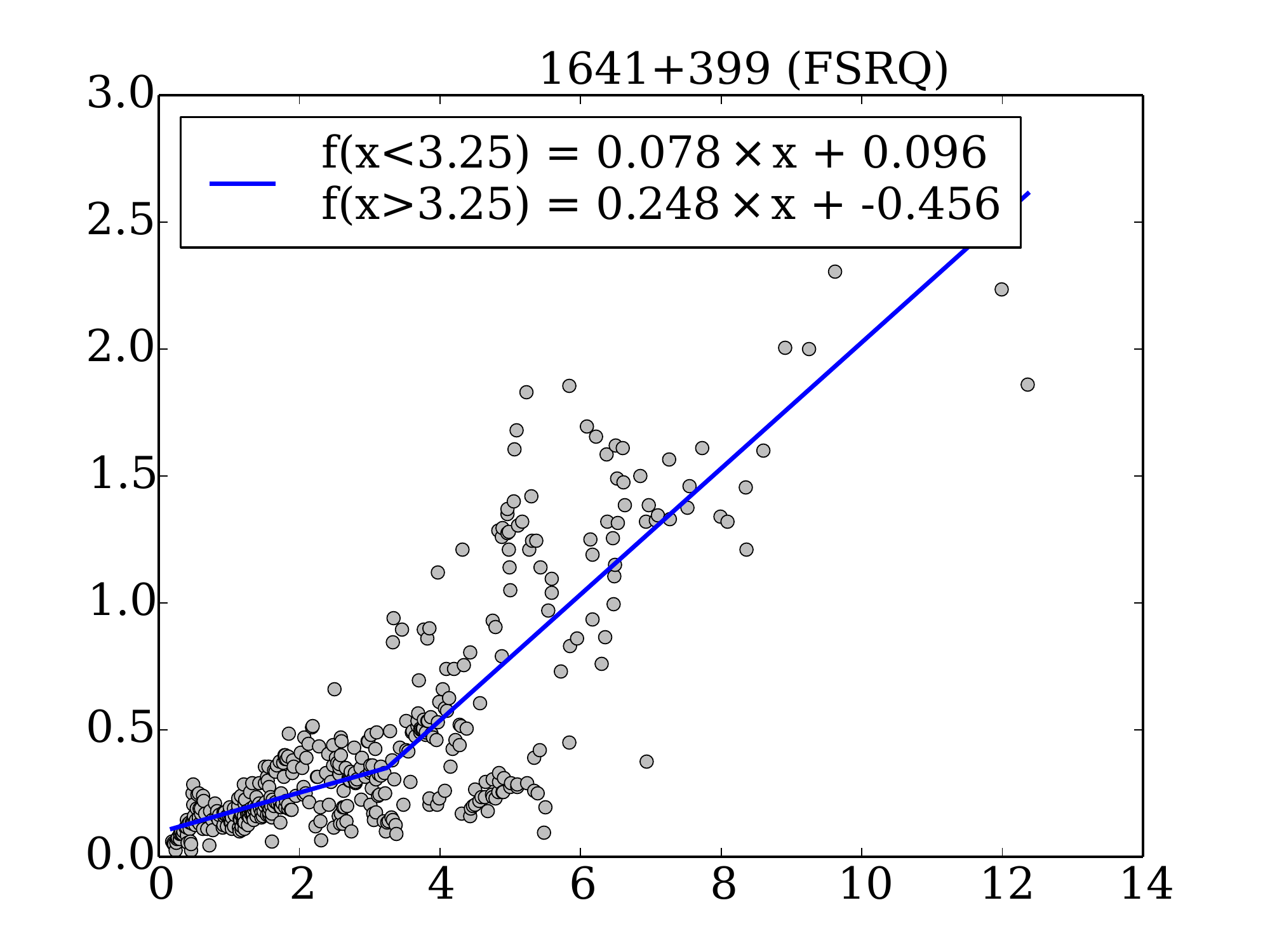}
\end{minipage}\hfill
\begin{minipage}[b]{0.33\linewidth}
 \centering \includegraphics[width=6.2cm]{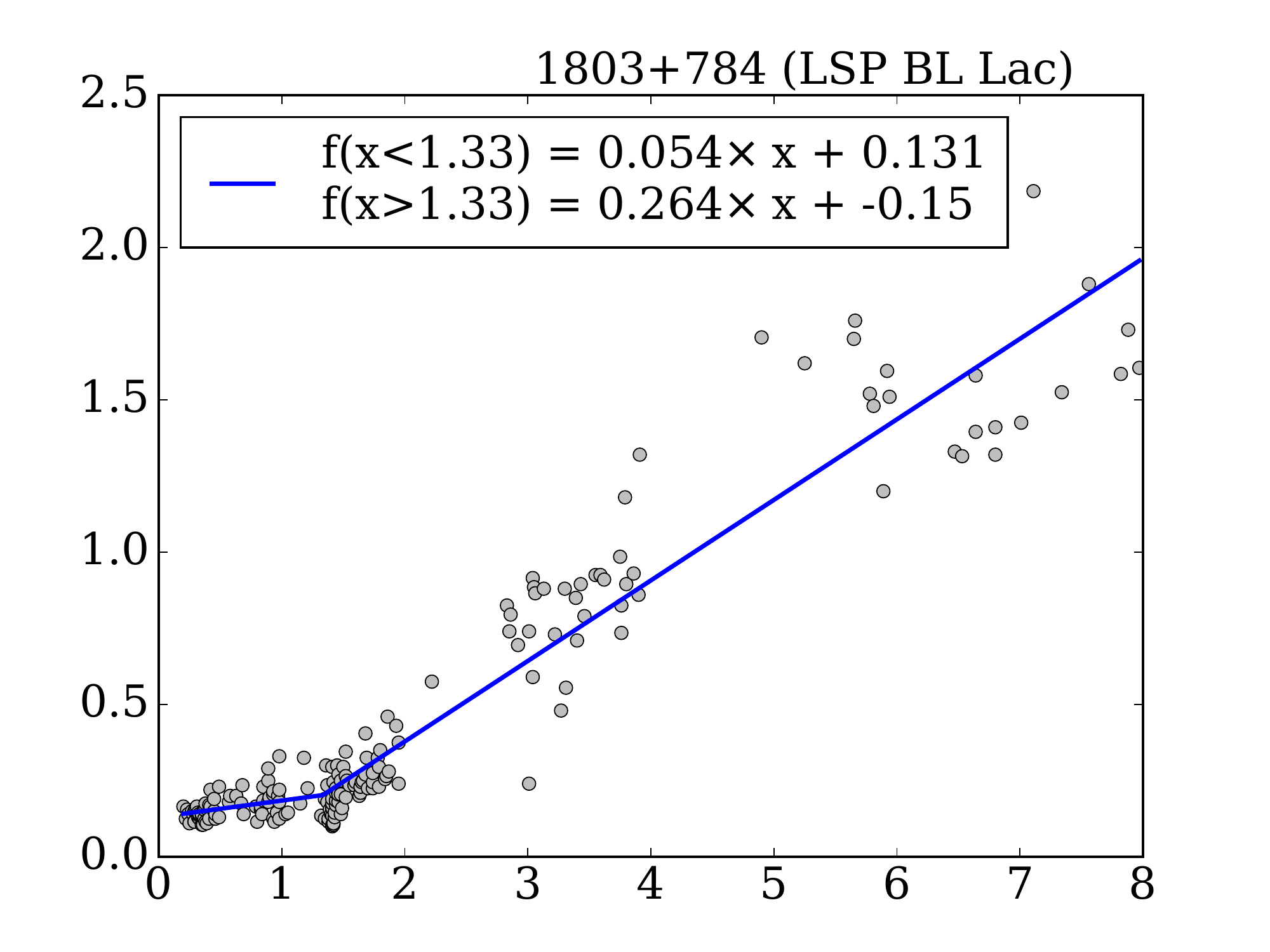}
\end{minipage}\hfill
\begin{minipage}[b]{0.33\linewidth}
 \centering \includegraphics[width=6.2cm]{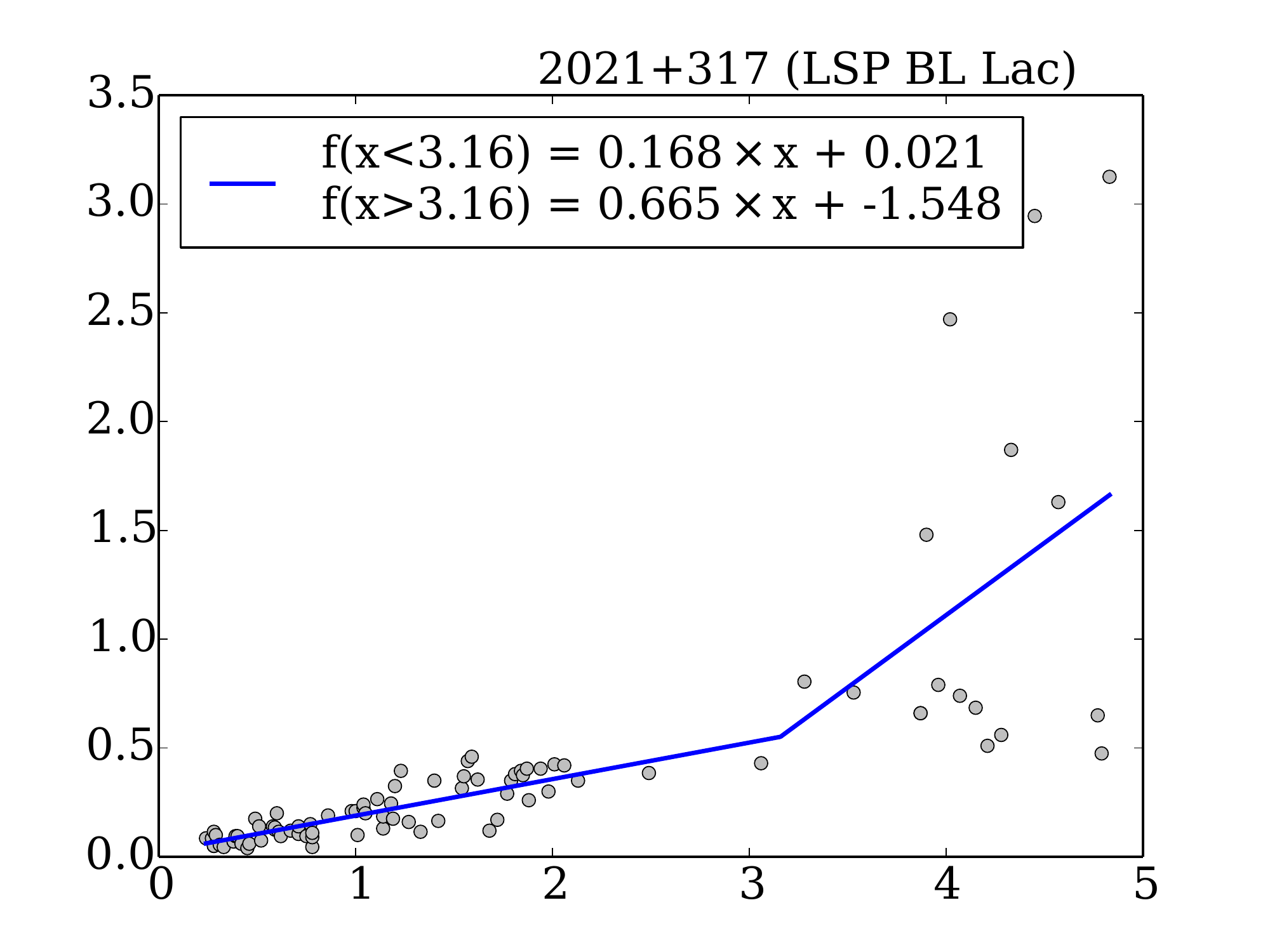}
\end{minipage}\hfill
\end{minipage}
 \put(-520,290){\rotatebox{90}{\makebox(0,0)[lb]{\large{Transverse size [mas]}}}}
 \put(-300,-10){\makebox(0,0)[lb]{\large{Core distance [mas]}}}
\end{center}
\end{figure*}

\begin{figure*}[h]
\begin{center}
\begin{minipage}[b]{\linewidth}
\begin{minipage}[b]{0.33\linewidth}
 \centering \includegraphics[width=6.2cm]{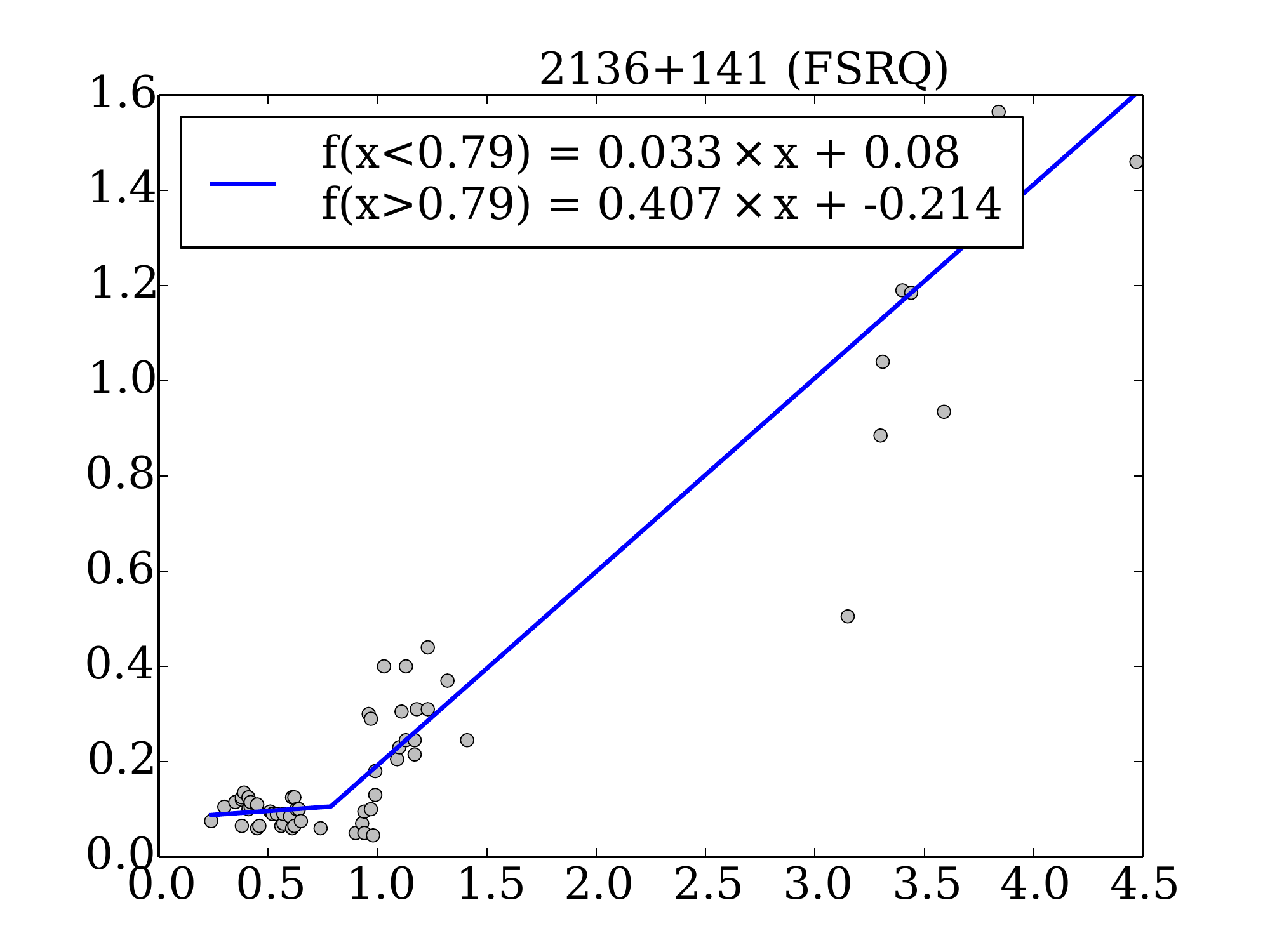}
\end{minipage}\hfill
\begin{minipage}[b]{0.33\linewidth}
 \centering \includegraphics[width=6.2cm]{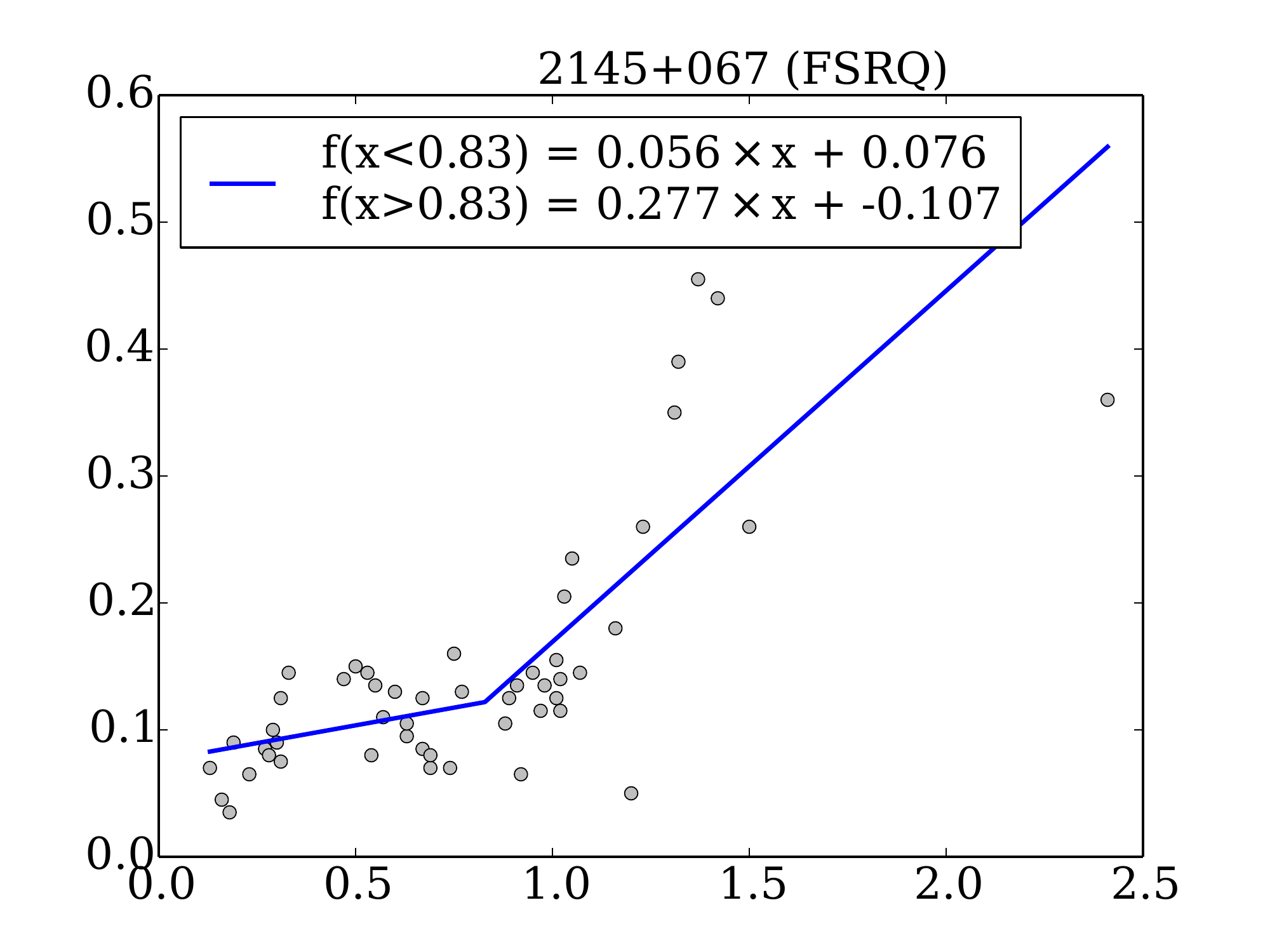}
\end{minipage}\hfill
\begin{minipage}[b]{0.33\linewidth}
 \centering \includegraphics[width=6.2cm]{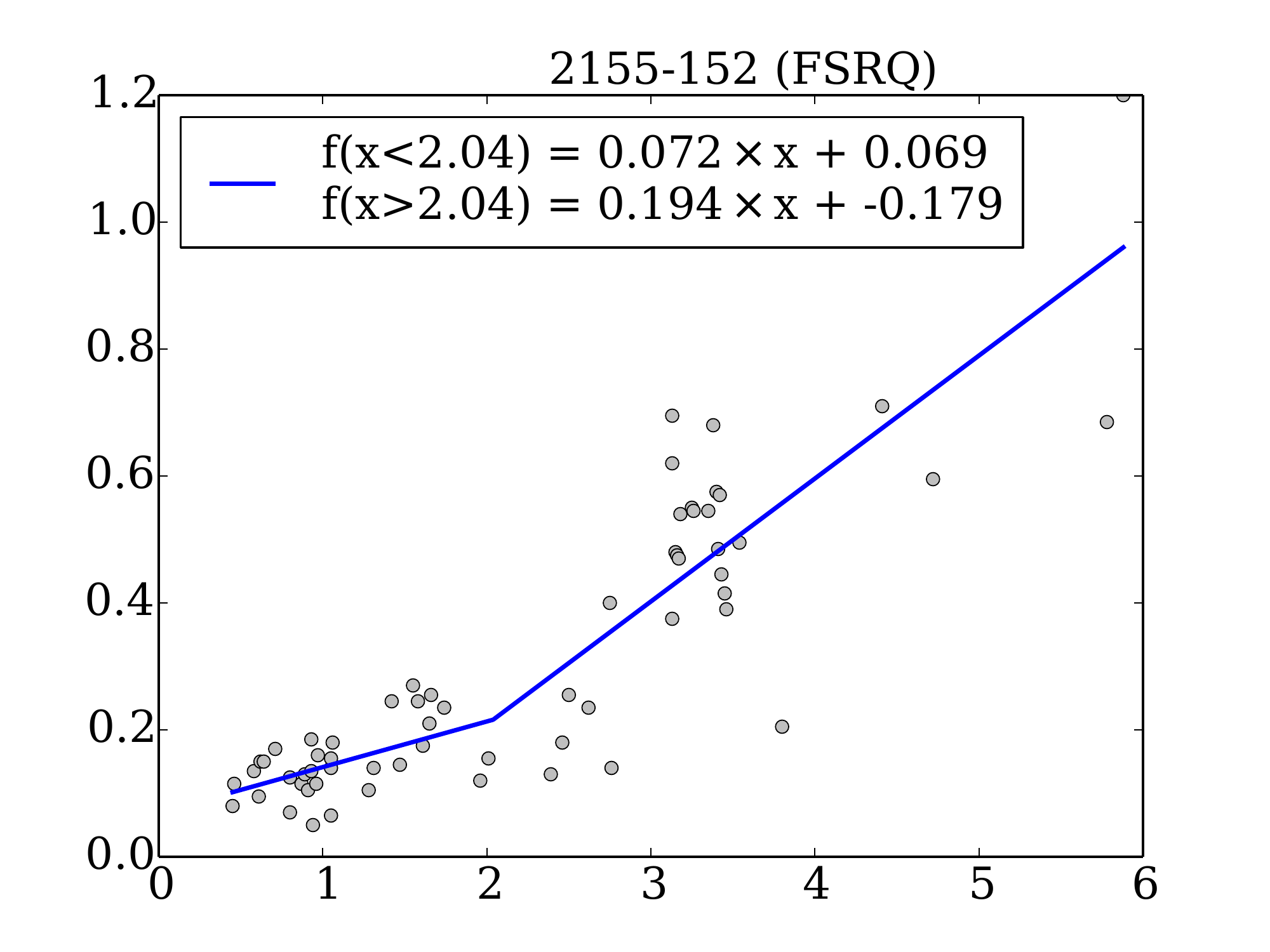}
\end{minipage}\hfill
\begin{minipage}[b]{0.33\linewidth}
 \centering \includegraphics[width=6.2cm]{Figures/Ouverture_jets_VLBI/jet_interne_2200+420.pdf}
\end{minipage}\hfill
\begin{minipage}[b]{0.33\linewidth}
 \centering \includegraphics[width=6.2cm]{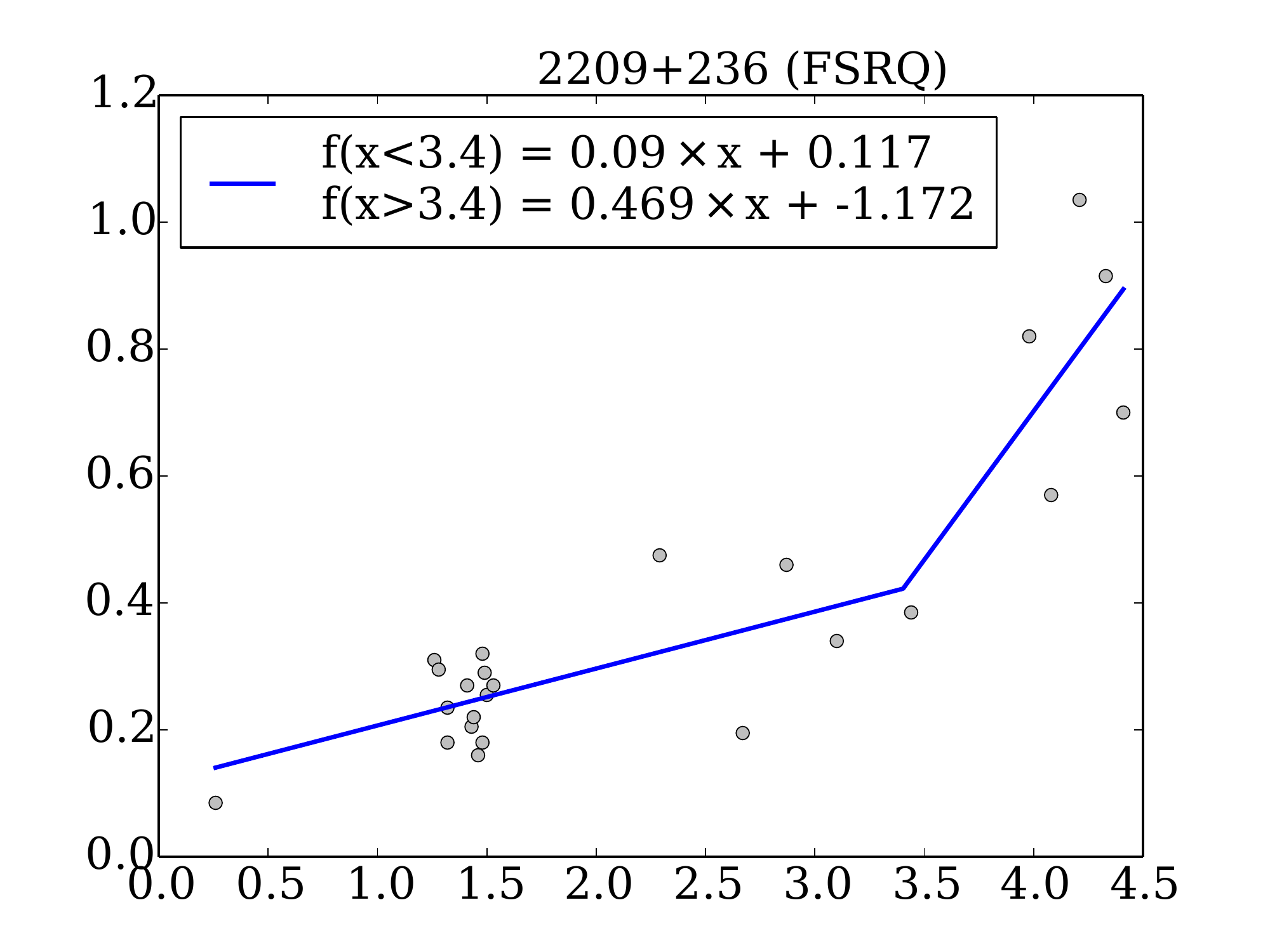}
\end{minipage}\hfill
\begin{minipage}[b]{0.33\linewidth}
 \centering \includegraphics[width=6.2cm]{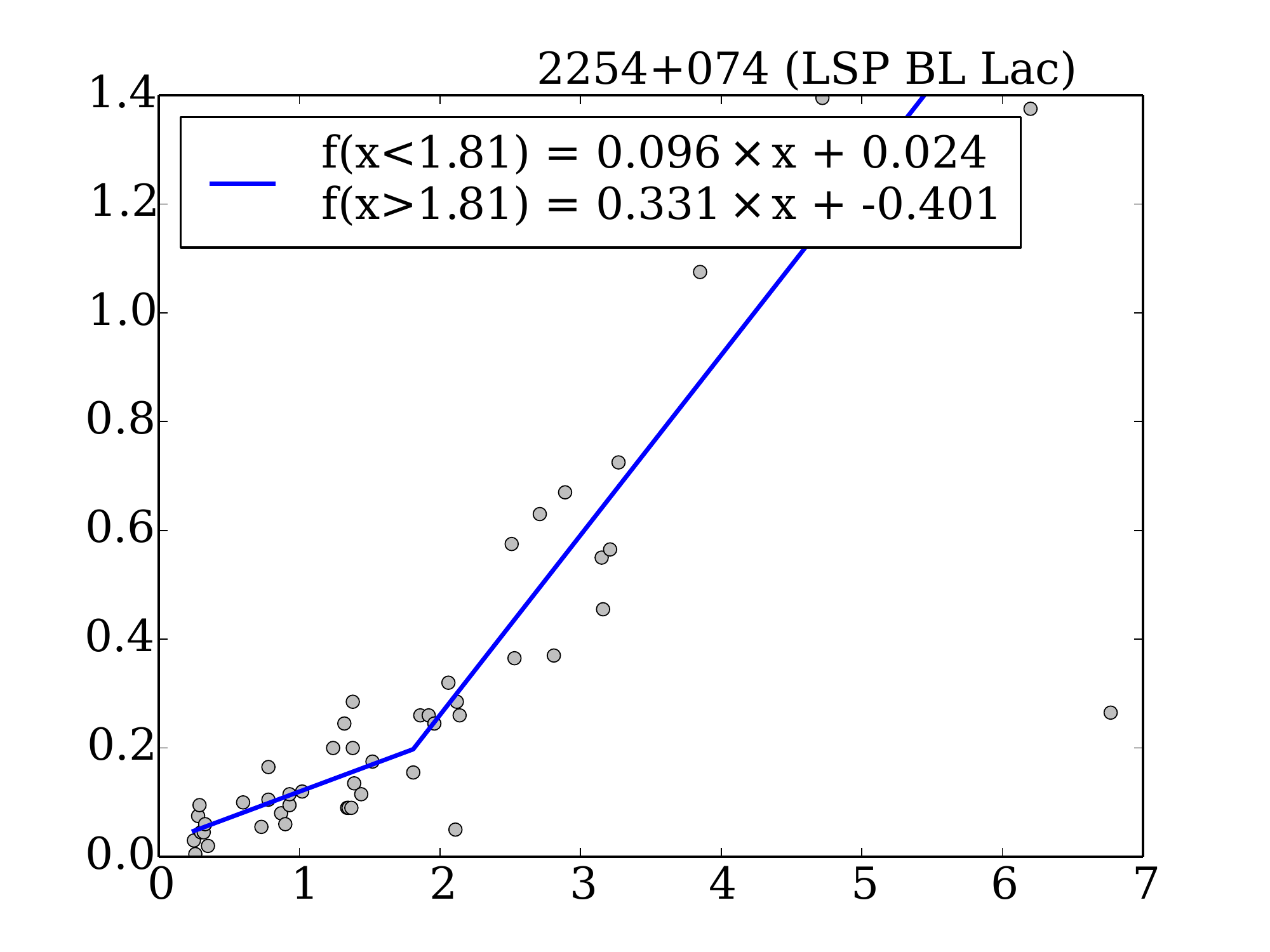}
\end{minipage}\hfill
\begin{minipage}[b]{0.33\linewidth}
 \centering \includegraphics[width=6.2cm]{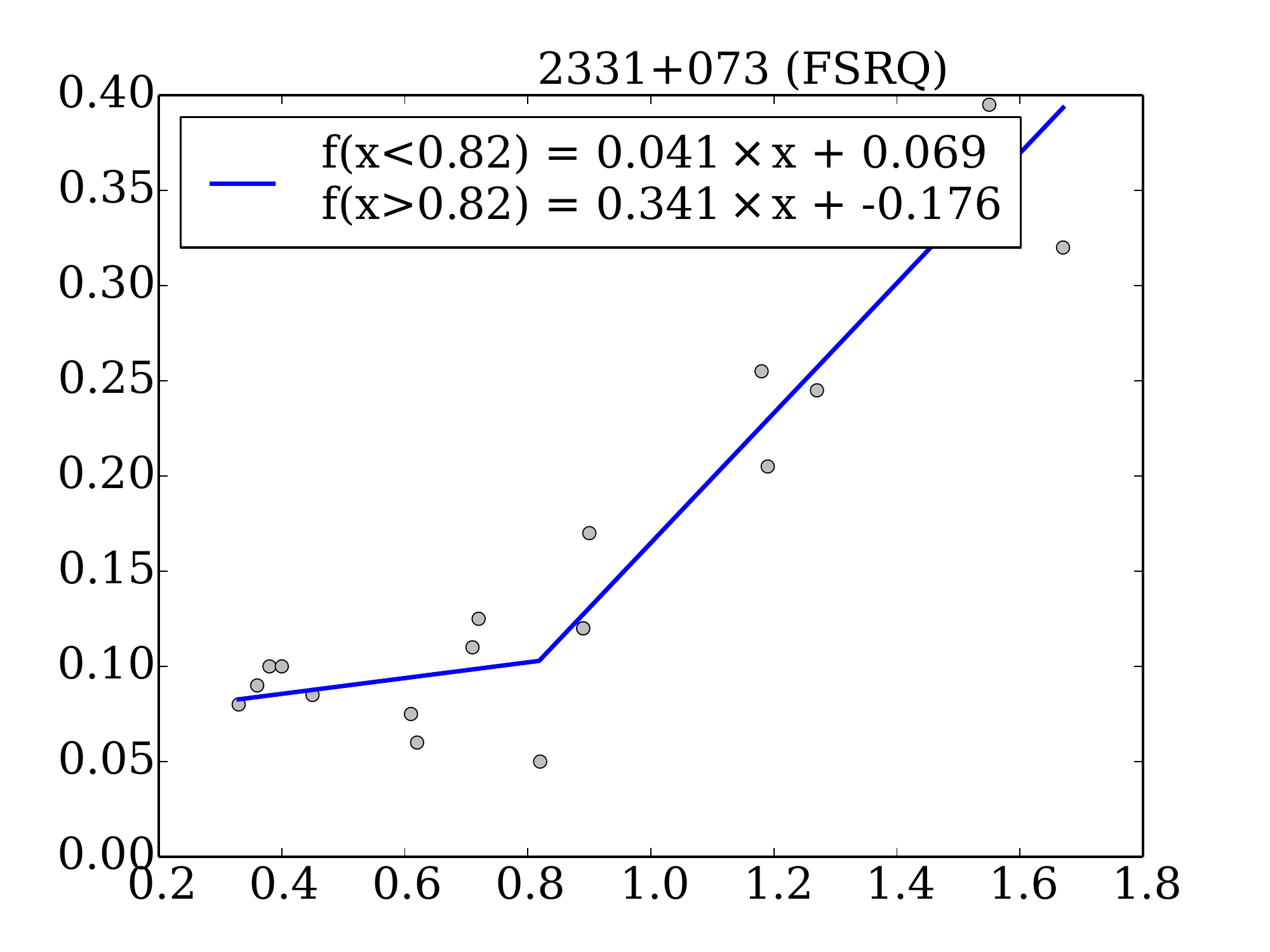}
\end{minipage}\hfill
\begin{minipage}[b]{0.33\linewidth}
 \centering \includegraphics[width=6.2cm]{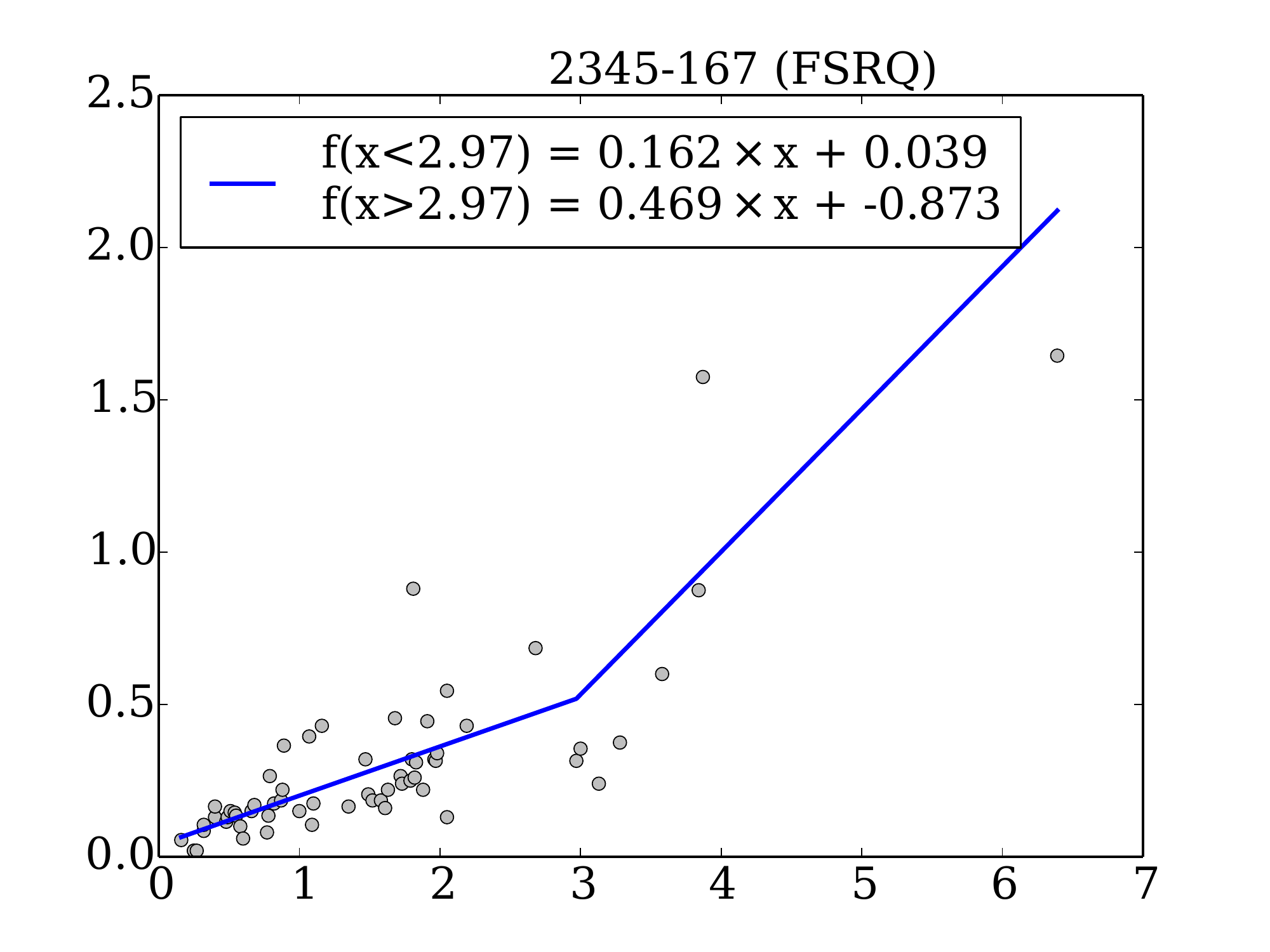}
\end{minipage}\hfill
\begin{minipage}[b]{0.33\linewidth}
 \centering \includegraphics[width=6.2cm]{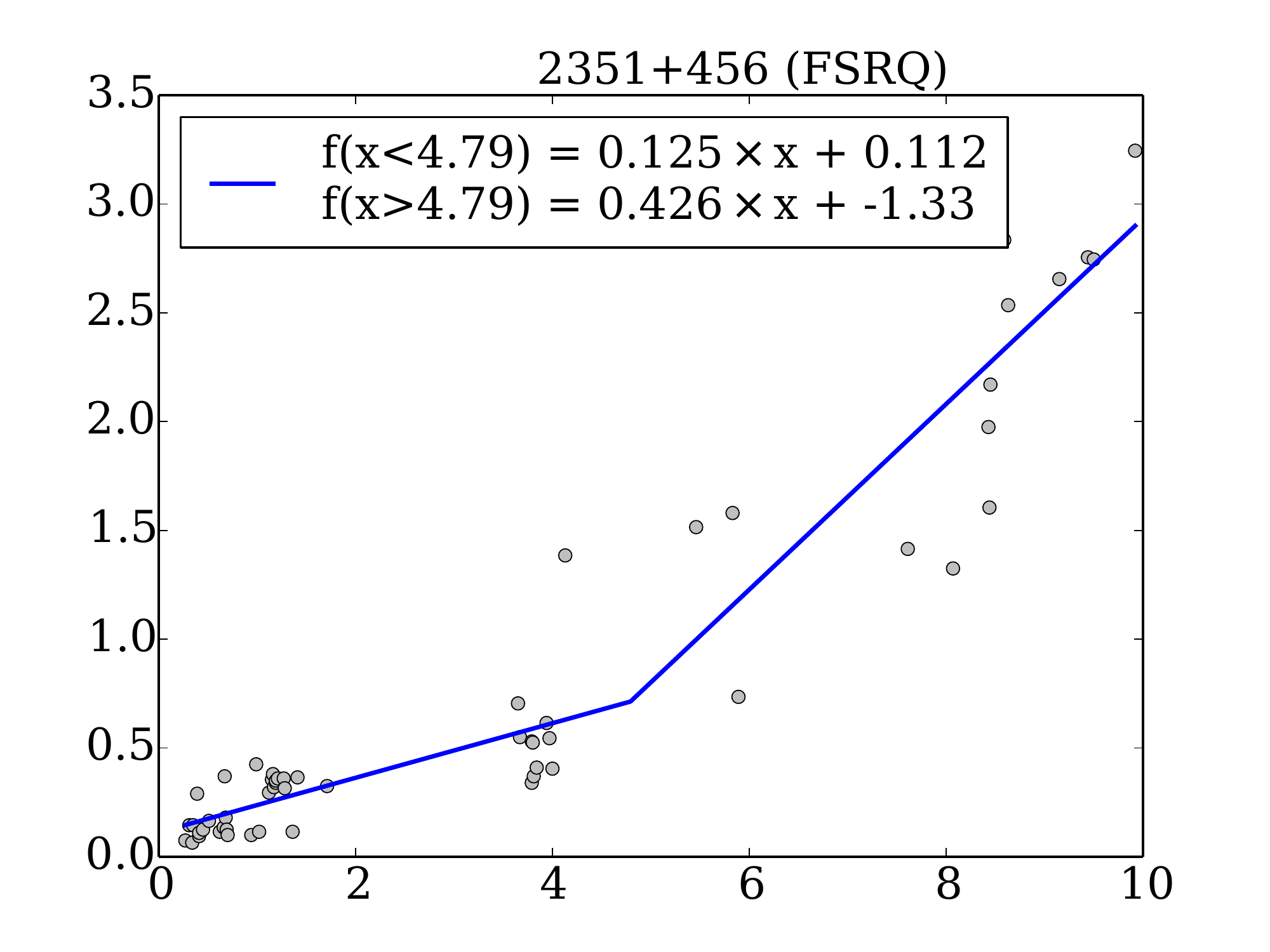}
\end{minipage}\hfill
\end{minipage}
 \put(-520,150){\rotatebox{90}{\makebox(0,0)[lb]{\large{Transverse size [mas]}}}}
 \put(-300,-10){\makebox(0,0)[lb]{\large{Core distance [mas]}}}
 \caption{Projected transverse size of the knots with their distance from the core fitted by a broken linear function (see Sect.~\ref{Sec:Aperture of VLBI inner jets}).}
\end{center}
\end{figure*}

\end{appendix}

\end{document}

%% file: TwoCjet2D_HD_lib.tex
\bibliographystyle{aa}